\documentclass{article}

\usepackage{amsmath}

\usepackage[pdftex]{hyperref,graphics} 
\usepackage{color} 
\begin{document}

\title{Nonlinear Classical Fields} 

\author{V. Radchenko\\email:{ rudchenkotmr@ukr.net}} 

\date{} 

\maketitle

\parskip 4 pt 

\abstract\large{{We regard a classical 
field as medium and so the additional 
parameter, the velocity of field, appears. 
If the one regards as potential  
then all self-energies become finite. 
Electromagnetic, mechanical, pionic and 
somewhat gluonic fields are regarding.}} \\

K e y w o r d s: classical fields, nonlinear fields

\tableofcontents{}

\section{Foundation}

Internal contradictions of classical field theory 
are well known. Mainly these are the infinite 
self-energies of Coulomb and Yukawa fields. 
In 1912 year G. Mie  considered the electron as state 
of electromagnetic field \cite{q} with long-range 
aim to eliminate the infinite self-energy
of Coulomb field.
His work gives strong impulse for 
development of field theory and  
it has been created many works in this 
direction. But always the models were 
contradicted at least to one of 
general physical principles. 
For example, in Born -Infield model 
the analyticity principle was not fulfilled. 
In work of a French physicist \cite{w} near 
1972 year the current of a  
field was 
constructed but without C-symmetry. 
Those hindrances come because it is impossible 
to build the fourvector of 
electromagnetic current within framework of 
electromagnetic field parameters only. 
The proofs of this impossibility can be found in  
most quantum electrodynamics textbooks, for example 
Landau-Lifschitz book (fourth paragraph). 

Hence one of the ways for construction 
of a model for electromagnetic field 
without internal contradictions is 
taking into account an interaction of 
electromagnetic field or with external particles, today 
it is usual approach, or with other fields. 
In area of quantum field theory the electromagnetic, 
week and strong fields were jointed to single model where the 
fields indirectly, via particles, interact between themselves. 

In classical field theory as 
well as in area of low energy nuclear physics 
the infinite self-energies of fields 
remain. In this article the extension
of classical field theory is considering, 
we aim to get an escape from infinite 
self-energies of classical fields. Using toolbox contains some not common 
elements - Clifford algebra, coherence condition, almost polynomial wave function. 
Short review of the ones is at the end of article.

Main idea is following. In general case any physical field has 
non-zero  density  of the mass. 
Therefore,  any field has  
additional parameter, 
$\color{blue}{U}$, it is the 
fourvector of the field velocity. 
The velocity, or fourvelocity for relativistic 
system, are usual parameters in physics, typically 
they are regarding as properties 
of the particle. About field velocity always 
implicitly is assumed that it is 
equal to light velocity. This is 
doubtful because, for example, the 
electrostatic 
field of a resting particle has the three 
velocity equal to zero. We will regard 
the fourvelocity of a field as  
local parameter and so 
$\color{blue}{U=U(x)}$. Only in case when 
the field is the continual variety 
of point-like not interacted between 
themselves particles the condition 
$\color{blue}{U^2 \equiv 1}$ is valid.  
In other words, the field is considering 
as special medium with two local 
parameters, which are the potential 
and fourvelocity of the field.  

However, as parameter the fourvelocity 
exists for any physical object and in 
any case it is essential quantity. At 
construction of any theoretical model 
for physical object, in today physics 
this means the lagrangian building, this 
parameter needs take into account. If 
this has done then either the fourvelocity 
is external parameter, similar 
to forces in Newtonian mechanics, or 
it is internal parameter. In last case 
the simple way is visible to reach up 
the complete system of differential 
equations. It is the 
consideration of $\color{blue}{U(x)}$ 
itself as a potential of some field. Then 
kinetic term,  
$\color{blue}{(\nabla U)^2 }$, in the  
lagrangian of a fields removes all 
problems with 
setting the full and close 
equation system. Below the local fourvelocity 
of a field is regarding as a 
potential of some field which we call w-field. 

Let us attempt to comprehend the 
physical meaning of this w-field. 
The tensions (forces) of the w-field are following 
(Clifford algebra \cite{e}, \cite{s}, \cite{a} always is using) 
\[\color{blue}{\nabla U=\nabla\cdot U+\nabla\wedge U}\] 
\[\color{blue}{\nabla\cdot U=\partial_0 U_0+\overrightarrow\nabla\cdot\vec U}\] 
\[\color{blue}{\nabla\wedge U=-\partial_0\vec U-\overrightarrow\nabla U_0+i_c \overrightarrow\nabla\times\vec U}\] 
In case  $\color{blue}{U^2=1}$ the quantity  
$\color{blue}{-\partial_t\vec U}$ is 
usual mechanical acceleration with 
opposite sign. Hence the w-field forces 
contain inertial forces. In typical case the square of fourvelocity is not 
equal to unit. The examples are: for scalar photons (these are electrostatic 
fields) $\color{blue}{U^2>0}$ , for longitudinal photons (these are magnetic fields) $\color{blue}{U^2<0}$. 
We may feel almost certainly  that by 
physical meaning the w-field is bearer 
of inertial (typically virtual) 
forces and expect that these forces 
create barrier which do not permit the
concentration of a field to point-like 
object with infinite self-energy as 
it take place within standard  theories for   
interaction of the particles via fields. In my opinion, 
all infinities of field theories has emerged from or 
self-interaction of the field or because of absence any interaction. 
For this reason below the self-interaction parts of the lagrangian 
will be discarded, instead the interaction of the field with own shadow 
field, that we call w-field, is regarding; faster of all few fields have 
common w-field by means of which they interact.     

Internal contradictions of classical 
field theory 
are sufficient for introducing of new 
essence. If we do not wish to do this 
then w-field may be regarded as representation,
of course incomplete, of week forces in 
area of classical physics. For simplest 
approach, used not only in classical physics, 
the fourvelocity of the field is an external parameter.  

If take into account the existence of 
w-field then it is not hard to build 
the lagrangian of fields (in other 
terminology - the Lagrange density 
of fields). 

\section{Nonlinear electromagnetic field}
\subsection{Lagrangian, current, Maxwell's equations} 

The lagrangian \cite{r} of the pair interacting between themselves 
fields can be written as following 
\[\color{blue}{L=L_e+L_u+L_{int}}\] 
We regard the electromagnetic field with 
its shadow, the w-field, in area that 
does not contain any particle. As 
usually, this area is the whole space-time. 

Fourpotential of electromagnetic 
field, $\color{blue}{A(x)}$, with 
tensions 

\[\color{blue}{F=\nabla\wedge A=\vec E +i_c \vec H}\]
is restricted by Lorentz gauge condition $\color{blue}{\nabla\cdot A=0}$ 

The lagrangian of free electromagnetic field is well known, namely 
\[\color{blue}{ 
L_e=\frac{F^2}{8\pi c}} 
\] 
Here in general case it needs denominate 
the Dirac conjugation above of one multiplier, 
for simplicity this sign is omitting. 

The lagrangian for free w-field let us take in similar form 
\[\color{blue}{L_u \sim (\nabla U)(\nabla U)}\] 
For lagrangian of free electromagnetic field the quadratic dependence upon tensions stems from experimental Coulomb low. But for free w-field 
such form of the lagrangian is the assumption and it is strong assumption. 

At first look about the interaction lagrangian nothing is known; in general case  it is true but for electromagnetic field it is known enough. 
The fourvector 

\[\color{blue}{ 
\frac{\delta L_{int}}{\delta A}=J} 
\] 
is the density of current for electromagnetic 
field. For simplicity below we call this 
fourvector or current, or jet. It is the 
source of electromagnetic field. 

From Bohr correspondence principle the 
jet is linear function 
of fourvelocity and at all transformations 
the current has exactly the 
properties of fourvelocity. General 
form of such quantity is following 

\[\color{blue}{J=c_1 U+c_2(FU-UF)+c_3FUF}\] 
where $\color{blue}{c_n}$ are scalar 
functions which are independent upon 
velocity four vector. From gradient 
symmetry of electromagnetic field 
the ones depend on electromagnetic 
field tension $\color{blue}{F}$ only. 

The phase of any physical quantity is 
relative. There can be done some deviation from 
usual description of phase transformation. Namely, 
within Clifford algebra we take the global phase as 
$\color{blue}{exp(-i_c\varphi)}$. Because the matrices 
$\color{blue}{i_c\gamma_{\mu}+\gamma_{\mu}i_c=0}$ those condition 
gives two restrictions. First limitation is 
$\color{blue}{c_2 \equiv 0}$. 
Second restriction is for 
argument of the coefficients - in general case  $\color{blue}{\vec E\cdot \vec H}$ 
is not zero so the coefficients 
$\color{blue}{c_n}$ depend on the fourth degree of tensions. 

For system without charged particles the condition 
$\color{blue}{c_1(0)=0}$ is 
valid. For usual transverse waves 
the current of the field is equal to zero 
that at once confirms this restriction. 

Then in first not trivial approach the 
jet of electromagnetic field can be written 
as
\[\color{blue}{ 
J=\frac{c}{4\pi  g}FUF}\] 
where the constant $\color{blue}g$  
determinate the scale of 
electromagnetic potential. This jet of 
electromagnetic field 
needs to compare with usual, 
$\color{blue}{eU}$, jet of a particle 
if the questions about C-symmetry arise. 

For current of w-field, 
$\color{blue}{{\delta L_u}/{\delta U}}$, 
all above arguments are true. Only 
the restriction on first coefficient 
is questionable, the interaction 
lagrangian may contain the term 
$\color{blue}{L_u\sim U^2}$. For 
simplicity the self-interaction of w-field 
is neglected, only the interaction of w-field with 
electromagnetic jet holds. 

Fourvelocity is dimensionless 
quantity so it is convenient to take the potential of electromagnetic field in dimensionless form as 
\[\color{blue}{A(g,x)=gA(y=\frac{x}{a})}\] 
where $\color{blue}{g,a}$ are the 
scales of the potential and length. 

Correspondingly, the simplest action 
for electromagnetic field, which has 
the w-field as shadow and does not contain 
any particle, is following 

$$\color{blue}{S=}$$
$$\color{blue}{=\frac{e^2}{8\pi c}\int\left [(\nabla F)^2+k^2(\nabla U)^2+2A\cdot (FUF)+qU\cdot(FUF)\right] d^ 4 y}$$ 
where all quantities after integration 
sign, including the variables of 
integration, are dimensionless. 

This action does not contradict with 
any general physical principle. 
The scale invariance of this lagrangian is 
general property of any model for fields 
without external particles. By construction, 
the validity area of this lagrangian 
is $\color{blue}{O(F^4)}$ and, because 
the scale parameter is external quantity, 
for description of macroscopic as well 
as microscopic systems the model 
can be applied.   

Owing to constriction the jet is 
not variable by electromagnetic potential 
and the Maxwell nonlinear equations for 
dimensionless quantities are following: 

\noindent in four-dimensional form
\[\color{blue}{\nabla F=FUF}\]
or, the same, component-wise
\[\color{blue}{\nabla\cdot F=FUF}\]
\[\color{blue}{\nabla\wedge F=0}\]
or, the same, in usual three-vector form 
\[\color{blue}{-\overrightarrow\nabla\cdot\vec E=u_0(E^2+H^2)+2\vec u\cdot(\vec H\times\vec E)}\] 
 $$\color{blue}{-\partial_0\vec E+\overrightarrow\nabla\times\vec H=}$$
$$\color{blue}{=\vec u(E^2+H^2)+2u_0(\vec H\times\vec E)-2\vec E(\vec u\cdot\vec E)- 
2\vec H(\vec u\cdot\vec H)}$$ 
\[\color{blue}{\partial_0\vec H+\overrightarrow\nabla\times\vec E=0}\] 
\[\color{blue}{\overrightarrow\nabla\cdot \vec H=0}\]
There always exist the solutions with 
potentials equal to constants anywhere. 
By physical meaning the ones are the vacuum 
states of fields. 

The bevector of electro\-magne\-tic ten\-sion,  
Max\-well equa\-ti\-ons 
and  Lo\-rentz gauge condition are not 
changeable at transformation 
\[\color{blue}{A \to A + constant}\] 
Therefore for pure electromagnetic 
system the value of electrostatic 
potential on infinity is free number. 

Using the coherence condition this
equation system can make to be closed. 

\subsection{Closing  equations of the field} 

Simplest state of electromagnetic  
field is that where the w-field is in 
vacuum state. In this case, 
taking into account the equality
 \[\color{blue}{A\cdot \big(FUF\big)=U\cdot \big(FAF\big)}\]
the equation for fourvelocity is 
\[\color{blue}{\nabla^2U=0}\]
Because this is the isolated 
equation the coherence condition take in form 
\[\color{blue}{\int (\delta U\cdot(FAF+qFUF)) d^4 y=0}\] 
\[\color{blue}{\int \delta_A J\cdot(2A+qU) d^4 y=0}\] 

Next states we call the coherent 
states. In this case the equation for 
fourvelocity is 
\[\color{blue}{k^2\nabla^2U=FAF}\] 
with restrictions 
\[\color{blue}{\int \left[(2A+qU)\cdot\delta_AJ+qU\cdot\delta_U J\right] d^4y=0}\] 
Note, that here $\color{blue}{\delta_A}$ as well as $\color{blue}{\delta_U}$ are the variations of the solutions of field equations. 
All other states involve in equation 
system the self-interaction of w-field, 
for simplicity we discard the ones here.

In this way the system of differential 
equations is completed. 

For applications let us regard firstly the nonlinear Coulomb field. 

\subsection{Nonlinear Coulomb field} 

This is spherically symmetrical electrostatic 
field. Dimensionless potentials and the space 
variable of the field denominate as following 
\[\color{blue}{s(x)=A_0(x),\;u(x)=U_0(x),\;{  } 
x=\frac{a}{ R}}\] 
In this case the Maxwell equation is following 

\[\color{blue}{s''=us'^2}\] 

This electrostatic field has nonzero 
electric charge and non-zero density 
of electric charge as well as nonzero mass what are the new 
phenomena not only for theory of classical fields. 

In case of a system with full positive electric charge equal to unit the solution for tension is following 
\[\color{blue}{ 
s'=exp\int_{s_0}^s  u(s) ds =[1-\int_0^x u(x) dx]^{-1}} 
\] 
In physical area, it is the area of 
big distances, the electrostatic potential is following 
\[\color{blue}{ 
s=s_0+x+s_2\frac{x^2}{2}+s_3 \frac{x^3}{6} +..} 
\]
or for physical potential of the system 
with electric charge 
$\color{blue}{e}$ 

\[\color{blue}{ 
A_0=\frac{es_0}{a}+\frac{e}{ R}+\frac{eas_2}{2R^2}+..}\]
In this expression the coefficients $\color{blue}{s_i}$ depend  upon interaction 
constant of free w-field and from vacuum 
potential of  electrostatic  field. 
So implicitly the vacuum potential of 
electrostatic field is observable. In this point the 
nonlinear model differ from linear one where vacuum potential is free number. Of course, for different states these are different constants. 

When the w-field is in vacuum state the equation 
for scalar velocity is the free field 
equation, namely
\[\color{blue}{u''=0}\] 
\[\color{blue}{u=u_0+bx}\]
Any physical field has finite 
self-energy. Hence the constant 
$\color{blue}{b=0}$ 
and electrostatic potential in this state is 
\[\color{blue}{\displaystyle s=s_0-\frac{1}{
 u_0}ln(1-u_0 x);\;\; u_0<0}\] 
Coherence conditions for these states 
are 
\[\color{blue}{\int (s+qu_0)s'^2 \delta u_0 dx=0}\] 
\[\color{blue}{\int (2s+qu_0)s'\delta s' dx=0}\]
then 
\[\color{blue}{u_0(s_0+qu_0)=1}\] 
\[\color{blue}{u_0(2s_0+qu_0)=3}\] 

For system with positive 
electric charge the scalar velocity  
$\color{blue}{u_0=-1}$.  
So $\color{blue}{s_0=-2}$ and 
$\color{blue}{q=-1}$. This is very 
exotic state, below only the coherent 
states of electromagnetic field are 
regarding . 

For coherent states the equation and solutions for scalar velocity are 
\[\color{blue}{k^2 u''=s s'^2}\] 
\[\color{blue}{ 
u=-\frac{s}{ k}+2\sum_n\frac{1}{ s-s_n}}\] 
with the equations 
\[\color{blue}{ 
s_n=2k\sum_{i\neq n}\frac{1}{s_n-s_i}} 
\] 
for determination of Hermitian 
numbers $\color{blue}{s_n}$, while the 
electrostatic tensions of these 
fields are 
\[\color{blue}{ 
s'\sim \prod_{0}^{N}(s-s_n)^2 exp(-\frac{s^2}{2k}); \;\;s'(0)=1} 
\] 
so there are the variety of states similar to Glauber states.  

For these coherent states must be $\color{blue}{|\varphi_{\infty})|>|\varphi_N|}$  
then the self-energy of free w-field is finite number.

From physical reason, any motion on 
infinity is free motion. 
Correspondingly, the scalar velocity,  
$\color{blue}{u}$, on infinity has one 
of three possible values, namely  
$\color{blue}{u_+=\{-1, 0, +1\} }$. 
However, more general restriction, $\color{blue}{U^2>0}$ 
is possible.
   
For states with positive electric 
charge  $\color{blue}
{du/ds <0}$ 
and so in case $\color{blue}{u_+=+1}$ 
these states have complicated 
electromagnetic structure because 
the density of electric charge change 
the sign in internal area of field. Note, those
signs stem from convention on the signs in 
lagrangian and can be changed.

Another example is the nonlinear 
electrostatic field of light nucleus. 
In nucleus the scalar velocity, 
$\color{blue}{u}$, depends upon 
parameters of the electromagnetic and
strong fields. In light nucleus the 
motion of nucleons is governed 
by strong interaction mainly. The 
distribution of nucleons is almost 
constant and their velocities are small 
compare with light velocity. Correspondingly,
the scalar velocity of electromagnetic 
field in light nucleus  
is constant for first approach.  
In other worlds, in equation for 
determination of electrostatic  
potential the scalar velocity 
is external 
parameter which is equal to zero in external 
area and it is the constant in internal 
area of nucleus. However, the parameter
$\color{blue}{u}$ is not necessary 
equal to unit because a nucleon moves  
in cloud of virtual bosons but not 
in empty space. If $\color{blue}{b}$ 
is the radius of light nucleus then 
the electrostatic tension is 
\[\color{blue}{E=\frac{eZ}{ R}\qquad{  u_0=0 }\qquad  R>b }\] 
\[\color{blue}{E=\frac{d}{R(R+b)} \qquad u_0=const \qquad R<b}\] 
By  physical meaning the
constant $\color{blue}
{d}$ is the 
polarization of internal nucleus medium. 

For that reason the distribution of 
electric charge in light nuclei at 
small distances essentially differs 
from the one in heavy nuclei. 

Remark that for nonlinear Coulomb 
field the Earnshaw theorem can be not 
valid because the laplacian of 
electrostatic potential is not equal to zero. 

For theory application let us regard the electron levels 
in nonlinear, and linear for comparison, Coulomb 
field and nonlinear electromagnetic waves.

But firstly the possible quantization of electric charge 
emerging from discrete type of coherent states is 
considering. 

\subsection{The quanta of electric charge}

In the expression for tension of electrostatic field which 
is in coherent state, the integration constant was taken in 
the form that ensure equal electric charge for any of those 
states. 

Now the integration constant is keeping as arbitrary number 
and, for convenience, let us redefine the tension as 
$\color{blue}{\varphi=h\sqrt{k}}$.
In this case the expression for h-tension is following
\[\color{blue}{h'=\frac{C}{R^2}k^N\prod_{i=0}^N(h-h_{i})^2exp(-\frac{h^2}{2})}\] 
Correspondingly, the electric charge of N-state is following
\[\color{blue}{q_{N}=Ck^N\prod_{i=0}^N(h_{\infty}-h_{i})^2exp(-\frac{h_{\infty}^2}{2})}\] 
As next step of simplification the multiplier 
\[\color{blue}{Ck^Nexp(-\frac{h_{\infty}^2}{2})}\] 
put equal to unit and so 
 \[\color{blue}{q_N=\prod_{i=0}^N(h_{\infty}-h_{i})^2}\] 
with condition $\color{blue}{q_0=1}$ because the interaction 
constant of free electric field, the elementary charge 
$\color{blue}{e}$, was previously factored out  
from action as scale of charge. 

Next simplification enters into the simplification 
process  naturally. In classical as well as 
in quantum physics the commonly accepted convention is that 
the value of vacuum potential of electrostatic field can be 
taken equal to zero; we apply this convention as first 
approximation. In result the electric 
charge of N-state is determined by hermitian numbers only, namely 
\[\color{blue}{q_N=\prod_{i=0}^N(h_{i})^2}\]
Hermitian numbers can be found easily solving the equations 
\[\color{blue}{h_{i}=\sum_{s\neq i}^N \frac{1}{h_i-h_s}}\]
If $\color{blue}{h_i}$ is the root of above equations then 
$\color{blue}{-h_i}$ also is the solution. Correspondingly, 
any state with odd $\color{blue}{N}$ has single root equal 
to zero and so $\color{blue}{q_{odd}=0}$ but mass of these 
states is not zero. 

For even states the value of electric charge at low 
$\color{blue}{N}$ is following
$$\color{blue}{q_0=1,\;q_2=1,\;q_4=9,\;q_6=225}$$ 
Hence there are two states with equal charges and different 
masses, it can be relevant to problem of electron-muon mass 
difference. Moreover, the strength of interaction between 
electromagnetic and w-field, the constant $\color{blue}{g}$, 
can be as positive as negative number that means 
the possible existence of two states with different masses 
and equal charge for any $\color{blue}{q}$. 

Concerning charge $\color{blue}{q=9}$ we are thinking that 
pure electromagnetic objects with charge $\color{blue}{q>1}$ 
are absent in the nature. However, the atomic nuclei with 
$\color{blue}{Z=9}$ exist and can have the tracks of possible 
$\color{blue}{q=9}$ single, not $\color{blue}{1+1+1..}$, 
electric charge. For example, the lower excitations of nucleus 
$\color{blue}{18F}$ are divided by too small gaps compare to 
neighboring self-jointed nuclei. More of fifty years ago 
J. P. Elliott and B.H. Flowers explained those puzzle evoking 
three-nucleon interaction. Below in this article, the lower 
excitations of $\color{blue}{18F}$ were calculated in the 
frame of shell model grounded on nonlinear pionic field. 
Nevertheless, the searching for tracks of $\color{blue}{q=9}$ 
electric field can be useful.

All in all it is wondering but the situation is extraordinary 
as itself 
as because the condition $\color{blue}{h_{\infty}=0}$ is not 
valid. Little deep consideration of those condition is relevant; 
for this the qualitative behavior of the potential on the 
axis $\color{blue}{R}$ is making below. 

From physical reason the potential of coherent state runs 
to infinity when $\color{blue}{R\to 0}$ and we cannot assign 
the boundary condition in those point. So let us take an 
point $\color{blue}{R_0\ll 1}$ and boundary condition 
$\color{blue}{h(R_0)<-|h_N|}$. For definiteness the even 
negative valued  potentials are regarding, there the set 
of hermitian numbers is following 
$$\color{blue}{\{-|h_N|,\;...,\;-|h_2|,\;|h_2|,\;...\;|h_N|\}}$$
Because $\color{blue}{h'\geq0}$ anywhere the potential growth 
at moving from $\color{blue}{R_0}$ to big distances but, due to 
Gaussian term in the tension, go to saturation - the 
growth stop\-ped quickly below line 
$\color{blue}{-|h_N|}$. The gap between $\color{blue}{h}$ and 
$\color{blue}{-|h_N|}$ can be as small as big, it depends on 
the initial value $\color{blue}{h(R_0)}$ of the potential. 
At suitable choice of initial value the potential can be 
approaching to line $\color{blue}{-|h_N|}$ on the arbitrary 
small distance; for special case $\color{blue}{h(R_0)=-|h_N|}$ 
it is $\color{blue}{h\equiv h_N}$. However the potential, 
which is beginning below of hermit line, cannot reach or 
intersect that line. 
So in the saturation area another initial point 
with another initial value of potential  in the area 
$\color{blue}{-|h_N|<h<-|h_{N-2}|}$ can be taken and 
previous situation will be reiterating; and so on up to line 
$\color{blue}{-|h_2|}$. After $\color{blue}{h=-|h_2|}$ line 
the potential can run up to $\color{blue}{|h_2|}$ as well as 
stopped below of axis $\color{blue}{R}$, it is matter of 
chosen point and initial  condition above of 
$\color{blue}{-|h_2|}$ line. 

The restriction $\color{blue}{h\neq h_i}$ follows from physical 
reason also; the scalar part of velocity fourvector is 
$$\color{blue}{u_0=-h+2\sum_i \frac{1}{h-h_i}}$$ 
therefore in the case $\color{blue}{h=h_i}$ the
self-energy of 
free w-field, $\color{blue}{1/2(u_0')^2}$, became infinite. 
However, due to $\color{blue}{h_{n}\to-h_{n}}$ symmetry of 
hermit numbers those infinities are spurious.  

Coherent  states  have structure similar to geological 
stratum - the layers of charged electrostatic fields divided 
by thin film of the photon or emptiness. In adjacent layers 
the density of electric charges, $\color{blue}{u_0(h')^2}$, 
are opposite. Due to charge conservation low the layers are 
stable; it looks as some kind of spherical symmetric 
condenser. 

Above of R-axis the potential is positive number, the 
running from zero to $\color{blue}{|h_2|}$ line can be 
formal because the sign of tension can be changed and so 
the mirror part of negative valued potential appears. 
But for even states the line 
$\color{blue}{R}$ is not hermitian line so the two mirror 
states can be stable if the area $\color{blue}{-|h_2|<h<|h_2|}$ 
is empty. 
 
Note, unless the vacuum was created the small excitations in 
the form of free electrons are presented anywhere. For 
this reason the full coherent state which is produced 
by jointed parts of positive 
and negative tensions can be more stable than 
states with  tensions of equal sign.  

Because $\color{blue}{h_\infty \neq 0}$, the odd-states 
also have small electric charges. 

It is query, always the energy needing for removal the 
piece of charge from 
the layer into free space is less of electron mass; for 
example in the atom with $\color{blue}{Z>137}$ the binding 
energy of some electrons can exceed the mass of free electron. 
But typically 
those energy is less of electron 
mass and so the square of field fourvelocity vector is 
positive number, namely
$$\color{blue}{U^2=[(U_0)^2-(\vec U)^2]>0.}$$  
But in general case the condition $\color{blue}{\vec U=0}$ 
can be satisfied 
only locally so as electrostatic as magnetic 
fields can be presented in coherent states of nonlinear 
Coulomb field.  

In this way, the variety of coherent states  
can be found and no end of this stuff. 
Their existence depends strongly from boundary 
conditions imposed on the potential, the value of interaction 
constant of free w-field, etc. 

Additioanal remark concerns the electromagnetic masses of 
coherent states: 
they are determinated by lagrangian of the field; the energy 
of free w-field is infinite due to singularities 
in $\color{blue}{U_0}$ in the points $\color{blue}{h_i}$. 
This defect can be removed adding to 
lagrangian suitable shadow term, the "shadow" means that this 
additional term is under coherence condition and so is unvariable 
quantity. Correspondingly, the equations 
for nonlinear electromagnetic field will be unchanged while 
the mass of the state became finite. 

Introducing additional terms in the lagrangian of a field is 
typical but complicated road. Below, in next subsection, the 
estimation of lepton masses is making via more limple appproach. 

Printing scratch is aiming to prompt the young 
physicists for working in the realm of nonlinear fields. 

\subsection{Electron-muon masses and Coulomb field}

 The origin of electron mass was discussed since 1904 year, 
see \cite{42} and references therein. The discovery in 1936 year of 
the muon, weighted twin of the electron, produced puzzle well known as "electron-muon mass difference". 

The anticipation and detecting of tau lepton \cite{43} 
complicated the situation which, in fact, now is considering 
as unsolvable. 

In the article \cite{44}, 
dated by 2016 year, M. J. Tannenbaum printed about solution of the 
puzzle: I wonder how much longer we will have to wait for that! 

The aim of this article is demonstrate that in the frame of nonlinear 
Coulomb field model, below cited NCF, the leptons mass 
difference is comprehensible. 

The tensions of NCF in dimensionless variables can be 
written as 
\[\color{blue}{E_N(h)=\prod_{i=0}^N\frac{(h-h_i)^2}{(h_f-h_i)^2}exp(-h^2/2+h_f^2/2)}\]
were: $h$ is dimensional potential of NCF; $E_N(h)$ is electric 
tension in N-state; $h_i$ are Hermit numbers; the constant $h_f$ 
is, in same sense, pseudo-infinity because when potential at moving 
from lower values reached this point the 
tension  $E_N(h_f)$ became equal to unity and so the electric 
charge gathered below of $h_f$ is equal to unity. The introduction 
of this 
quantity is useful because of following reasons. The system of 
NCF equations has two additional solutions: the trivial one 
with electrostatic potential and scalar velocity equal to zero 
and half-trivial one with constant electrostatic 
potential not equal to zero, $h==h_0$ and, correspondingly, 
constant scalar velocity $v==v(h_0)$. In the point $h_f$ we can 
sew above printed NCF solution with half-trivial one. In this 
case the whole electric charge of any N-state is equal to 
unity, the potential is continual function  while 
tension has abrupt in $h_f$ point; together with Gaussian 
dependence of the tension upon potential this falls  
permits the consideration of NCF states 
as particles.  However, the abrupt arise some questions 
about  electron-electron interaction. 

Hermit numbers $h_i$ are determined by equations 
\[\color{blue}{h_i=2\sum_{s\neq i}^N\frac{1}{h_i-h_s}}\] 
So the states of NCF are labeled by Hermit numbers $N$.

In the center-of-mass frame the energy and mass of a field 
coincide numerically. Correspondingly, the mass of NCF in N-state 
can be written as 
\[\color{blue}{m_N=\int_{h_-}^{h_+}\big[E_N(h)+
({du_N(h)/dh})^2E_N(h)-}\]
\[\color{blue}{-g\times h\times u_N(h)\times E_N(h)\big]dh}\]  
by physical meaning the first two terms are kinetic energies of 
electrostatic and w- fields correspondingly, third term is the 
interaction energy between w- and electrostatic fields;  
$\color{blue}{g}$ 
is interaction constant. The multiplication 
sign introduced to ease the reading of those formula. The 
expression for scalar velocity of NCF in N-state is following
\[\color{blue}{u=-h+2\sum_{i}^N\frac{1}{h-h_i}}\] 

Hermit numbers divided h-space on few fields. It is easy check 
numerically that the mass of states stemmed from Hermit numbers 
is unstable. Only in the intervals 
$\color{blue}{\{-\infty, h_+<-max |h_i|\}}$ the masses are stable; 
correspondingly $h_-=-\infty$ while upper point in the mass 
integral is the point of sewing 
usual and half-trivial solutions of NCF. Because of Gausian 
dependence the $-\infty$ can be taken equal to $-100$, this 
is unessential; maybe at other situations the half-solutions  
of NCF can be relevant on the small distances.
The determination of upper point of integration is the problem. 

For resolving of the one let us consider  
additional set of numbers, $\color{blue}{hq(q_x)}$; they are the roots 
of the equations 
\[\color{blue}{\prod_{i=0}^N(h-h_i)^2=q_x},\] 
by physical meaning $q_x$ are some pseudo-charges.

For N=2 we take $q_x=1$  because at simplest consideration of 
electric charge quanta in previous subsection  the charge of 
this state is equal to unity.
So the set of numbers for N=2 state is following
\[\color{blue}{N=2;\;|h_i|=\{ 1\};\;q_x=1,\;|hq|=\{0,\sqrt 2\}}\]
The odd states are discarded for reason that in simplest 
case, $h_{\infty}=0$, 
they have zero charge and we do not have reasons for assignment 
the number $q_x$ there.  

At simplest consideration  N=4 state has $q_4=9$ electric 
charge so the value $q_x=9$ is taken for this state; of course,   
by construction all states have whole electric charge $q=1$ 
independently from choice of $q_x$ constant. Those complexity 
emerged because of correlation between distribution of charge 
and mass of the field.  The introduction of pseudo-charges 
 $q_x$ determinate the upper point of mass density integration 
but it is obviously - it is enough arbitrary step.   
For N=4 state the  set of needing numbers is following  
\[\color{blue}{N=4,\;|h_i|=0.74196...; 2.33441...,\;q_x=9}\]
\[\color{blue}{\;h_+=2.44948974}\]

The value and sign of interaction constant $\color{blue}{g}$ 
is unknown. In case  $\color{blue}{|g|=1}$ 
the masses of N=0 state are: 
\[\color{blue}{m_0(g=-1)=0;\;m_0(g=+1)=2.5}\]
However, the objects with zero mass 
and not zero electric charge were not observable. We consider 
the solution of this obstacle as  $\color{blue}{|g|\neq 1}$
 
For simplicity let us take  $\color{blue}{g>0}$ and assign 
for electron, as having lightest mass,  
the $\color{blue}{m_0(-g)}$ state while for next leptons with 
not zero mass  $\color{blue}{m_N(+g)}$ states. 
The minimization in $\color{blue}{g}$ of relative masses of
muon-electron couple and tau-electron couple together by SciPy 
tools returns 
$$\color{blue}{g=0.99122686}$$

With electron mass me=0.511 MeV the hierarchy of lepton masses, 
in square brackets are observable masses in MeV, is following:
\[\color{blue}{me=0.511\;[0.511];\;muon=101.48\;[105.658]}\]
\[\color{blue}{tau=m_2=1712.7\;[1777];\;m_4=56970.5\;[?]}\] 

Qualitatively, in nonlinear Coulomb field the hierarchy of 
massive leptons can be observed.

Quantitatively, this is only toy model. Creator know 
complete  knowledge. 
Nevertheless, searching the fourth lepton 
will be needing; however, after preliminary job of theorists, especially 
working in hep-physics area. But, looking back, if last ten years 
were not enough to explain the electron-muon mass difference, 
maybe future hundreds years will be enough.    

\subsection{Electron levels in linear and nonlinear Coulomb fields } 

The spin effects are essential here and 
Dirac equation must be using. 
Electromagnetic field itself is four 
vector field but it is joining with 
spin framework. For example, 
electrostatic field of proton in 
general case is the sum of two terms with 
spins equal to 1/2 and 3/2. 

Nevertheless, for calculation the first is hydrogen atom because 
for this atom the deformation of electrostatic field 
caused by proton structure can be ignored and in NIST 
tables the hyperfine splitting is removed. 
Dirac equation for electron in electromagnetic field 
is
\[\color{blue}{(i\nabla -eA)\Psi =m\Psi}\] 
\[\color{blue}{c=1,\;{ }\hbar =1}\] 
Let us go to usual three-spinors, 
separate the angle dependence of wave 
functions and convert two equations 
for two radial wave functions into 
one equation. Then the Schroedinger-like equation 
appears as 
\begin{gather*}
\color{blue}{ 
F''+\frac{2}{R}F'+\frac{F'V'}{W-V}=}\\
\color{blue}{=\left[\frac{l(l+1)}{R^2}-\frac{kV'}{{R(W-V)}}-E^2+2EV-V^2+m^2\right]F} 
\end{gather*}
where $\color{blue}{E, V, m, j, l}$ are 
the energy, potential energy, mass, 
moment and orbital moment of the 
electron;
$\color{blue}{W=E+m}$; and 
$$\color{blue}{k=\{-l,\;(l+1)\}}$$ for 
$\color{blue}{j=\{l+1/2,\;l-1/2\}}$. The equation for logarithmic 
derivative of wave function, $\color{blue}{f(R)=F(R)'/F(R)}$, is more 
convenient, those is 
\begin{gather*}
\color{blue}{ 
f'+f^2+\frac{2}{ R}f+\frac{V'}{W-V}f=}\\
\color{blue}{=\frac{l(l+1)}{R^2}-k\frac{V'}{R(W-V)}-E^2+2EV-V^2+m^2 }
\end{gather*}
It is Riccati equation, in this form the analytical properties of wave function are more realizable; below the term Riccati function refers to logarithmic derivative of wave function.

\subsubsection{Electron levels in linear Coulomb field}
In the case of linear field the denominator 
\[\color{blue}{W-V=W+\frac{\alpha}{R}}\] 
has single zero in the point 
$\color{blue}{R_0=-\alpha/W}$ and two terms 
in the equation for wave function are singular. 
Thus we take the radial part of electron wave functions as
\[\color{blue}{\Psi\sim R^B(R-R_0)^D\prod_{n=1}^N(R-R_n)exp(AR)}\]
One of emerging algebraic equation, 
$\color{blue}{D^2-2D=0}$, has two solutions. 
The case $\color{blue}{D=0}$, which 
corresponds to 
cancellation of the equation singularities, 
refers to well known situation available in 
all quantum mechanics textbooks, there the 
variety of main quantum number is \{1,2,...\}. 
If $\color{blue}{D=2}$ the wave function 
has double zeros and main quantum number 
runs the variety \{3,4,...\}. So these 
novel states with main quantum number 
$\color{blue}{n>2}$ are degenerated with the 
first ones while the states without or with 
one single zero of wave function are not degenerated. 
External electromagnetic field 
can take off this degeneration. For some 
reason the implicit assumption about 
singularities cancellation is used commonly, 
the existence of the solutions 
with double zeros of the electron wave function 
were not considered, at least this is in 
accessible for me articles. 
Below we will regard the usual states only. 

Binding energy of the electron is
\[\color{blue}{\varepsilon =m\Big[1-\frac{B+N+1}{\sqrt{(B+N+1)^2+\alpha^2}}\Big]}\]
where extended moment $\color{blue}{B}$ depends upon  full electron moment 
$\color{blue}{j}$ as 
\[\color{blue}{B(j)=-1+\sqrt{(j+1/2)^2-\alpha^2} }\]
Correspondingly, the main quantum number is  
\[\color{blue}{n=N+j+1/2}\]
Excluding $\color{blue}{N=0,\;\;j=l-1/2}$  states, which cannot exist because they contradict to restriction at $\color{blue}{R\to -\alpha/W},$  all other states with equal $\color{blue}{(n,\;j)}$ and opposite parity are degenerated.  

\subsubsection{H I levels in nonlinear field}

Remind that only coherent states of electromagnetic field are regarding.

We wish to find the compact expression for binding energy of the electron in nonlinear Coulomb field which contains all quantum numbers explicitly. This 
is piecemeal work because the location of an approximation for potential with nice analytic properties is not unique.    

Fortunately, enough general approach to description of spin-orbital interaction is there. Indeed, nonlinear potential is slowly deformed 
linear ones and both are monotonic functions. Correspondingly, in each of spin-orbital term only single pole exists. In linear case the pole is in the point $\color{blue}{R_0=-\alpha/W_0}$, in case of nonlinear potential the location of pole is slowly shifted what approximately can be written as
\[\color{blue}{R_s=-\alpha/W}\] 
With approximation of nonlinear potential the situation is more complicated. Via known spin-orbital term the potential can be restored as 
\[\color{blue}{V=V_\infty +W\Big[1-{\Big(1-R_s/R}\Big)^{-\alpha/{(WR_s)}}\Big)\Big]}\]
with $\color{blue}{R_s<0}$  for obvious reason. This effective potential has unusual property - in the interval $\color{blue}{R_s<R<0}$ it become complex number unless the degree is integer number; for linear field those degree is unity while the existence of polynomial potentials is under question. 

On the big distances the first not trivial expand of the potential into series contains $\color{blue}{\sim R^{-2}}$ term which produce unusual properties of wave function due to presence of strong singularities in the square of the potential; as first step the terms of $\color{blue}{R^{-3},\;R^{-4}}$ in the square of potential energy are neglected while potential energy in out area of the field  is  taken as 
\[\color{blue}{V_{out}=-\frac{\alpha}{R}+\alpha^2\frac{a}{R^2}}\]
where the value of potential energy on the infinity is jointed to full electron energy and $\color{blue}{a>0}$ because $\color{blue}{u_0(R\to \infty)=-1}$ for states with positive density of electric charge on big distances. 

Unknown parameter of spin-orbital term is approximated as
\[\color{blue}{\frac{\alpha}{WR_s}=-1-\alpha^4bW_0}\] 
where $\color{blue}{b}$ is unknown constant, $\color{blue}{W_0}$ is corresponding quantity of linear case.  For coherence and effective potentials, comparing  the 
series on the big distances, we get the restriction $\color{blue}{a=-\alpha^2b/2}$ so $\color{blue}{b<0}$; such sign means that effective and coherence potentials are incompatible on the small distances. 

Electric charge of nonlinear field is distributed over whole space and so in internal area of the field partial renormalization of electric charge appears.  
For this reason it is desirable involving into job the pieces of the potential on small distances. 

Addressing to small distances, there little more insight is demanding, we assume well analytic 
properties of potential. Because at small distances the nonlinear potential growth the simplest series in this area is
\[\color{blue}{\frac{1}{\varphi(R)}=c_1R+c_2R^2}\]
Correspondingly, potential energy in internal area of the field is approximated as
\[\color{blue}{V_{in}=\beta\Big(\frac{1}{R}-\frac{1}{R+\gamma}\Big)},\]
from physical reason both constants are positive numbers.

Now both parts of the potential need sew together in an unknown point. Instead of 
this, because the singularities are carrying main part of information about potential, we joint these parts as
\[\color{blue}{V=-\frac{\alpha}{R}+\alpha^2\frac{a}{R^2}+\beta\Big(\frac{1}{R}-\frac{1}{R+\gamma}\Big)}\] 
but in the square of potential the cross-term, $\color{blue}{\sim \alpha^2a\beta}$, will be discarded because internal part of the potential is the single bit dividing of which is incorrect; only $\color{blue}{\sim \alpha^2,\;\beta^2}$ terms are counting.   

Correspondingly, Riccati function can be written as
\[\color{blue}{f=A+\frac{B}{R}+\frac{D}{R+\gamma}+\sum_{n=0}^{N}\frac{1}{R-R_n}}\] 
where $\color{blue}{A,\;B,\;D,\;\gamma,\;R_n}$ are constants. Note that $\color{blue}{R_s}$ pole was not separated so only usual part of spectrum is calculating and for $\color{blue}{n>2}$ it can be found more lines then it is printed in the data tables.  
 
Unknown constants of wave function can be found at full solving of equations and will depend upon parameters of potential - in our case they are the constant $\color{blue}{a,\;\beta }$; it is cumbersome job. However, that can come round. Because spin-orbital energy decrease on the infinity as $\color{blue}{1/R^3}$ and coherent stated of potential have two free parameters the both of ones will be involved in the solutions, this circumstance permits the consideration of constants 
$\color{blue}{b,\;\beta}$ as free parameters.

The expression for binding energy is 
\[\color{blue}{\varepsilon =m\Big[1-\frac{B+N+1+\beta^2}{\sqrt{(B+N+1+\beta^2)^2+\alpha^2}}\Big]}\]
while extended, or effective, orbital moment is
\begin{gather*}
\color{blue}{B_{\pm}=-1+\frac{\alpha^2bW_0}{2}}+\\
\color{blue}{+\sqrt{(j+\frac{1}{2})(j+\frac{1}{2}\mp\alpha^2bW_0)-\alpha^4bE_0-(\alpha-\beta)^2}}
\end{gather*}
where $\color{blue}{\pm}$ signs are referred to $\color{blue}{j=l\pm 1/2}$ states;  under square root some unessential terms were discarded for best viewing from where the nonlinear effects arise. 

Reduced mass of the electron  and fine structure constant are
\[\color{blue}{m=0.5107207446\cdot10^6eV, \;\; \alpha=7.297352568\cdot10^{-3}}\]
The calibration of transition energies in three points by sagemath.org tools gives
\[\color{blue}{bm=-0.0133,\;\;\beta=0.00161}\]
but those must be checked because of unclear ability of available numerical tools. I am checking the calculation with ApCalc but this helps slowly.  
  
The issue, in form of tables for lower transitions, is following. 
Quantum numbers of the electron
\[\color{blue}{n=N+j+1/2;\;\; N;\;\;l;\;\; j}\]
 in final state are printed; initial state is (1; 0; 0; 1/2); calculated values are in blue while observable, which were taken from NIST tables\cite{u}, in black; transition energies are in electron-volts. 

$$ $$
\begin{tabular}{|c|c|c|c|}
\hline
n&N\;\;l\;\;j&\color{blue}{calculated}&data\\
\hline
1&0\;\;0\;\;1/2&\color{blue}{0}&0\\
\hline
2&1\;\;1\;\;1/2&\color{blue}{10.1988056}&10.19880570432\\
2&1\;\;0\;\;1/2&\color{blue}{10.1988104}&10.19881007922\\
\hline
2&0\;\;1\;\;3/2&\color{blue}{10.19885376}&10.1988510686\\
\hline
\end{tabular}
\[\]
\begin{tabular}{|c|c|c|c|}
\hline
n&N\;\;l\;\;j&\color{blue}{calculated}&data\\
\hline
3&2\;\;1\;\;1/2&\color{blue}{12.08749346}&12.0874931306\\
3&2\;\;0\;\;1/2& \color{blue}{12.0874949}&12.0874944326\\
\hline
3&1\;\;2\;\;3/2&\color{blue}{12.08750629}&12.0875065498\\
3&1\;\;1\;\;3/2&\color{blue}{12.0875077}&12.08750657\\
\hline
3&0\;\;2\;\;5/2&\color{blue}{12.0875120}&12.0875110\\
\hline
\end{tabular}
\[limit\;\;\color{blue}{13.5984344};\;\color{black}{13.598434005136 }\]
Ionization energy cannot be calculated for unknown value of the potential on the   infinity.  

In some sense the result is somewhat worse compare with previous variant of this article  but here it is possible the calculation of spectra for any quantum number that can be useful and, that is essential, without invoking of mass forces.  Effective potential itself can have interest for theoreticians. 

Anyone can make the calculation of deuteron spectra, the results have the same  exactness as for hydrogen atom but for He II it is not case
   
Before consideration of He II levels let us look into possible restrictions on the variety of approximations for nonlinear Coulomb potentials.  

\subsubsection{Wilds} 
There are many corrections that have to be counted for. Let us look, without calculations, what happens if the potential and its square in external area are modeling up to  
$\color{blue}{R^{-4}}$ terms and then applied to whole space. In this case the wave function will contain additional multiplier, $\color{blue}{exp(-C/R),\;\;ReC>0}$. Correspondingly, in the point $\color{blue}{R\to 0}$ two additional restrictions on the parameters of wave function arise but only one new parameter is introduced. Thus the equations will define one of free constants in the potential. The constant $\color{blue}{\beta}$ can be chosen for the solution. From physical reason the constants $\color{blue}{\beta/\alpha,\;Cm/\alpha}$ are,   typically, small numbers and can be, at first approach, discarded in the equations  for energies and effective moment. In this way we return to previous, somewhat simple, solution. 
However, there is drastically another solution for effective moment because the hard  restriction  $\color{blue}{B>-1/2}$ now is not indispensable. Indeed, the multiplier $\color{blue}{exp(-C/R)}$, in spite of constant $\color{blue}{Cm}$ smallness, gives finite norm of wave function in case $\color{blue}{B<-1/2}$  and  another solution for effective moment can be written as 
\[\color{blue}{B=-1+\alpha^2bW_0/2-\sqrt{(j+1/2)(j+1/2\mp \alpha^2W_0)+..}}\]      
In this case the main quantum number is 
\[\color{blue}{n=N-j-1/2}\]
and runs variety $\color{blue}{n=\{0,1,2,..\}}$. In this new structure each shell 
contains infinite number of states which can be labeled by or moment or radial quantum number; the $\color{blue}{j=1/2}$ state has minimal while $\color{blue}{j=\infty}$ state has maximal binding energy. Excluding $\color{blue}{n=0}$ shell the average binding energies and so the transitions from $\color{blue}{n=1}$ shell almost coincide with the ones of usual models. Another properties has $\color{blue}{n=0}$ shell - there binding energies eat away almost whole mass of the electron. Such spectra seems unbelievable. But who know what the world is, below of atomic physics the interaction with strong coupling is typical.

At least we somewhat learn about possible approximations of electrostatic  potentials. There only  $\color{blue}{1/R,\;1/R^2}$ terms can be presented after expand of the potential into series on the big distances and applying the result to whole space.  The terms of type $\color{blue}{1/(R-R_x)}$ are allowed but they are mimicked by internal and spin-orbital parts of the potentials. 

\subsubsection{He II lines}
 Extending to He II the expressions for hydrogen spectra we do not get satisfactory result and
 the reason of this is clear -  the structure of nucleus must be taken into account. 
 Nevertheless, the energy $\color{blue}{E_1=m+\varepsilon_1 }$ calculated with found parameters for hydrogen atom but with reduced electron mass for He II and with $\color{blue}{\alpha \to 2 \alpha,\;\beta \to 2\beta}$ can be used as first approach to more precise computations.

As tool to understand electrostatic structure, at least of light nuclei, the properties of extended Yukawa potential can be used.

On the small distances extended Yukawa potential runs to zero such quickly that in  this area the one become flat, there the motion of the nucleons is free; as a result the protons in nuclei are redistributed near the nucleus surface (more exactly - away from the center of pionic field) and form some type of potential bar (or well) to external electrons. Now electrostatic  potential is not monotonic function and the approximation of spin-orbital  interaction, which works well for proton, is not valid. We discard previous nonlinear part of spin orbital terms entirely from the equation for radial part of wave function. In this case the question has arisen - from where the 
splitting of parity degenerated states emerges; our answer is that for plus-states the electric bar while for minus-states the electric wall can be the solution, for effective potential it is possible. However this approach itself produce the question.  

 The distribution of electric charge in whole space do not vanish in any case and, for simplicity, the constant $\color{blue}{\beta}$ is holding in the expressions for energy and effective moment without calculation.  For this reason and taking into account previous remarks we approximate potential and its square as 
\[\color{blue}{V=-\frac{\alpha}{R}+\frac{\alpha^3C}{R[(R-d)^2+k^2]}}\]
 
\[\color{blue}{V^2=\frac{\alpha^2}{R^2}-\frac{2\alpha^4 C}{R^2[(R-d)^2+k^2]}}\]
There $\color{blue}{C,\;d,\;k}$ are fitting constants, only the constant 
$\color{blue}{\lambda=C/d}$ works at calculation of the levels. 

The solution of Riccati equation for some of wave function parameters is following: 
\noindent
$\color{blue}{2\alpha d+E(d^2+k^2)=0}$, it is the result of complex poles cancellation; 
\noindent
$\color{blue}{res(1/R)=-2\alpha E}$ so the expression for binding energy of the electron is the same as for linear field but we hold, without detailing calculation, the constant $\color{blue}{\beta}$ in the solutions for energy and 
moment as free parameter;
\noindent
$\color{blue}{res(1/R^2)}$ contains additional term $\color{blue}{-\lambda\alpha^3 E_1}$ where we approximated $\color{blue}{E\to E_1}$ for disconnection of the equations for effective moment and energy.   

In this case the expression for binding energy of the electron is 
\[\color{blue}{\varepsilon =m\Big[1-\frac{B+N+1+\beta^2}{\sqrt{(B+N+1+\beta^2)^2+\alpha^2}}\Big]}\]
while for effective moment it is
\[\color{blue}{B_{\pm}=-1}\\
\color{blue}{+\sqrt{(j+1/2)^2-(\alpha-\beta)^2 \mp \lambda\alpha^3 E_1}}
\]
\noindent Formally it is as if additional term $\color{blue}{\pm\alpha^3\lambda E_1 /R^2}$ is presented in the potential energy.

Remark that now additional underpinning of unusual approach to levels splitting arises. 
Wave function zeros $\color{blue}{R_i}$ , which  are transformed to poles of Riccati function, can be grouped into two sections: non-relativistic zeros with $\color{blue}{Re( R_i)>0}$ which corresponds with $\color{blue}{j=l+1/2}$ states and relativistic ones $\color{blue} {Re(R_i)<0}$ which match the $\color{blue}{j=l-1/2}$ case. So the constant  $\color{blue}{d}$ can be taken with different signs and  only potential, without spin-orbital terms, can splits the levels. 

With reduced mass of the electron $\color{blue}{0.51072078*10^6\,MeV}$ the calibration within three 
 $\color{blue}{j=l-1/2}$ lines gives 
\[\color{blue}{\beta=0.0000687,\;\lambda m=-1.3375}\]
The comparison of calculated and taken from nist.gov tables (however, in the tables theoretical, not measured, lines are shown) is following 
$$ $$
\begin{tabular}{|c|c|c|c|}
\hline
n&N\;\;l\;\;j&\color{blue}{calculated}&data\\
\hline
1&0\;\;0\;\;1/2&\color{blue}{0}&0\\
\hline
2&1\;\;1\;\;1/2&\color{blue}{40.81302917}&40.8130290720\\
2&1\;\;0\;\;1/2&\color{blue}{40.80308574}&40.813087144\\
\hline
2&0\;\;1\;\;3/2&\color{blue}{40.81378263}&40.8137558670\\
\hline
\end{tabular}
\[\]
\begin{tabular}{|c|c|c|c|}
\hline
n&N\;\;l\;\;j&\color{blue}{calculated}&data\\
\hline
3&2\;\;1\;\;1/2&\color{blue}{48.37129537}&48.3712953026\\
3&2\;\;0\;\;1/2& \color{blue}{48.37131213}&48.3713126042\\
\hline
3&1\;\;2\;\;3/2&\color{blue}{48.37151024}&48.371510118349\\
3&1\;\;1\;\;3/2&\color{blue}{48.37151867}&48.37151047236\\
\hline
3&0\;\;2\;\;5/2&\color{blue}{48.3715874}&48.371581834744)\\
\hline
\end{tabular}
\[limit\;\;\color{blue}{54.41776314};\;\color{black}{54.4177631(2) }\]
The coincidence of the limits induces to some questions.

\subsubsection{Levels of Li I }		

Since N. Bohr, because of 
experimental physicists achievements, 
the ato\-mic spectra did not have to 
become simple for theoretical calculations. 
To obtain the correspondence with data 
even in case of few lines of simple atom 
the needing work is huge, see \cite{pachucki} 
and references therein. Of course, the 
computations are true if using models 
are true.

However, the spherical symmetrical potential energy,
$\color{blue}{V(R)}$, 
of the valence electron in the atom, 
without relying on a model, on big distances 
can be decomposed  in series 
\[\color{blue}{V=V_0+\frac{V_1}{ R}+\frac{V_2}{ R^2}+..}\]

Then, regarding $\color{blue}{V_i}$ as free 
parameters, anyone can perform a fit of the 
data and in this way to check the analyticity 
principle directly. I had done this 
for Li I 2s1nl states, which are 
simplest between the ones, with usage 
of coherent states that means  
the independence only of two fitting coefficients. 
It has been fruitlessly.

What the matter is? The analyticity principle 
is the hardest base of theoretical physics. 
The hint for escaping was found in interesting 
work \cite{k} where thermodynamic 
formalism is applying to pure mathematical 
system. The solution is: global, $\color{blue}{0<R<\infty}$, 
analyticity is absent, at least the single 
point, $\color{blue}{R=R_0}$, exists where the 
electrostatic potential fails to be analytic. 
Physical underpinning of such situation can 
be found easy. For clearness we regard 
LiI 2s1nl states only. The density of 
negative electric charge produced by 
two electrons on s-shell is 
$\color{blue}{\sim exp(-\alpha R/R_0)}$. 
Correspondingly, in area $\color{blue}{R>R_0}$ 
the valence electron is moving in 
some kind of field gas created by cloud 
of virtual scalar photons while in 
area $\color{blue}{R<R_0}$ 
the motion is in some kind of field 
liquid. So the atom is the system with 
internal phase structure. 

As tentative example we take the potential energy of last electron 
in Li I as sum of three terms, $\color{blue}{V=V_1+V_2+V_3}$; in whole 
the potential energy is  
\[\color{blue}{V=-\frac{\alpha}{ R}+\alpha\frac{a}{ R^2}+\alpha\frac{b}{R-R_0}}\]
where two first terms are referring to linear and main nonlinear parts of Coulomb interaction while third term represent the interaction of valence and core electrons. Last term is some effective potential energy because on the big distances, those correspond to big quantum numbers, $\color{blue}{V\neq\alpha/R}$ unless the constant $\color{blue}{b}$ runs to zero at big quantum numbers or $\color{blue}{b\sim 1/E}$. We choose the first variant because the second case is the mask of the first one. You are aware that in this nonlinear model the field itself, instead of  particle, is the carrier of electric charge and so the local electric charge, which is different for potential and tension, is constant only on the infinity. In case of hydrogen and He II this can be ignored for big distance between electron and nucleus while in Li I the electrons are the nearest.  

In the square of the potential the interplay of nonlinear terms is discarded and this quantity is taken as 
\[\color{blue}{V^2=\frac{\alpha^2}{ R^2}-2\alpha^2\frac{b}{ R(R-R_0)}+\alpha^2\frac{b^2}{(R-R_0)^2}}\]
Here is the question about the term $\color{blue}{V'/(W-V)=V_{SO}}$, which was labeled as spin-orbital term but for atom the spin-orbital splitting of the levels is unessential compare with orbital shift of the levels; the $\color{blue}{V_{SO}}$ has to be counted. At least for lower states the point of singularity is essential so we approximate spin-orbital term as
$$\color{blue}{V_{SO}\simeq \frac{k1}{R-R_0}}$$
where $\color{blue}{k1}$ is the constant to fit.

The solutions of Riccati function is searching as
\[\color{blue}{f=A+\frac{B}{ R}+\frac{D}{R-R_0}+\sum \frac{1}{ R-R_n}}\]
that leads to algebraic equations for 
determination of unknown constants, 
$\color{blue}{E,\;A,\;B,\;D,\;R_n,\;R_0}$, including  
$\color{blue}{R_0}$ because it is the one of the wave 
function knots.

The expression for binding energy, omitting the vacuum energy, is following 
\[\color{blue}
{\varepsilon=m\Big[1-\frac{B+N+1+D+k1}{\sqrt{(B+N+1+D+k1)^2+\alpha^2(1-b)^2)}}\Big]}\] 
where the expressions for other coefficients are
\[\color{blue}
{B=\frac{1}{2}\Big[-1+\sqrt{(2l+1)^2-4\alpha^2+8\alpha bE1}\Big]
}\]
\[\color{blue}
{D=\frac{1}{2}\Big[1-k1-\sqrt{(1-k1)^2-4\alpha^2a^2}\Big]}\]
there, for decoupling of energy and orbital moment, the value of electron energy was reduced to the one in linear Coulomb field.

In D-coefficient the solution with 
minus sign of square root is selected 
by analogy with hydrogen levels to avoid 
the discussion on the hidden levels. 

In this way we have three dimensionless 
parameters, $\color{blue}{a,\;b,\;k1}$, 
with obscure unknown vacuum energy. 
For avoiding this complication, 
but mainly because the ground and first 
excitation states are on the same shell,  
the limit of transition energies extrapolated 
in NIST data table, namely
\[\color{blue}
{\varepsilon_{limit}=5.3917149511\;eV},\]
is taken as energy of ground state. 

Because $\color{blue}{b}$ decrease when quantum numbers of valence electron 
increase we take simplest parametrization of this parameter, namely
\[\color{blue}{b\to \frac{(-1)^lb1}{(N+l+1)(l+1)}}\]
where the $\color{blue}{(-1)^l}$ no has any numerical significance 
but permits do not plunge into discussion on the signs of the parameters. Of course, this tentative model is not shaped but the time scheduled 
to work as physicists ended up to next year.   

For calibration few lower state are selected. Sagemath utility 'minimize' with 'bfgs' alghorithm return following values of fitting parameters 
\[\color{blue}{ma=-16.8677,\;b1=0.02455,\;k1=0.05533}\]
Some of experimental 
and fitted transition energies in eV are following
$$ $$
\begin{tabular}{|c|c|c|}
\hline
n,l & calculated & exp\\
\hline
0,1 &\color{blue}{1.847860}&1.847860\\
\hline
2,0 &\color{blue}{3.373126}&3.373126\\
\hline
1,1 &\color{blue}{3.8384}&3.834258\\
\hline
0,2 &\color{blue}{3.87814}&3.878613\\
\hline
\end{tabular}
$$ $$
\begin{tabular}{|c|c|c|}
\hline
3,0 &\color{blue}{4.3393}&4.340942\\
\hline
2,1 &\color{blue}{4.524}&4.521648\\
\hline
1,2 &\color{blue}{4.540723}&4.540723\\
\hline
0,3 &\color{blue}{4.547}&4.54157\\
\hline
\end{tabular}
$$ $$
\begin{tabular}{|c|c|c|}
\hline
4,2 &\color{blue}{5.11399}&5.11391\\
\hline
5,2 &\color{blue}{5.179104}&5.17898\\
\hline
\hline
\end{tabular}

In any case here is good deal of theoretical work. Even students can be involved in job, the flat type of education system is well for some limit. 

But there, for spectra of simplest atoms, exists the puzzle which is the existence of hidden (hid- for shortness) states which are presented theoretically but not in the data tables.   

\subsubsection{What about applications is}
In my opinion experimental searching of hid-states is worthwhile; for example look at Li I case. For D-coefficient the second solution has plus sign of square root, namely 
 \[\color{blue}
{D=\frac{1}{2}\Big[1-k1+\sqrt{(1-k1)^2-4\alpha^2a^2}\Big]}
\]
Hence here are two, shifted on the one step, ladders of states which can be or not be degenerated; splitting of the degeneracy depends on the value of fitting parameters. For small constant $\color{blue}{k1}$
 previous value of fit-constants can be used. In this case the transition from  usual ground state to ground hid-state, is  
\[\color{blue}{(1,0)_{usual}-(1,0)_{hid}=3.267\,eV};\]
 the (1,0)-hid and (2,0)-usual states are mates; those carry on the ladder of excitations. In case of different energies the wave functions of usual and hid-states are orthogonal. So, for example, the electron transition from hid-ground state to usual one can be produced by external field which carry the electron from hid-ground state to some upper usual state; then the electron itself follows to usual ground state.

This simple example is enough to encourage for searching of hid-stats, even if both sets are degenerated. 

The measurements of spectra have long story. It is unbelievable that hidden states were not observable. Maybe hid-states were missing, the history of physics has enough errors. However, it is more plausible that, just because of possible applications, those is a security game. But for theoretical works the true manners are biblical ones: after lighting a candle it is not covered by jug.

\subsection{Nonlinear electromagnetic waves} 

For classification of electromagnetic 
field states the signs of the invariants 
\[\color{blue}{E^2 -H^2;\; \vec E \cdot\vec H;}\] 
are using. In nonlinear model somewhat 
another classification of the states 
is more convenient. Indeed, the square 
of electromagnetic jet is
\[\color{blue}{J^2 \sim [(E^2 -H^2 )^2 +(\vec E\cdot\vec H)^2]u^2 }\] 
Therefore, the states of the field 
are distinguishable via 
$\color{blue}{u^2=u_0^2 -\vec u^2  }$ 
sign. 

The states with 
$\color{blue} {u^2\equiv 0}$ 
contain the usual electromagnetic 
waves. In case 
$\color{blue} {u^2 >0}$ the field 
has electric charge, these are the 
electric states of the field. 
If $\color{blue} {u^2 <0}$ then these are the  
magnetic states of field, the ones have the
electric charge equal to zero. 

Correspondingly, at least the local 
coordinate systems exist where: the 
charged states have positive energy 
and zero impulse; the magnetic states 
have zero energy and not zero impulse. 
Of course, these did not mean that the 
magnetic states are moving with super light 
velocity. The example is the usual 
electric current in usual conductor. 
For magnetic and electric states the 
attribute 'velocity' 
has different physical meaning. Here the 
remarks are relevant. In classical electrodynamics 
the charge but not electromagnetic energy is conserved quantity that is 
repairing by invoke the dissipation of field energy on the particles. 
In nonlinear model, pure electromagnetic systems with zero as well 
as nonzero charge and without particles can exist. Maybe someone 
carefully regards the energy conservation low for 
coupled w- and electromagnetic fields.        

Typically, the electromagnetic waves 
are states without electric charge 
and with periodical phase. Thus 
nonlinear electromagnetic waves are 
magnetic states of the field. 

For description of nonlinear waves it 
is conveniently to choose the 
coordinate system where the scalar 
part of electromagnetic potential is 
equal to zero. From Maxwell nonlinear 
equations follow that the four vectors 
$\color{blue} {A, u}$ are collinear 
then in appropriate coordinate system 
both scalar potentials are equal to 
zero, 
$\color{blue} {A_0=0, u_0=0}$. 

It is conveniently take the four 
potentials of the field for flat 
electromagnetic waves as following 

\[\color{blue} {A=A(x)\gamma_y exp\{i (\omega t -k z)\}}\] 

\[\color{blue} {u=-u(x)\gamma_y exp\{i (\omega t -kz)\}}\] 

With such choice of potentials the four tension of electromagnetic field is given by expression 

\[\color{blue}{F=\left(-i \omega Ae_y +ii_c kAe_x +i_c A' e_z \right) exp\{ ( i(\omega t- kz )\} }\] 
what is more complicated form compare 
with usual description of the vector 
field without usage of Clifford algebra. 
However, the needing equations 
contain the amplitudes of potentials 
so this complication has no matter. If we 
wish divide the tension into electric 
and magnetic parts then the suitable phases 
need to take. The phase multiplier in 
potentials creates same theoretical 
trouble because this convert the 
vector field in mixture of vector 
and pseudo-vector fields. For 
simplicity we go round of that by usual 
manner - rewriting the definition of
jets as $ \color{blue}{FuF\to Fu^+ F,\;FAF\to FA^+ F}$.  

Simplest waves are those where the 
w-field is in free state. 
Nevertheless, we consider the 
coherent states of the field. 

For plainness, we regard only slow 
waves (put $\color{blue}{\omega =0}$) 
and denominate the dimensionless 
amplitudes of the potentials as following 

\[\color{blue}{A(kx)=g\sqrt {k_{int}} p(s)}\] 

\[\color{blue}{gv(s)=\sqrt {k_{int}} u(kx)}\] 

\[\color{blue}{s=kx}\] 
where $\color{blue}{k_{int}}$ is the interaction constant of free w-field. 

In this case the 
equations for dimensionless amplitudes is following

\[\color{blue}{p'' -p=v(p'^2 -p^2)}\] 
\[\color{blue}{v''-v=p(p'^2 -p^2 )}\] 
The symmetry $\color{blue}{p(-x)=p(x)} $ 
and boundary  $\color{blue}{p'(\infty)=0,\; p(\infty)}=constant$ 
conditions are taking that is suitable 
for paramagnetic waves.  
The last condition looks surprisingly 
because then the density of field 
energy on infinity contains a constant 
terms. However, the appropriate choice 
of interaction constant $\color{blue}{q}$ 
in the lagrangian of electromagnetic 
field takes off the problem.
In paramagnetic waves the local 
currents are parallel therefore, on 
ground of Ampere low, these waves are 
stable. 

\noindent Few simple exact solutions of these 
equations exist. First is tri\-vial  
  $\color{blue}{p\equiv v\equiv 0}$ which correspond to pure vacuum state of field. Second is 
$\color{blue}{p\equiv v\equiv 1}$, 
because the phase of fields is not 
zero these are the usual  waves with 
fixed constant amplitudes. The 
solution  $\color{blue}{p\sim 
exp(\pm x),\;v\sim exp(\pm x)}$ 
represent the free states of field, 
in this case the equations 
$\color{blue}{p'' =p, v''=v}$ are 
equations of free field and the 
interaction between electromagnetic 
and w-field is absent. 

Hence at least the free coherent 
nonlinear electromagnetic waves exist 
in the model. In these waves the 
field is concentrating near surface 
$\color{blue}{x=0}$ and they did not have 
the internal structure along x-axis. 

Apparently, the  more 
complicated waves are here. For their 
detecting consider the example. Let 
us take the simplest connection, 
$\color{blue}{v\equiv p}$, between 
electromagnetic and w-field. By  
physical meaning the velocity 
parameter, $\color{blue}{v}$, is the 
polarization of vacuum with nonlinear 
dependence upon electromagnetic potential. The solution 
$\color{blue}{p\equiv v}$ correspond 
to linear connection between the 
polarization of vacuum and potential of electromagnetic field.  
For this case the first integral is 
\[\color{blue}{p'=\pm \sqrt{ p^2+C\, exp(p^2)}}\] 
and here are periodical solutions 
for potential if the first integration 
constant, $\color{blue}{C}$, is small 
negative number - then under square 
root expression is positive in area 
$\color{blue}{p_- <p<p_+}$. With 
conditions $\color{blue}{p'(0)>0}$ 
the amplitude of potential increase at 
moving along s-axis and reach the 
value $\color{blue}{p=p_+}$ in same 
point $\color{blue}{s=s_1}$. After 
this point we may or put 
$\color{blue}{p\equiv p_+}$, or 
change the sign of the derivative. In 
last case the amplitude grow down to 
value $\color{blue}{p=p_-}$ in 
point $\color{blue}{s=s_2}$. These 
circles may be repeated not once but 
on big distances need to put  
$\color{blue}{p\equiv p_+ }$, 
or $\color{blue}{p\equiv p_-} $. The 
situation is similar to usual 
trigonometric states where 
$\color{blue}{p'= \sqrt{1-p^2 }}$, 
therefore $\color{blue}{p(x)=sin (x)}$, 
or $\color{blue}{p(x)\equiv \pm 1}$. 
The energy of these states is the sum 
of the bits that gives additional chance 
for stability of these waves.

In this way, the states of nonlinear 
electromagnetic field in the form of 
nonlinear waves exist on the paper. 
These states have richer structure 
compare with usual electromagnetic waves. 
For their existence the external 
mechanical walls are not demanding.  

Nonlinear waves interact with 
external electromagnetic field. 
Indeed, in this model all fields are 
interacting, however, not with 
themselves but with w-field. If take 
into account that any external field 
has fixed phase then it is easy 
to build the simplest lagrangian for 
interaction of nonlinear wave with 
external electromagnetic field. 
Because the usual electric jet 
itself is magnetic state we expect 
that all effects observable at 
spreading of usual electromagnetic 
waves via mechanical medium can be 
observable at spreading of nonlinear 
waves via an external electromagnetic 
field. 

Repeat, nonlinear electromagnetic waves are reaches object.

\section{Fluid as mechanical field} 
\subsection{Introduction} 

Here the word 'fluid' means a 
continual isotropic homogeneous mechanical 
medium. 

In mechanics, the description of 
continual states is grounded on 
Newtonian lows. At this approach from impulse conservation 
low and phenomenological properties of a system the Navier-Stokes 
equations are constructed. Such method 
extends the particle 
dynamics in area of field objects.  

Besides, the fluid regard as a field 
of mechanical shifts and the 
lagrangian formalism employ as 
framework of fluid dynamics, \cite{t} and \cite{y}. 
The lagrangian formalism is general 
method for description of any 
field. However, when this method is 
taken for description of mechanical 
continual system then the main 
property of any field lose of the sight. 
The property is the spreading of 
internal interaction in any field from point 
to point with finite velocity. 

Internal mechanical interactions 
in a fluid are transmitting with 
velocity of sound. This property is 
bringing into play if the Lorentz, 
not Galileo, transformations with 
parameter $\color{blue}{c}$, which is the 
velocity of sound in fluid, are employed for 
coordinate system changes. In this 
article such road is chosen for 
description of the fluid. 

The Clifford algebra, \cite{e}\cite{a}\cite{s}, 
with standard lagrangian formalism take 
as tool for delineation the dynamics of 
fluid. The short review of Clifford 
algebra properties is in the end of 
the article. The choice of Clifford's 
formalism is forced by situation in 
high energy physics where for 
description of fermions the space-time 
assumed, however implicitly, to be 
the Clifford algebra but for 
description of the bosons the 
space-time is taken as vector variety. This is hard for comprehension. Because the algebra contain the vector space it seems more consistently put into use the Clifford algebra anywhere, including the mechanical systems.  

For going in this way the some mixture 
of Euler and Lagrange variables is 
convenient. If in point 
$\color{blue}{\vec x}$ at time 
$\color{blue}{t}$ is the particle, 
which at time $\color{blue}{t_0}$ 
was in point $\color{blue}{\vec x_0}$, 
then  three-vector of mechanical shift 
$\color{blue}{\vec \xi(t,\vec x)=\vec x-\vec x_0 (t,\vec x)}$ 
together with quantity $\color{blue}{\xi_0=c(t-t_0(t,\vec x))}$ 
regard as the space and  time parts of the shift 
four vector $\color{blue}{\xi (x)}$ 

More detailing this looks as  

\[\color{blue}{\xi=\xi_0 \gamma_0 +\xi_n \gamma_n}\] 

\[\color{blue}{x=x_0 \gamma_0 +x_n \gamma_n}\] 

 \[\color{blue}{x \gamma_0 =x_0 +\vec x}\] 

  \[\color{blue}{\xi \gamma_0=\xi_0  +\vec \xi }\] 

When absolute time is taking - this 
is usual representation for 
mechanical systems because the 
fundamental interactions are 
transmitting 
with velocity of light - then the 
time is independent upon the space 
points and time's part of the shift 
four vector becomes known (taking   for simplicity $\color{blue}{t_0=0}$) 

\[\color{blue}{\xi_0=ct}\] 

The space variety also becomes absolute in this case. 
Due to existence of fundamental interactions the relative velocities in fluid may exceed the velocity of sound. Of course, they always are less of light velocity. 

\subsection{Fluid parameters} 
It is well approach regard any 
mechanical system as variety of 
particles. The interaction between 
particles is so small that the ones 
are on mass shell and each particle 
move on trajectory $\color{blue}{x(s)}$. 
The tangent four vector, $\color{blue}{u(s)}$, 
to trajectory of the particle 
which has the mass  $\color{blue}M$ 
and four impulse $\color{blue}{P}$ is 
\[\color{blue}{P=Mu}\] 
\[\color{blue}{u=\frac{dx}{ds}}\] 
\[\color{blue}{\displaystyle {u \gamma_0=\frac{1}{\sqrt {1-\displaystyle{\frac{v^2}{  c^2}}(1+\frac{\vec v}{ c}) }}}}\] 

For continual state it needs to put $\color{blue}{M=\rho d^3 x}$ where $\color{blue}{\rho}$ 
is the density of the mass and  $\color{blue}{d^3 x}$ is  small volume. Then two independent 
parameters exist. These are the four scalar $\color{blue}{m=\rho \sqrt{1-{v^2/c^2}}}$ 
and the four vector of the mass flow $\color{blue}{J_m}$ 

\[\color{blue}{J_m \gamma_0=\rho(1+{\frac{\vec v}{ c})}}\] 

In mechanical interactions the mass is conserving quantity so 
\[\color{blue}{\partial_t\rho +\overrightarrow\nabla\cdot( \rho\vec v)=0}\] 

Other set of parameters is the 
relative shifts (these are the
deformations), 
which  in local limit are following 
\[\color{blue}{\nabla\xi =s+f}\] 
\[\color{blue}{s=\nabla\cdot\xi =\partial_0\xi_0+\overrightarrow\nabla\cdot\vec\xi}\] 
\[\color{blue}{f=\nabla\wedge \xi=-\partial_0\vec\xi -\overrightarrow\nabla\xi_0+i_c\overrightarrow\nabla\times\vec\xi}\] 
where $\color{blue}{c\partial_0= \partial_t}$ and all matrices are four-dimensional. 

In case of absolute time the deformations are following 
\[\color{blue}{s=1+\overrightarrow\nabla\cdot \vec\xi}\] 
\[\color{blue}{f=-{\frac{\vec v}{ c}}+i_c \vec h}\] 
\[\color{blue}{\vec h=\overrightarrow\nabla\times \vec\xi}\] 
where $\color{blue}{\vec v}$ is the local velocity of fluid, 
all matrices are four-dimensional. 

If not relativistic formalism is using then 
\[\color{blue}{\overrightarrow\nabla\vec \xi =\overrightarrow\nabla\cdot\vec\xi +i\vec h}\] 
where all matrices are two-dimensional.
Hence the appearance of the pseudovector $\color{blue}{\vec h=\overrightarrow\nabla\times\vec\xi}$
as a part of deformations is 
inevitable in field model of a fluid. 
As standard in usual model of fluid 
the variable  
$\color{blue}{\overrightarrow\nabla\times\vec v}$  
is taking for one of parameters. 

The deformations are field quantities. 
They are pure mathematical 
but not physical objects. In other 
words, these parameters do not 
exist in nature. It is because 
$\color{blue}{m\xi}$  but not $\color{blue}{\xi}$ itself may 
exists as real quantity. We get out 
these variables with goal to 
make the construction of fluid 
lagrangian more evident. 

Remark, the coincidence of parameter 
$\color{blue}{\vec v}$ for both  
$\color{blue}{\nabla\xi}$ and  
$\color{blue}{u}$ fields permit 
do not count the shadow w-field separately.  

The speed of fundamental 
interactions spreading is equal to 
velocity of light. 
Hence the snapshot of stream can be done and the picture of  current lines will 
be appearing. On a line of current the 
interval is space-like and so the tangent 
three-vector for a line is unite 
pseudovector $\color{blue}{i{\vec v/ v}}$ . 
For simplicity below the existence of such additional parameter 
is ignored. Remark that in work of a 
soviet physics near 1972 year a unit 
vector was introduced as an external 
parameter for description of solid 
state. 

\subsection{Lagrangian of the fluid} 

The lagrangian of a single field can 
be written as sum of the free 
field lagrangian and the self-interaction lagrangian 
\[\color{blue}{L=L_0+L_{sint}} \] 
Here  $\color{blue}{L_0}$ is the 
lagrangian of free field. This means 
that external and self-interaction 
forces are switching off. The 
lagrangian of free fundamental 
field is the square form of field 
tensions (if the ones disappear on 
infinity) 
\[\color{blue}{L_0=kF^2}\] 
\[\color{blue}{k=\;{constant}}\] 
For fluid the condition  of internal forces 
decreasing when the distances increase is 
valid. From  Hook low the fluid 
tensions are linear functions from 
deformations. However, in fluid the 
parameter  $\color{blue}{k}$  is not 
constant. Indeed, this parameter has 
the dimension of energy density. 
Because self-interaction is switching 
out only the quantity 
$\color{blue}{mc^2=c^2\rho\sqrt{1-v^2/c^2}}$ 
has required dimension. And from Bohr 
correspondence principle - this is the 
one of general physical principles - it 
needs to take  $\color{blue}{k\sim c^2\rho}$.  Hence for fluid the lagrangian of free field (in Hook's low area) is following 
\[\color{blue}{L_0=\frac{1}{ 2}mc^2\overline{\nabla(s+f)}\nabla(s+f)=\frac{1}{2}mc^2(s^2-f^2)}\] 
\[\color{blue}{\nabla=\frac{1}{ c}\gamma_0\partial_t-\gamma_n\partial_n;\;\; \nabla\gamma_0=\frac{1}{ c}\partial_t-\overrightarrow\nabla}\] 
Not mechanical interaction created 
the mass of fluid so, because of 
Lorentz invariance, the quantity 
$\color{blue}{m}$ is external 
parameter. Also $\color{blue}{m\xi}$ 
is entire quantity and, foregoing, 
the fluid motion is independent upon 
the shifts directly. From these 
reasons $\color{blue}{m}$ is not variable by 
shift four vector and so 
$\color{blue}{m}$ is 
considering as a function of space-time 
point, $\color{blue}{m=m(x)}$, not as 
function of deformations, $\color{blue}{m\neq m(\xi)}$.  
This simplification is the example of coherence condition usage. 
Correspondingly, the four dimensional 
equation for ideal fluid is 
\[\color{blue}{\nabla \big[mc^2(s+f)\big]=0}\] 
Because the parameters 
$\color{blue}{s,\;f}$ are taken with 
equal coefficients the compression modulus 
is equal to $\color{blue}{1/(c^2\rho)}$, this is not in bad agreement with data and  so acceptable for first approach.  

In any case the self-interaction lagrangian can be written as 
\[\color{blue}{L_{sint}=\xi\cdot J}\] 
where four vector   
$\color{blue}{J}$ is not variable 
quantity. By physical meaning this 
four vector represent the 
self-interaction forces in fluid.  
As norm for fluid the ones do not 
depend directly upon shifts and so 
$\color{blue}{J\sim J_m}$.  From 
physical reasons these forces 
disappear when  $\color{blue}{f\to 0}$ 
even if the pressure is not equal to zero. 
Correspondingly, 
\[\color{blue}{J\sim fJ_m}\] 
or, more widely, 
\[\color{blue}{J=af\cdot J_m+bf\wedge J_m}\] 

In this way the equation of  fluid 
dynamics in four dimension form is 
\[\color{blue}{\nabla\big[mc^2(s+f)\big]=a f\cdot J_m +b f\wedge J_m }\] 
Unknown scalar functions 
$\color{blue}{a,\; {} b }$ depend 
on the deformations and have  the 
dimension of acceleration.  Below for 
simplicity it is assuming that these 
functions are equal to constants. As 
with the compression modulus it is 
possible to take, as more symmetrical 
for first approach, $\color{blue}{b=-a}$ 

Two essential differences with standard approach are here. 

In any field theory both  the time 
and the space  derivatives appear 
symmetrically, in usual model of a fluid 
the degree of the ones differ. 

The parameter $\color{blue}{k=mc^2}$ 
is not constant and so the 
pseudo-four-vector part  of 
field equation in general case is not 
zero as it is must be  for fundamental 
fields. For example and comparison, 
the lagrangian of free electromagnetic 
field contains the pseudoscalar term 
$\color{blue}{\vec E\cdot\vec H}$. It 
has no matter is 
$\color{blue}{\vec E \cdot \vec H}$ equal to  
zero or not, this part of lagrangian 
create the pseudo-four-vector 
dynamical equation. However, the one 
coincide with kinematic restrictions 
$\color{blue}{\overrightarrow\nabla\cdot\vec H=0}$, 
$\color{blue}{\partial_t\vec H+\overrightarrow\nabla\times\vec E=0}$. 
For fluid, because of 
$\color{blue}{m=m(x)}$, the 
kinematic and the dynamical 
pseudo-four-vector equations do not 
coincide. In a field models the 
invariance of lagrangian at time and 
space inversions are jointed so the whole 
pseudovector part of the lagrangian may be equal to zero what bring some kinematic restrictions for shifts. 

\subsection{Equations of the stream} 
For going to three-dimension equations of motion it needs to take the absolute time, 
then multiply the basic equation on 
matrix $\color{blue}{\gamma_0}$ and 
put together the terms with equal O3 
 properties. It is conveniently 
 denominate 
\[\color{blue}{p_c=-c^2 m (1+\overrightarrow\nabla\cdot\vec\xi)=(p-c^2\rho)\sqrt{1-\frac{v^2}{c^2}}}\] 
where the note of hydrostatic 
pressure,  $\color{blue}{p}$,  is clear. 
In result  the equations of stream 
have following form  
\[\color{blue}{\partial_t p_c +c^2\overrightarrow\nabla \cdot (m\vec v)=\frac{a}{c}\rho v^2}\] 

\[\color{blue}{\partial_t(m\vec v) +\overrightarrow\nabla p_c +c^2\overrightarrow\nabla\times(m\vec h)= 
-\frac{a}{ c} \rho (\vec v +\vec h\times \vec v)}\] 

\[\color{blue}{\overrightarrow\nabla\cdot(m\vec h)=-\frac{b}{ c^3}\rho \vec v\cdot\vec h}\] 

\[\color{blue}{\partial_t (m\vec h)-\overrightarrow\nabla\times(m\vec v)=\frac{b}{ c} \rho \vec h}\] 

\[\color{blue}{m=\rho \sqrt{1-\frac{v^2}{ c^2}}}\] 
Below the velocity of sound take 
for unit of speed. 
From Le Chatelier principle the constant $\color{blue}{a}$ is positive and the constant $\color{blue}{b}$ is 
negative numbers. Usual boundary 
conditions are valid with one 
supplement. Because any mass move 
with finite speed the singular solutions 
for velocity must be rejected. For example, if the axial symmetrical stream 
has singularity  $\color{blue}{v=q \;{ } ln(\sqrt{x^2 +y^2})}$   with q=constant 
then self-energy of this state is finite but because at small distances the velocity growth to infinity we must put q=0. 

The elimination of unusual parameter $\color{blue}{\vec h}$ from equations  of stream is possible. Using both kinematic \[\color{blue}{\overrightarrow\nabla\cdot\vec h=0;\;\;\;\partial_t \vec h-\overrightarrow\nabla\times \vec v=0;}\] 
and dynamical pseudoscalar and pseudovector equations, we get the following connection between $\color{blue}{\vec h}$  and the usual variables of fluid dynamics 
\[\color{blue}{\vec h(b\rho-\partial_tm)=\vec v\times\overrightarrow\nabla m }\] 

Consequently, the vicinities of points \[\color{blue}{\partial_t\big(\rho\sqrt{1-v^2}\big)=b\rho}\] regularly are the regions of unstable flow. 

After the pseudovector 
$\color{blue}{\vec h}$ excluding the 
equation system contains only usual 
quantities and is closed. However, 
reduced system is closed only 
formally, in general case the 
compatibility of its solutions with 
kinematic restrictions 
for pseudovector 
$\color{blue}{\vec h}$ needs to check. 
Or another restriction for extra 
equations elimination brings into being. 

Other essential difference 
with usual model is that in general 
case a stream is not continual. When 
velocity of stream reach the value of 
the sound velocity then the phase of 
quantity 
$\color{blue}{m}$  is changing 

\[\color{blue}{\rho\sqrt{1-v^2}\to i\rho\sqrt{v^2 -1}}\] 
and we are going in area which is space-like for  mechanical but time-like 
for fundamental interactions.  
For example, the three-vector equation 
in space-like area is 
\[\color{blue}{\partial_t(m\vec h)-\overrightarrow\nabla\times(m\vec v)= 
-a\rho(\vec v+\vec h\times\vec v)}\] 
Both intuitively and formally, this is 
abeyance area because here is the 
solution $\color{blue}{f=0}$ which is trivial for usual zone but unexpected for 
space-like area. 

The mass conservation low is valid 
anywhere if the relative  
velocities in stream are  less 
essentially of  light velocity. 
Remark once again, the density of mass is external parameter. 

Because here the sound velocity always 
is the constant it is somewhat unusual 
thermodynamic situation, nevertheless, it is real in many cases. 
Not thermodynamic but field 
approach was using, however, the 
scalar 
equation after integration gives the 
thermodynamic connection. 

The $\color{blue}{m\xi}$, not 
$\color{blue}{\xi}$ itself, exist in 
nature so initially  the 
$\color{blue}{m\xi}$ needs take as 
potential of mechanical field. 
Nevertheless, because of mass 
conservation low, it is possible 
the dividing of the mass from other variables.

Of course, this is the simplest field model of fluid.  

For example regard few streams. Only the short equation system, because of its formal fullness, is using. 

\subsection{Ideal fluid} 

In this case all internal 
interactions ignored and the equations for extended pressure $\color{blue}{p_c}$ and impulse $\color{blue}{m\vec v}$ 
are isolated wave  equations 
\[\color{blue}{(\partial_t^2-\Delta)p_c=0}\] 
\[\color{blue}{(\partial_t^2-\Delta)m\vec v=0}\] 
Elementary motion is the stationary 
stream of uncompressed fluid 
restricted by two sheets with gap 
$\color{blue}{x=[0,L]}$. For 
geometry 
\[\color{blue}{\vec v=v(x)\vec e_z}\] 
and with boundary conditions 
$\color{blue}{v(0)=0;\;v(x_1)=v_1}$ 
the velocity of stream is 
\[\color{blue}{v^2=\frac{c^2}{ 2}\Big(1-\sqrt{1-k^2x^2}\Big)}\] 
Such flow is possible if the integration constant $\color{blue}{k}$ is small, $\color{blue}{kL<1}$. 
Moreover, the stream becomes unstable 
if the velocity $\color{blue}{v_1}$ 
is big enough. Indeed, 
if the forces are absent the motion 
of a particle on straight line is 
going with zero acceleration, the 
similar motion of ideal fluid take 
place with zero gradient of pressure 
along the velocity direction. Then 
the pressure in this flow has no the explicit dependence on space point, it is 
\[\color{blue}{p=c^2\rho+q\frac{c^2\rho}{\sqrt{1-v^2/c^2}}}\] 
where, because from data in any fluid 
$\color{blue}{p<c^2\rho}$, the 
integration constant 
$\color{blue}{q}$ is less of zero.
(The discrepancy with Bernoulli low 
is caused by simple expression 
for compression modulus).   
Correspondingly, for big velocities 
the pressure becomes negative that 
means the instability of flow. 
These properties of simplest stream 
are similar to the ones of the plane Couette flow, however the last is being for not ideal fluid. 

For pipe stream the situation is similar so sometimes the coaxial but not simple tubing will be best for a fluid moving. 

In case of the stream with geometry 
\[\color{blue}{\vec v=v(x_\perp)\vec e_\varphi;\;\;\vec x_\perp \cdot \vec e_\varphi=0 }\]
which stands, for example, to 
different rotations in atmosphere, 
the solutions are:

or \[\color{blue}{v=0},\] 

or \[\color{blue}{v\sqrt{1-v^2}=Ax_\perp +\frac{B}{ x_\bot}}\] 
Correspondingly, in general case as 
the cyclones as the anticyclones are 
concentrated, due to constrain 
$\color{blue}{v<c}$, in the some 
finite areas, out of the ones the 
fluid is or immovable or there is the 
whirl in whirl structure of flow.  

Ideal fluid is the first face of the 
any. At least qualitatively, the 
consideration of this system reveals 
the main features of real fluids. So 
it is possible to expect the 
emergence as $\color{blue}{ h}$, 
they are created by accelerations, 
as $\color{blue}{ v}$, they become 
due to interaction spreading 
finiteness, singularities in the 
streams. Because 
$\color{blue}{f=-\vec v+i\vec h}$ it 
is questionably that the 
approximation of 
$\color{blue}{a,\;b}$ functions 
by means of constants is valid for 
unstable stream. 

In Newtonian mechanics any interaction spread with infinite velocity and 
it is well approach if the relative 
speed of the particles is more less 
of light velocity. But in fluid 
dynamics certainly it needs take 
into account the 
field conception on point to point 
spreading of the mechanical 
interaction with velocity of sound. 
This may be make up using the field 
theory tools as it 
is done in this article, or using 
the methods of relativistic mechanics, 
or by other way, but this needs doing. 

\section{Nonlinear pionic field}
\subsection{Potentials of pionic field}

In low energy area the atomic nuclei 
can be regarded as consisting of nucleons, 
those are the protons and neutrons. The  
nucleons are surrounded by pionic field 
and interact with other nucleons via 
their common pionic field.

Yukawa introduced the conception of pionic 
field into physics and found the expression 
for simplest potential of this field as 
\[\color{blue}{p\sim \frac{exp(-mR)}{R}}\]
where $\color{blue}{m}$ is the mass of free pion.
 There is other solution of linear Yukawa equation, 
namely
\[\color{blue}{p\sim \frac{exp(mR)}{R}}\]
Commonly the last solution is regarding as unphysical one 
but in fact the situation is reverse, this  
will be clear up in the next subsection. 

The history of application of Yukawa potential 
has rises and falls. Today in nuclear physics 
the chiral effective field theory is in fact 
the theory of pionic field.

We regard in-nuclei pionic field as 
classical nonlinear field, however within 
uncustomary frame. Because for hundred 
years the conception of selfinteracting fields 
 brings variety of infinities this 
approach is rejected. Instead the conception of 
inertial field, via which the fields interact between 
themselves, is explored. Of course, any theoretical 
model is the speculation, sometimes in their basic even 
commercial one. But for electromagnetic interaction the 
field of inertial forces, labeled 'w-field', shown  
to be  working tool. In addition to this remarks, 
at calculation of planet motion the Ptolemaic model in same 
aspects exceed the Newtonian one in the results and cost 
of calculation. So we cannot divide true and untrue models, 
any model is acceptable if it brings well results.

For pionic field the first step is the consideration 
of virtual particles cloud in a 
nucleus as a continual state. Then in 
general case the additional parameter, 
$\color{blue}{u(x)}$, exists; by physical 
meaning it is the 
local fourvelocity of virtual medium. 
The pion in the cloud is far off mass 
shall and for moving a pion to free 
space it needs to spend the energy 
no less of $\color{blue}{mc^2}$.   
Therefore, for pionic cloud 
$\color{blue}{u^2<0}$. Of course, 
this condition can be valid in many 
other cases. The cloud of virtual pions form the pionic field 
in nucleus. So in-nucleus pionic field has two 
local parameters - the potential $\color{blue}{\varphi(x)}$ and 
fourvelocity vector $\color{blue}{u(x)}$; both  are essential 
for description of the field. 

For quantitative description of  
virtual pionic cloud in classical 
physics it is need to build the 
lagrangian of the fields. On this stage the 
main constructing requirement is: 
the source of the field, it 
is the quantity 
\[\color{blue}{\frac{\delta L_{int}}{\delta \varphi}},\] 

\noindent must be the linear function of fourvelocity.
The pion is pseudoscalar isovector 
particle and so the pionic field has 
these properties. Simplest lagrangian of 
pionic field with its shadow w-field take 
as following (always using Clifford algebra  
\cite{e}, \cite{a} ,\cite{s}, the brief review 
of this algebra is at the end of article. 
\[\color{blue}{L\sim\frac{1}{2}(\nabla \varphi)^2+\frac{k^2}{2}(\nabla u)^2 +g\varphi(u\cdot \nabla)\varphi}\] 
\[\color{blue}{\nabla=e_0\partial_0 -e_n\partial_n}\] 
In this lagrangian the first and second terms are the 
lagrangians of free pionic and free w-fields. 
Today, for usual interactions, the square 
form of the ones is an axiom. 

Third term is simplest lagrangian 
for interaction between pionic and w-field. 
Note, the selfinteraction as pionic as w-field 
is discarded.  

Using the standard way of variation  
formalism we get the field equations 
in form 
\[\color{blue}{\nabla^2 \varphi =-g \varphi \nabla \cdot u}\] 
\[\color{blue}{k^2\nabla^2 u=\frac{g}{ 2}\nabla \varphi^2}\] 
For static spherical symmetrical field 
\[\color{blue}{\varphi=\varphi(R), \; { }  u \gamma_0 =u(R)\vec e_R}\] 
and after velocity parameter excluding 
the equation for pionic potential is 
\[\color{blue}{\varphi''+\frac{2}{ R}\varphi' =-g_1\varphi+c_1  \varphi^3}\] 
In 70's Heisenberg and teams work with such type of equations for pionic 
potential. With time that go out but in today high energy physics 
similar equations are using widely. However, the consideration of 
the equation which in right part contain only 
pionic potential return us to selfinteracting field. For this reason 
the upper equation is discarded. Instead the modification of standard 
variational task, 
labeled 'coherence condition', is explored. More details on the coherence condition 
is at the end of the article. 

\noindent {Following coherence condition is choosen:}

  in the interaction lagrangian the fourvelocity and 

  last  pionic term, which undergo differentiation, 

 are regarding as not variable quantities. 

\noindent{In this case the solutions of variation task are 
following:}

 w-field is in free state, and so      
\[\color{blue}{u(R)=\frac{c_{0}}{R^2}+c_{\infty}R};\]

the tension of pionic field is 
\[\color{blue}{\varphi'\sim \frac{1}{R^2}exp(-\frac{a}{R}-\frac{R^2}{b^2})}\] 
were $\color{blue}{c_0,\;c_{\infty},\;a,\;b}$ are integration constants. 
Hence we can declare that pionic field is of short-range 
type if the integration constant 
$\color{blue}{c_{\infty}\neq 0}$. But in fact such 
declaration is similar to political one, more exactly these 
forces are surface forces. 

In case when all interaction constants are independent quantities 
the coherence 
condition eliminate all solutions unless those constants  are connected 
between themselves. 

It was no found, maybe because no attempt was done, any track of w-field 
on big distances; for this reason we put  $\color{blue}{c_{\infty}=0}.$ 
In this simplest case the expression for pionic potential is
\[\color{blue}{\varphi\sim exp(-\frac{a}{R})}\]
where $\color{blue}{a}$ is interaction constant. 

This looks as result of standard approximation procedure: as 
if firstly pionic potential take equal to zero that return 
the potential of free w-field; that expression was inserted 
in the equation for pionic potential, in returns above 
printed  simplest pionic potential emerged. However, 
coherence  condition instead of perturbation procedure was 
applyed and this issue is explicit solution of variation 
task.        

Pionic field is pseudoscalar object. However, in Clifford algebra two 
different pseudoscalars exist at once: usual $\color{blue}{i}$ 
which also is 
presented in non relativistic quantum mechanics, 
 and chiral $\color{blue}{i_c=e_0e_1e_2e_3}$ which is 
relativistic quantity typically labeled as $\color{blue}{\gamma_5}$ 
matrix. For compatibility with quantum field theory we must take 
the simplest nonlinear potential of pionic field as
\[\color{blue}{\varphi\sim i_c exp(-\frac{a}{R})}\]
This is simplest potential of nonlinear pionic field, below 
referred as extended Yukawa potential.

\subsection{Nucleon in pionic field}
We do ordinarily - the nucleus is quantum system where the point-like 
nucleons are moving in classical fields. 

For description of the nucleon moving in the external field 
the Dirac equation is relevant. 

The Dirac equation initially was 
grounded on Clifford algebra and for
inclusion of the interaction between the 
fermion and the field with any 
algebraic structure the problems do not 
appear. The 
equation for wave function of the nucleon 
in pionic field is following 
\[\color{blue}{(i\nabla -i_c V)\Psi=M\Psi}\] 
\[\color{blue}{V=G exp(-\frac{a}{R})}\] 
\[\color{blue}{\hbar=1,\;c=1}\] 

For stationary states with energy $\color{blue}{E}$ 
the equation for upper part of wave 
function is 

\[\color{blue}{(\Delta +V' \vec e_R )\Psi_{up}=(M^2 
-E^2+V^2)\Psi_{up}}\] 
where $\color{blue}{M}$ is the mass of free nucleon.
Obviously, this wave function is not 
simple factorisable term, it is the 
sum of two terms with different 
algebraic structure and differential 
equations for radial parts of wave 
function have forth order. By physical 
meaning the term 
$\color{blue}{V'\vec e_R}$  represents 
the spin-orbital interaction. It is strange   
but two new issues - quadratic dependence of 
nucleon potential energy on the pionic potential  
and spin-orbital interaction of the nucleon in 
pionic field - where not counted for today. 

The squaring of pionic field potential, that emerge 
due to chiral nature of pion, means that 'true' Yukawa 
potential is unphysical one for producing only repulsive force.
Instead, increasing on big distances Yukawa potential can have 
matter for as nuclear as hep-physics. 

Spin-orbital interaction in nuclei needs for separate consideration.

\subsection{Spin-orbital interaction }

For angle part of wave function 
there are two independent 
matrix-solutions \cite{e} with opposite 
parities, $\color{blue}{\vec e_RS
(\theta, \varphi)}$ and $\color{blue}
{S(\theta,\varphi)}$, they are  
connected by angle part of the gradient operator 
\[\color{blue}{\vec\nabla=\vec e_R\partial_R+\frac{T}{ R}}\]
as 
\[\color{blue}{T\vec e_R S=
kS;\;\;TS=(2-k)
\vec e_R S}\]
where the coefficient $\color{blue}{k=\{-l,\;l+1\}}$ for $\color{blue}{j=\{l+1/2,\;l-1/2\}}$.
When spin-orbital interaction is switched off the states of the nucleons 
are parity degenerated.

Upper part of wave function can be written as 
\[\color{blue}{\Psi_{up}=F_1(R)\vec e_RS+F_2(R)S},\]
the dependence of $\color{blue}{\vec e_RS}$ on the 
jointed Legendre polynomials is 
\[\color{blue}{\vec e_RS=\big[P^m_l(\theta)+i\vec e_\varphi P^{m+1}_l(\theta)\big]exp(im\varphi\vec e_z)}\]
The equations for radial parts of wave function are 
\[\color{blue}{F_1''+\frac{2}{R}F_1'+V'F_2=\big[\frac{k(k-1)}{R^2}+m^2-E^2+V^2\big]F_1}\]
\[\color{blue}{F''_2+\frac{2}{R}F_2'+V'F_1=\big[\frac{(k-1)(k-2)}{R^2}+m^2-E^2+V^2\big]F_2}\] 
Finding of formal solution of those coupled equations is easy, let us put  
\[\color{blue}{F_2=cF_1}\]
where $\color{blue}{c}$ is unknown constant or, in general case, function. 

In this case we can deal with single equation, namely
\[\color{blue}{F''+\frac{2}{R}F'=\big[\frac{l(l+1)+s_{\pm}}{R^2}+m^2-E^2+V^2\big]F}\]
where \[\color{blue}{s_{\pm}=\pm(j+1/2-\sqrt{(j+1/2)^2 +(R^2V')^2}}\]
the signs $\color{blue}{\pm}$ are for $\color{blue}{j=l\pm 1/2}$ states  
and, if the potential is known, the quantity $\color{blue}{(R^2V')^2}$ 
can be approximated numerically by constants. 

However, similarly to scalar parts of fourvector quantities such as relativistic energy 
or electrostatic potential, the pionic potential is of odd-time number; 
those numbers can be taken with as plus as minus signs. So in the previous 
formula we can replace  $\color{blue}{s_{\pm}\to -s_{\pm}}$, the experiment 
or explicit convention can fix the sign of spin-interaction constant.

\subsection{Shell model}

In wide use the shell model rely on the potential of nucleon-nucleon 
interaction, the one is extracted at analysis of NN scattering data or 
taking the Wood-Saxon as well as other phenomenological potentials. 
Then go to solution of many body problem. 

In our approach the potentials of pionic field were found as self-sustaining 
objects; for calculation of binding energy of single nucleon, and so the 
energies of whole nucleus, there no need for struggle with many body 
problem, instead the dependence parameters of potential on the 
numbers of proton and neutron in the nuclei must be found.

Below extended Yukawa potential will be applied 
to description of light nuclei structure.  This potential 
is the solution of linear differential equation for field 
tension only. Hence the arbitrary constant, some kind of 
vacuum potential, can be added too,namely  
\[\color{blue}{\varphi =G_0\varphi_{\infty}\Big(1+f(p,n)(exp(-\frac{a}{R}\Big))}\]  
where $\color{blue}{G_0}$ is interaction constant of the nucleus 
with pionic field and at once the depth of potential well; in 
general case this constant can carry isospin index that is ignored 
because here proton and neutron are regarding separately. The 
function   
\[\color{blue}{f(p,n)=\Big(1-c\big[(-1)^p+(-1)^n\big]/q(p,n)^{(1/3)}\Big)}\] 
represent the well known empirical fact - nucleon binding energy is 
biggest for nuclei with even number as neutrons as protons, less in 
case when one of those numbers is even but other number is odd, and 
more less if both numbers are odd; there $\color{blue}{c=\varphi_0/\varphi_{\infty}}$  is small unknown constant, 
$\color{blue}{\varphi_0,\;\varphi_{\infty}}$ are potential 
values at $\color{blue}{R\to 0, \;R\to \infty}$, respectively. Of course, it is only testing parameterization 
for  $\color{blue}{f(p,n)}$. 

Correspondingly, we relabeled 
\[\color{blue}{G_0\varphi_{\infty}\to G}\]  
- it is new depth of pionic well. The meaning of  parameter 
\[\color{blue}{q(p,n)}\]
will considering below.

You must be conscious that potential of any nonlinear field do not 
contain any arbitrary additive constant. Such constant 
in the expression for pionic potential appeared because 
w-field was taken in free state. In the result the equation for 
pionic field contains only the tensions and so the potential 
acquired additive arbitrary constant.  This constant in extended 
Yukawa potential is the simplest step to take into account 
the real non-linearity of pionic field equations.     

Not only pionic field, also other fields contribute to 
forces acting on the nucleons in nuclei; they are Coulomb 
and vacuum gluonic fields, as well as somewhat else. 

The influence of electrostatic field on the structure 
of light nuclei is negligible and we discard this field.

 About vacuum gluonic field the remarks are relevant:

in case of extended Yukawa potential the preliminary 
fit of binding energies of lightest nuclei shown overestimation of the ones. 

relying on A. V. Thomas \cite{Thomas}, in MIT bag model an 
vector field shifts down the energies of nucleon in nuclear medium;

These reasons push us to include  vacuum potential of gluonic field  to acting in the nuclei. 

Concerning somewhat else that can be found anywhere by anyone. 

\subsection{Approximation of the potentials}
 Main influence on 
the structure of light nuclei is that of pionic field gives.
On small distances extended Yukawa potential quickly decrease, 
the forces act only on the nucleons which are near surface 
$\color{blue}{R=a}$; it is just the property of liquid drop, some kind of 'asymptotic freedom'. 
In other words, the pionic field forces have surface nature. For this 
reason the quantity $\color{blue}{q=Ga}$, which by physical meaning is 
the charge of pionic field, can be written as 
\[\color{blue}{q\sim(A-1)^\frac{2}{3}}\]
where $\color{blue}{A=p+n}$ and $\color{blue}{p,\;n}$ are the number of protons and neutrons in the nucleus. 

The expression for pionic potential was found 
in the mass center of $\color{blue}{}{(A-1)}$ nucleons 
that we are taking into account parametrize the charge for nuclei 
with equal number of protons and neutrons as 
\[\color{blue}{q_0=k\frac{A-1}{2\sqrt{pn}}(A-1)^{2/3}}\]
where $\color{blue}{k}$ is fitting constant. 
For $\color{blue}{n\neq p}$ nuclei the intuition works worse; chosen 
expression for pionic field charge is following  
\[\color{blue}{q=q_0\Big(1-k1\frac{|p-n|}{\sqrt{min(p,n}}\Big)^{2/3}\Big(\frac{2\sqrt{pn}}{p+n}\Big)^{2/3}}\]
The condition $\color{blue}{q(2,9)=0}$, which means that $\color{blue}{10He}$ nuclide is the last upper one, determinate unknown 
constant as $\color{blue}{k_1=0.202}$. The last multiplier was 
selected by hands to ensure suitable first excitation in $\color{blue}{}4H$ nucleus. 

With increasing of nucleon numbers the extended Yukawa 
potential on the not small distances became similar to 
Wood-Saxon as well as to smooth box potentials.

Aimed to reach the explicit expression for energy of the 
nucleons in pionic field the potential energy  
is approximated as
\[\color{blue}{V^2=\big(G^2-2G\frac{qf}{ R}+\frac{q^2(f+f^2)}{R^2}\big)}\]
This type potentials are known in math-ph as 
Kepler potential while in physics as Kratzer 
potential \cite{cc}, sometimes as Hellmann 
potential \cite{Hellmann},  
all names are true.

For vacuum potential of the fields the independence on the 
mass  number is assumed. But at moving to center mass of the nucleus we 
parametrize the subtraction as multiplication of two terms, the first one act for $\color{blue}{p=n}$ nuclei while the last expand the action to $\color{blue}{p\neq n}$ area, namely
 
$$\color{blue}{V_{subtr}=-2k_s\frac{(p+n-1)}{(p+n)}\frac{2\sqrt{pn}}{(p+n)}}\times$$
$$\color{blue}{\times\Big(1-0.202\frac{abs(p-n)}{\sqrt{min(p,n)}}\Big)^{2/3}\Big(\frac{2\sqrt{pn}}{p+n}\Big)^3}$$
The constant $\color{blue}{k_s}$ is searching from physical  
reason: the deuteron do not have any excitation; 
commonly this is explaining by small depth of potential well 
but then it is unclear why the resonances are not observed; as 
for me the absence as excitation as 
resonances means - the energy of fist excitation in the deuteron 
is equal exactly to value of subtraction. Asymmetric part of subtraction borrowed from expression for pionic field charge.

As somewhat else we take the fourvelocity of the nucleons in nuclei.  
The velocity of the nucleons in the nucleus is smallest compare with light 
velocity, so the fourvelocity of the nucleon  has almost zero space 
part, on the big distances it can be written as  
\[\color{blue}{U^2\simeq 1\;\; \to \;\; U\gamma_0\sim U_0;\;\;U_0\simeq c0+\frac{c1}{R}+\frac{\beta}{R^2}+...}\]
At least in first approach those quantity is decoupled from fourvelocity of 
gluonic field. We expect that self fourvelocity of the nucleus generate 
small forces acting on the mass of the nucleus. About those forces we know nothing 
but something can be assumed: vacuum constant will be jointed to gluonic 
subtraction; $\color{blue}{c1=0}$ because the existence of mass charge 
will be too strong assumption; the parameter $\color{blue}{\beta}$ can be not zero - this we suppose. Correspondingly, the small perturbation of 
nucleon mass due to motion of the nucleons can be written as    
\[\color{blue}{1+\frac{\delta M}{M}=1+\frac{\beta}{(N+l+1)^3(l+0.5)}}\]  
there averaging on the Coulomb functions was taken; constant 
$\color{blue}{\beta}$ is fitting parameter. This step is 
invoking to avoid the overloading of basic parameters, 
 $\color{blue}{G,\;k,\;c}$, as well as because their 
dependence on quantum numbers is unbelievable. 

It is exotic approach but the choice of corrections to main 
free parameters is unknown road also. 
The linearity equations for tension of simplest 
Yukawa potentials ensure many variants.   
For example, Luneburg-lens-like approaches, for details on 
 the ones see \cite{Ohkubo} and references therein, suggests 
 the consideration 
of the potential as sum  $$\color{blue}{V{(G,k)}+V{(g1,k1)}}$$ 
to which the constant potential can be added; in this case 
the approximation 
up to $\color{blue}{R^{-3}}$ term brings, in fact, 
pure phenomenological potential 
of Kepler type with many parameters that ensure well fit without 
involving unknown inertial 
forces. 

\subsection{Numerical results}
In this way, the  main unknown constants are: the depth of 
pionic well $\color{blue}{G}$ which is at once interaction 
constant of the nucleon with pionic field;    
 the charge $\color{blue}{q}$ with unknown constant 
$\color{blue}{k}$; the constant  $\color{blue}{c}$ which was 
introduced into vacuum potential $\color{blue}{f}$ to scale 
$\color{blue}{\varphi_0/\varphi_{\infty }}$ values and 
burdened by some kind of staggering; the constant $
\color{blue}{\beta}$ is unknown parameter in mass term. About staggering see \cite{Bertsch} and 
references there.    

For convenience the binding energy,  $\color{blue}{\varepsilon}$, of the 
nucleon in the nucleus is regarding as difference between 
the mass of free nucleon and the mass of nucleon in the nucleus, so it is 
positive number.

Taking into account the above remarks the binding energy of 
the nucleon in extended Yukawa potential is following:
\[\color{blue}{\varepsilon=V_{subtr}+\Big(1+\frac{\delta M}{M}\Big)\varepsilon_{pion}}\]
here $\color{blue}{\varepsilon_{pion}}$ is the energy of 
nucleon in pionic field without subtraction, first term   
refers to subtraction energy which is negative number. 

The nucleon energy in pionic field and extended orbital moment 
$\color{blue}{B}$ are given as   
\[\color{blue}{\varepsilon_{pion}=M_{eff}\Big[1-\sqrt{1-\frac{G^2q^2}{M_{eff}^2(B+N+1)^2}}\;\Big]}\] 
\[\color{blue}{B=\frac{1}{2}\Big[-1+\sqrt{(2l+1)^2+8q^2+4s}\Big]}\]
where $\color{blue}{M_{eff}=\sqrt{M^2+G^2}}$ is effective mass 
of the nucleon in pionic field;
the expressions for $\color{blue}{q,\;\delta M}$ are 
in previous subsection; $\color{blue}{M=938.918}$ is nucleon mass;
 all energies are in MeV. 

Concerning SO-interaction the essential remarks are relevant. 
Formally, the splitting of nucleon levels is alike in oscillatory 
and pionic potentials. The nucleon shells are marked by main 
quantum number $\color{blue}{n}$, each shell contains $\color{blue}{n+1}$ states.  

In pionic well main quantum number of the nucleon is 
$\color{blue}{n=N+l}$ 
were $\color{blue}{N,\;l}$  are radial and orbital moment quantum 
numbers, respectively. Lowest, with biggest energy, shell $\color{blue}{n=0}$ contains 
single state $\color{blue}{(0,\;0)}$; the next shell $\color{blue}{n=1}$ 
contains two states $\color{blue}{(0,\;1)}$ and $\color{blue}{(1,\;0)}$; 
the next $\color{blue}{n=2}$ shell contains three states 
$$\color{blue}{(0,\;2)>(1,\;1)>(2,\;0)}$$ and so one. However, 
this is not absolute labeling because with growth of pionic field 
charge the shells overlapping. 

Similar scheme is for labeling nucleon levels in oscillatory potential with 
spin-orbital interaction. The difference between two schemes for 
levels marking is  that levels splitting in the shell is emerged 
due to  radial quantum number for nucleons in  nonlinear pionic 
field while for nucleon in an abstract oscillatory potential due 
to SO-interaction; the formalism of the last was adopted from EM 
theory. 

If in nonlinear pionic well the SO-interaction will 
switched on the additional  splitting of levels emerged. This is 
not observed, hence the question about SO-interaction in pionic 
field arose. It is similar to situation in atomic physics were 
the exactness of measured lines for one- and many- electron 
atoms differ essentially; as consequence for today in many-electron 
atoms only orbital shift of the lines is visible because the  
cloud of electrons produce stronger effect compare with SO-part of 
nucleus electrostatic field. 

Due to non-relativistic motion nucleons in the nuclei 
we can conclude that the interaction, about which in the upper 
subsection was assumed as representation of SO-interaction in pionic field, is untrue conjecture. Instead we assume that term 
$\color{blue}{\vec e_R V'}$ in Dirac 
equation is creating additional orbital shift of nucleon levels. Maybe the 
spin-orbital splitting can be visible at much more exact measuring of nucleon lines. Correspondingly, from four solutions for SO-parameter  
 \[\color{blue}{s=\pm(j+1/2)+\big[\pm\sqrt{(j+1/2)^2+\lambda^2q^2}\big]}\]
one only must be chosen and it is the reason to conserve 
$\color{blue}{(N,l)}$ scheme for states; note that here the 
'plus-minus' signs near root are not correlated with first similar 
signs. 

Selected 'SO-coefficient' is
\[\color{blue}{s=(l+1)-\sqrt{(l+1)^2+\lambda^2q^2}}\]
were fitting parameter $\color{blue}{\lambda}$ is averaged value 
of $\color{blue}{R^2 V'}$. This parameter can depends from nucleon 
numbers in the nucleus; in this case unexpected phenomena will 
take place; for example at big, $\color{blue}{\lambda > q}$, 
values  of this 
parameter the S-states as well as another ones with small orbital 
momentum became unstable. It seems similar situation is for 
electron levels in many-electron atoms. 
But for simplisity we put $\color{blue}{\lambda=0.5}$ from simple geometrical reason.  

Certainly, this simplification is not valid for 
rising potentials, such as oscillator or 'unphysical Yukawa' potentials 
because the quantum averaging of the ones brings in the SO-coefficient the term 
which quickly rise with quantum numbers of the nucleon. 

Self-jointed, $\color{blue}{p=n}$, nuclei are regarding as consisting of proton-neutron pairs. 
The reasons are the absence of bound states in neutron-neutron system and observed spectra of alpha particle. 

The data are from \cite{i} and AME2012 mass evaluation(II), 
below they are printed 
in  black. 

For fixing four unknown constants, 
\[\color{blue}{G,\;k,\;c,\;\beta\;}\]
the set of calibration points is following:
 binding energies of deuteron [2.22456..], alpha 
particle  [28.29566..] and 6Li [31.99398..] together with 
first observable line of 6Li, [2.186].

 Chosen structure of those nuclei is:
single pn-pair in (0,0) state for deuteron, two pn-pairs 
in (0,0) state for 4He, two pn-pairs in (0,1) and single 
pair in (1,0) states for 6Li;  
those are labeled as 
\[\color{black}{2H=(0,0),\;4He=2(0,0);\; 6Li=2(0,1)+(1,0)}\]
Preliminary fit shown that first observable excitation of 6Li is 
single pair transition from (1,0) to (0,2) state, this transition 
is marked as $\color{black}{(1,0-0,2)}$; below all transitions  
and ground states for $\color{black}{p=n}$ nuclei are labeled in 
this manner while for $\color{black}{n>p}$ nuclei such labels are 
referred to single nucleons. Please, take these labeling in the  
attention.   
 
Because 'sagemath.org' became too wide we run SciPy 
optimization tools . Double minimization return following values 
$$\color{blue}{G=289.418...MeV,\;k=0.4146...,}$$
$$\color{blue}{c=-0.00183..,\;\beta=-0.00513..}$$

Note, the tail of constants can slowly vary that depend on 
the initial conditions and chosen tools.

We get away from standard road of statistics, mainly 
because four points were taken for four parameters fit, 
as if the toy model is exact theory. But it is too doubtful that 
'true' fit with few hundred free parameters and thousands 
of data point lighten the nuclei theory.  

\subsection{Z=1\; nuclei}
 
          $$\color{red}{2H}$$
Binding energy of the deuteron is

$\color{blue}{2.224564}\;\;\color{black}{\;[2.224566]}$

  $$\color{red}{3H}$$
The triton is unstable in week interaction while 
$\color{blue}{3He}$ 
is stable. Common explanation of this phenomenon rely on QED, but 
in the frame of classical field it can be caused by big binding 
energy of the neutron in $\color{blue}{(0,0)}$ state; also the 
calculation gives unsatisfactory result for triton binding energy 
if all nucleons are in lower state. So chosen structure is with 
single neutron in $\color{blue}{(1,0)}$ state, namely 
$$\color{blue}{2(0,0)+(1,0)=8.405}\;\;\color{black}{[8.482]}$$
Additionally, the excitations of triton were not found and so 
an complicated subtraction, or another mechanism, need find - now 
we discard this trouble. 

$$\color{red}{4H}$$
Selected ground state and transitions are following 
$$\color{blue}{3(0,1)+(1,0)=7.141}\;\;\color{black}{[6.88]}$$
$$\color{blue}{(0,1-1,0)=0.326}\;\;\color{black}{[0.3304]}$$
$$\color{blue}{(0,1-1,1)+(1,0,-2,0)=2.063}\;\;\color{black}{[2.077]}$$
$$\color{blue}{(0,1-2,0)+(0,1-1,0)=2.822}\;\;\color{black}{[2.846]}$$
     $$\color{red}{5H}$$
At present the spectral data absent and assumed structure of ground state is chosen as  
$$\color{blue}{3(1,0)+2(1,1)=6.757}\;\;\color{black}{[6.680]}$$
For such structure the excitations are possible, for example for first excitation 
can be two equivalent variants 
$$\color{blue}{(1,0-1,1)=0.837}\;\;\color{black}{[?]}$$
$$\color{blue}{2(1,1-1,2)=0.837}\;\;\color{black}{[?]}$$
For unknown for me reasons the works about 5h lines are absent  
while the works about 5H resonances are not 
rare, for example see \cite{ua}, \cite{laz}
    $$\color{red}{6H}$$
For this isotope the spectral data absent, supposed structure of 
6H is 
$$\color{blue}{2(0,1)+2(1,0)+2(2,0)=5.836}\;\;\color{black}{[5.76]}$$
If two last states are filled by neutrons the transitions from 
lower state are blocked - the reason to choice this structure. 

This is last isotope in Z=1 isotope chain.

\subsection{Z=2 nuclei}        
$$\color{red}{4He}$$
Ground state energy is 
$$\color{blue}{2(0,0)=28.29567}\;\color{black}{[28.29566]}$$
The data show clearly that whole nucleus takes part at lower excitation, 
that is treated as clustering of the deuterons  in 4He.
The ladder of pn-pairs excitation from ground state can be 
following
$$\color{blue}{(0,0-0,1)+(0,0-0,2)=20.217}\color{black}{[20.210]}$$
$$\color{blue}{(0,0-0,1)+(0,0-2,0)=20.918}\color{black}{[21.010]}$$
$$\color{blue}{(0,0-1,0)+(0,0-1,1)=21.888}\color{black}{[21.840]}$$
$$\color{blue}{(0,0-1,0)+(0,0-2,1)=23.282}\color{black}{[23.330]}$$
It is unclear why few lower possible transitions are absent. Of course, 
at perturbation as water as pionic drops the symmetric excitations 
are almost impossible, but more likely that complex mechanisms 
can be found to explain those suppression.  
    
Last stable state is (2,2) while the energy of next, (3,1), state is negative; 
correspondingly the first 'full' resonance is 
$$\color{blue}{2(0,0-3,1)=28.329}\color{black}{[28.310]}$$
The result is acceptable, nevertheless, maybe something 
is missing or made untruly.

General fault of this model is noticeable, it is the variety 
of extra, 
non observable transitions.  

For comparisons with usual models about 4He spectrum see, for example, 
\cite{dd} and references therein.
$$\color{red}{5He}$$
Acceptable configuration for ground state of this nucleus is
$$\color{blue}{(0,0)+4(0,1)=27.73}\;\color{black}{[27.56]}$$
$$\color{red}{6He}$$
Selected configuration for ground state is following 
$$\color{blue}{2(0,0)+(0,1)+(1,0)+2(0,2)=30.918}\color{black}{[29.271]}$$
while fit of transitions is following
$$\color{blue}{2(0,2-1,1)+(1,0-1,1)=1.782}\color{black}{[1.797]}$$
$$\color{blue}{(0,2-0,3)+(0,0-0,1)=5.570}\color{black}{[5.6]}$$
$$\color{blue}{(0,0-1,1)+(0,0-0,2)=13.896}\color{black}{[13.9]}$$
$$\color{blue}{2(0,0-1,1)+(0,2-1,1)=15.2553}\color{black}{[15.255]}$$
$$\color{red}{7He}$$
Chosen configuration for ground state is
$$\color{blue}{3(0,1)+4(1,0)=28.917}\;\color{black}{[28.8617]}$$
while for transitions it is
$$\color{blue}{(1,0-0,2)+(1,0-2,0)=2.955}\color{black}{[2.92]}$$
$$\color{blue}{2(1,0-0,2)+2(1,0-2,0)=5.910}\color{black}{[5.8]}$$
   $$\color{red}{8He}$$
Assumed ground state is
$$\color{blue}{4(0,1)+4(1,0)=32.188}\;\color{black}{[31.396]}$$
The fit of transitions is 
$$\color{blue}{3(1,0-0,2)=3.381}\color{black}{[3.1]}$$
$$\color{blue}{(0,1-1,0)+3(1,0-0,2)=4.515}\color{black}{[4.515]}$$
   $$\color{red}{9He}$$
The fit of ground state configuration is
$$\color{blue}{6(0,1)+3(1,0)=30.530}\;\color{black}{[30.141]}$$
For first transitions there are few variants, chosen is  
$$\color{blue}{(1,0-1,1)=1.173}\color{black}{[1.1]}$$
the next are
$$\color{blue}{(1,0-0,2)+(1,0-1,1)=2.230}\color{black}{[2.26]}$$
$$\color{blue}{(1,0-0,2)+(0,1-2,0)=4.131}\color{black}{[4.1308]}$$
$$\color{blue}{(0,1-1,1)+(0,1-2,0)+(0,1-1,0)=5.055}\color{black}{[5.055]}$$
$$\color{blue}{4(0,1-1,1)=7.627}\color{black}{[8.0]}$$
   $$\color{red}{10He}$$
Possible configuration of ground state is 
$$\color{blue}{3(0,0)+3(0,1)+4(1,0)=29.26}\;\color{black}{[29.98]}$$
For transitions it is 
$$\color{blue}{(1,0-1,1)=3.26}\color{black}{[3.24]}$$
$$\color{blue}{2(0,0-0,1)=6.782}\color{black}{[6.8]}$$
 \subsection{Z=3 nuclei}        
$$\color{red}{6Li}$$
The value of observable binding energy give reason for    
following structure of 6Li:
the lower, (0,0), shell is empty, 
two pn-pairs are in (0,1) state and one pn-pair 
is in (1,0) state. Just this structure was assumed 
at the calibration because it was clear that the energy of (0,0) 
state is big. Binding energy of this 
configuration is
$$\color{blue}{2(0,1)+(1,0)=
31.993987}\color{black}{[31.99398]}$$ 
The data shown that excitations of ground state contain only odd number of pn-pairs. 
In this case the assumed connection between calculated 
and observable excitations is 
$$(1,0-0,2)=\color{blue}{2.186}\color{black}{[2.186]} $$
$$(0,1-1,0)=\color{blue}{3.491}\color{black}{[3.562]}$$
$$(1,0-2,0)=\color{blue}{4.320}\color{black}{[4.312]}$$
$$(1,0-1,2)=\color{blue}{5.272}\color{black}{ [5.366]}$$
$$ (1,0-2,1))=\color{blue}{5.635}\color{black} {[5.65] }$$
$$ 2(0,1-1,1)+(1,0-1,1)=\color{blue}{15.840}\color{black}{[15.8]}$$
$$ 2(0,1-1,1)+(1,0-0,3)=\color{blue}{17.96}\color{black}{[17.985]}$$
$$ 2(0,1-0,3)+(1,0-2,0)=\color{blue}{21.446}\color{black}{[21.5]}$$
First level with negative energy is (6,0) so the last not resonance 
excitation is
$$2(0,1-5,1)+(1,0-5,1)\color{blue}{=31.855}\color{black}{[?]}$$
in fact, this transition coincide with threshold 
of 6Li decay to 
nucleons and so  eliminate all next resonances; this is similar 
to feature of the deuteron.

Here possible but not observable excitations 
were not printed. 

Another model, grounded on clustering in 6Li nucleus, also bring well results \cite{Fraser}. Note, usually the 6Li regard as p-shell nucleus; that means two
protons and two neutrons are in S-state and form inert core while next proton 
and neutron are in valence P-state. In our model the structure of 6Li is reverse because the spectral data strongly 
support such structure.  

The things did not have to 
be so simple as the table shown.
By physical meaning the second observable 
transition, (0,1)-(1,0), is the excitation to isomeric state. 
Observable small width of this 
transition, 8.2 eV, confirms such 
viewing. Empty (0,0) state means that observable 
6Li can be not in lowest state; or empirical principle about minimum 
of potential energy has boundaries for application; or, because the  
w-field prevents as infra as ultra infinities of self-energies, 
the suppression of lowest states is action of inertial 
forces or 'SO' term.  The puzzle deserves as experimental as theoretical 
efforts to clarify the situation. 
$$\color{red}{7Li}$$
Natural configuration of ground state is 
$$\color{blue}{4(0,1)+2(1,0)+(0,2)=39.398}\color{black}{[39.2312]}$$
with this ground state suitable fit of 
transitions is following 
$$ (0,2-1,1))=\color{blue}{0.467}\color{black}{[0.477]}$$
$$ (0,2-0,3)+2(1,0-1,1)=\color{blue}{4.607}\color{black}{[4.630]}$$
$$(0,2-2,0)+(0,1-1,0)+(0,1-1,1)=\color{blue}{6.681}\color{black}{[6.680]}$$
$$(0,2-1,1)+2(0,1-1,1)=\color{blue}{7.407}\color{black}{[7.459]}$$
$$(0,1-1,2)+(0,1-2,1)=\color{blue}{9.634}\color{black}{[9.67]}$$
$$2(0,1-1,2)+(0,2-1,1)=\color{blue}{9.910}\color{black}{[9.850]}$$
However, dilute configurations of halo type 
but with not empty (0,0) state can bring 
well fit of data; it seems that this 
circumstance holds along all isotope 
chains.   

    $$\color{red}{8Li}$$
Possible configuration of ground state is 
$$\color{blue}{2(0,1)+4(1,0)+2(0,2)=41.629}\color{black}{[41.277]}$$
while for transitions it is 
$$ 2(0,2-1,1))=\color{blue}{1.018}\color{black}{[0.980]}$$
$$ (0,2-1,1)+(0,2-0,3)=\color{blue}{2.16}\color{black}{[2.255]}$$
$$ 2(0,2-0,3))=\color{blue}{3.304}\color{black}{[3.219]}$$
$$ (0,1-0,2))+(1,0-2,0)=\color{blue}{5.417}\color{black}{[5.400]}$$
      $$\color{red}{9Li}$$
Acceptable binding energy is 
$$\color{blue}{3(0,1)+(1,0)+5(0,2)=47.0}\color{black}{[45.340]}$$
while for transitions it is 
$$ 2(0,2-2,0)=\color{blue}{2.723}\color{black}{[2.691]}$$
$$ 4(0,2-1,1)+(0,1-1,0)=\color{blue}{4.147}\color{black}{[4.301]}$$
$$ (0,2-1,2)+(0,1-1,1)=\color{blue}{5.416}\color{black}{[5.380]}$$
$$ (1,0-2,0)+2(0,2-2,1)=\color{blue}{6.493}\color{black}{[6.430]}$$
$$ 5(0,2-1,3)+(1,0-1,2)=\color{blue}{16.06}\color{black}{[16.0]}$$
$$ 3(0,1-2,1)+(1,0-1,1)=\color{blue}{16.96}\color{black}{[17.1]}$$
$$ 3(0,1-2,1)+(0,1-1,1)=\color{blue}{19.03}\color{black}{[18.9]}$$
      $$\color{red}{10Li}$$
Acceptable binding energy is 
$$\color{blue}{3(0,1)+(1,0)+2(0,2)+
3(1,2)=45.355}\color{black}{[45.314]}$$
while transition energies are
$$ (1,2-2,1)=\color{blue}{0.224}\color{black}{[0.210]}$$
$$ (0,2-1,1)=\color{blue}{0.484}\color{black}{[0.470]}$$
$$ 3(1,2-2,1)=\color{blue}{0.672}\color{black}{[0.670]}$$
$$ 2(1,2-2,1)+2(0,2-1,1)=\color{blue}{1.395}\color{black}{[1.370]}$$
$$ (0,2-0,3)=\color{blue}{1.597}\color{black}{[1.6]}$$
$$ (1,2-1,3)+(0,2-0,3)+2(1,2-2,1)=\color{blue}{2.938}\color{black}{[2.820]}$$
      $$\color{red}{11Li}$$
There chosen ground state corresponds 2n-halo structure
$$\color{blue}{4(0,1)+2(0,2)+3(1,1)+2(0,3)
=46.716}\color{black}{[45.709]}$$
Lower lines adjusted as
$$ (0,3-1,2)+(1,1-1,2)=\color{blue}{1.276}\color{black}{[1.266]}$$
$$ (0,2-2,0)+(1,1-2,1)=\color{blue}{2.441}\color{black}{[2.474]}$$
$$ (1,1-0,3)+(1,1-1,2)+2(0,2-0,3)=\color{blue}{3.701}\color{black}{[3.70]}$$
It is last lithium isotope with measured lines. 
  
Last lithium isotope presented in database is 13Li while in this 
model 14Li is last isotope in Z=3 chain, so theoretical 
parametrization as subtraction as dripline parameters must be 
improved because with electric charge growth this discrepancy  
will increase. 

Because with growth of nucleon number the 
suitable choice even for first transition 
is uncertain below the fit of few lower lines is printed.  

\subsection{Z=4 nuclei}        
      $$\color{red}{8Be}$$
Because of 6Li structure, the natural configuration for ground state 
of 10Be is: (0,0) state is empty; each of (0,1) and (1,0) states 
contain two pn-pairs.  For this configuration the
binding energy of 8Be is
$$2(0,1)+2(1,0)=$$
\[\color{blue}{=56.687}\color{black}{[56.5]}\]
The energies of lower states are well distinguished, the difference between 
levels is no less of 0.4MeV. As for 6Li the natural restriction on 
the allowed transitions is that the excitations of single or
three pn-pairs are possible. In this case the lowest line is
\[\color{blue}{(1,0-1,1)=3.098}\color{black}{[3.030]}\]
while the next ones are
$$2(1,0-1,1)+(0,1-1,0)=$$
\[\color{blue}{=11.321}\color{black}{[11.35]}\]
$$2(1,0-1,1)+(0,1-2,0)=$$
\[\color{blue}{=16.628}\color{black}{[16.626]}\]
$$2(1,0-1,1)+(0,1-0,3)=$$
\[\color{blue}={16.876}\color{black}{[16.921]}\]
This is acceptable result. The suppression of (1,0-0,2) transition
is natural behavior but that it is reversed to 6Li situation. 
 Among the lot of possible odd transitions those were selected 
as giving best fit.

At main parameters searching the contribution of electrostatic 
interaction to binding energies of 4He and 6Li was not 
extracted, the binding energies were taken 
just from data tables; for this reason binding energies of Li 
isotopes were fitted exactly. For next nuclei the exact fit of 
binding energies is wrong, we must select the 
configuration of ground states which give somewhat big values than 
experimental ones are.  
      $$\color{red}{9Be}$$
Chosen ground state of 9Be is
$$3(0,1)+2(1,0)+4(0,2)=\color{blue}{59.664}\color{black}{[58.154]}$$
while four lower lines are
$$2(1,2-1,1)=\color{blue}{1.653}\color{black}{[1.684]}$$
$$3(0,2-1,1)=\color{blue}{2.479}\color{black}{[2.459]}$$
$$(0,2-2,1)=\color{blue}{2.736}\color{black}{[2.78]}$$
$$2(1,0-1,1)=\color{blue}{3.088}\color{black}{[3.049]}$$
      $$\color{red}{10Be}$$
Chosen ground state is 
$$3(0,1)+2(1,0)+3(0,2)+2(1,1)=\color{blue}{65.769}\color{black}{[64.976]}$$ 
In this nucleus are adjacent lower states, (2,0)=3.542, (0,3)=3.497 with slowly different energies; this circumstance is regarding as source almost degenerated lines. Suitable fit of lower lines is
$$(0,1-0,2)=\color{blue}{3.393}\color{black}{[3.368]}$$
$$2(1,1-2,1)+(0,2-2,0)=\color{blue}{5.947}\color{black}{[5.958]}$$
$$2(1,1-2,1)+(0,2-0,3)=\color{blue}{5.991}\color{black}{[5.989]}$$
$$2(1,1-2,1)+(0,2-1,2)=\color{blue}{6.275}\color{black}{[6.271]}$$
      $$\color{red}{11Be}$$
Commonly this nucleus is regarding as 
having one neutron halo. In nonlinear 
pionic well there is transition
$$\color{green}{(2,4)-(2,5)=0.318}\color{black}{[0.320]}$$
and so 11Be can have height one nucleon 
halo; also the hierarchy of levels,  
(2,4), (5,0), (3,3) etc., prevent 
appearance additional small transitions. 
Because in this nucleus (2,0), (0,3) states 
reversed but still produced 
enough small transitions both these states 
are empty. For these reason suitable 
ground state configuration take as  
$$3(0,1)+2(1,0)+3(0,2)+2(1,1)+(2,4)=$$
\[\color{blue}{67.016}\color{black}{[65.477]}\] 
while transitions can be following
$$\color{blue}{(2,4-2,5)=0.318}\color{black}{[0.320]}$$
$$\color{blue}{(2,4-2,5)+(1,0-1,1)
=1.861}\color{black}{[1.783]}$$
$$\color{blue}{(2,4-2,5)+2(1,1-2,0)=2.676}\color{black}{[2.654]}$$
$$\color{blue}{(2,4-2,5)+2(1,0-1,1)=3.404}\color{black}{[3.400]}$$
However, the structure of the nuclei is the 
question even at small number of nucleons.
      $$\color{red}{12Be}$$
Acceptable ground state configuration is
$$2(0,1)+2(1,0)+6(0,2)+2(1,2)=$$
\[\color{blue}{69.922}\color{black}{[68.648]}\] 
the fit of transitions is 
$$\color{blue}{2(1,2-1,3)=2.091}\color{black}{[2.102]}$$
$$\color{blue}{(1,2-2,2)+(1,0-1,1)=2.737}\color{black}{[2.702]}$$
$$\color{blue}{(1,0-1,1)+(1,0-1,2)=4.589}\color{black}{[4.560]}$$
$$\color{blue}{2(0,2-1,1)+2(0,2-0,3)=5.7}\color{black}{[5.7]}$$
Here last isotope is 17Be, observable dripline 
is 16Be.                      
\subsection{Z=5 nuclei}        
$$\color{red}{10B}$$
Chosen structure of 10B is 
$$(0,1)+(1,0)+(0,2)+(1,1)+(2,1)$$
\[\color{blue}{=64.948}\color{black}{[64.75]}\]
 
Because any clustering, except of deuteron itself, is absent those 
structure is unusual but other variants of 10B structure 
bring troubles 
at fitting of the excitations. Another 
reasons \cite{Zuker},\cite{Ormand} and \cite{Hirsch} also push to  
exotic structure of 10B.

Suitable fit of observable lines is following
\[\color{blue}{(1,0-0,2)=0.748}
\color{black}{[0.718]}\]
\[\color{blue}{(2,1-2,2)=1.761}\color{black}{[1.740]}\]
\[\color{blue}{(0,2-1,1)=2.259}\color{black}{[2.154]}\]
$$(1,1-2,0)+(1,0-0,2)=$$
\[\color{blue}{=3.592}\color{black}{[3.587]}\]
For last printed line other variant of transition exist, but 
the same situation is typical for this model. 

Via two next excitations there are observable transitions with tiny gaps between 
lines which still are not connected with transitions to 
upper states. Sparse structure of 10B ensure the fit some 
of those almost contiguous group of lines
      $$\color{red}{11B}$$
Chosen ground state is 
$$2(0,1)+3(1,0)+2(0,2)+4(1,1)$$
\[\color{blue}{=78.814}\color{black}{[76.205]}\]
The nucleons in (1,1) state form an cluster which begins the work at big excitations. 
The fit of lower lines is following
$$(1,0-0,2)+(1,0-1,1)=\color{blue}{2.143}\color{black}{[2.125]}$$
$$(1,0-2,0)+(1,0-1,1)=\color{blue}{4.454}\color{black}{[4.445]}$$
$$(0,2-0,3)+(0,2-2,0)=\color{blue}{5.000}\color{black}{[5.020]}$$
$$2(1,0-0,2)+(0,1-2,0)=\color{blue}{6.728}\color{black}{[6.742]}$$
$$4(1,1-1,2)+(1,0-0,2)=\color{blue}{6.820}\color{black}{[6.792]}$$
But even for first line as well as for ground state there are few variants.
      $$\color{red}{12B}$$
Selected ground state is extremely symmetric, namely 
$$4(1,0)+4(0,2)+4(1,1)=$$
\[\color{blue}{80.116}\color{black}{[79.575]}\]
Lower lines are fitted as  
$$3(1,0-0,2)=\color{blue}{0.943}\color{black}{[0.953]}$$
$$(1,1-1,2)=\color{blue}{1.623}\color{black}{[1.673]}$$
$$(1,1-0,3)+(1,1,2,0)=\color{blue}{2.650}\color{black}{[2.620]}$$
$$(0,2-1,2)=\color{blue}{2.803}\color{black}{[2.723]}$$
$$(0,2-2,1)=\color{blue}{3.411}\color{black}{[3.389]}$$
There four nucleons in (0,2) state act as 
cluster to prevent low excitations from this state. 
      $$\color{red}{13B}$$
In some sense selected ground state is the anti-core one, namely 
$$(0,1)+4(1,0)+4(0,2)+4(0,3)=$$
\[\color{blue}{86.306}\color{black}{[84.453]}\]
Suitable fit of lines is following
$$2(0,3-0,4)+(0,3-1,2)=\color{blue}{3.446}\color{black}{[3.482]}$$
$$3(0,3-2,1)+(0,3-1,2)=\color{blue}{3.528}\color{black}{[3.534]}$$
$$2(0,3-0,4)+2(1,0-0,2)=\color{blue}{3.676}\color{black}{[3.681]}$$
$$2(0,3-0,4)+(0,3-2,0)+(0,3-1,2)=\color{blue}{3.708}\color{black}{[3.712]}$$
Maybe the measurements confirm the reality of this ground state. 
      $$\color{red}{14B}$$
For best fit of lines the state (0,3) is taken empty otherwise 
many small excitations will be observed; chosen structure for 
ground state is 
$$(1,0)+7(0,2)+4(1,1)+2(2,0)=$$
\[\color{blue}{86.482}\color{black}{[85.422]}\]
The fit of all lines presented in the 
data table is
$$(2,0-2,1)=\color{blue}{0.767}\color{black}{[0.740]}$$
$$(1,1-2,0)=\color{blue}{1.408}\color{black}{[1.380]}$$
$$(2,0-2,1)+(0,2-1,1)=\color{blue}{1.880}\color{black}{[1.860]}$$
$$(2,0-1,2)+(2,0-2,1)+(0,2-1,1)=$$\[\color{blue}{2.077}\color{black}{[2.080]}\]
$$(0,2-0,3)=\color{blue}{2.317}\color{black}{[2.320]}$$
$$(2,0-2,1)+(1,1-2,1)=\color{blue}{2.943}\color{black}{[2.970]}$$
For next Z=5 nuclides the NNDC database does not contain the spectra.
In this model the last nuclide is 21B while the last observable is unbound 19B.

\subsection{Z=6 nuclei}        
In this model overlapping of shells begins at light nuclei. That is essential for   (1,0) and (0,2) state becase small transitions between these states can exist; correspondingly the (1,0) state is  or filled up and so the cluster of 
alpha-particle type appears or half filled or it is empty - that depends from the  value of lowest excitation.      
    $$\color{red}{12C}$$
The structure of 12C is taken as
$$(0,1)+2(1,0)+3(1,1)=$$
\[\color{blue}{94.976}\color{black}{[92.162]}\]
that slowly match usual alpha-clustering structure of 12C but brings more 
natural order of the lines. Other structure with single p-n pair 
in the (0,3) state also can be acceptable. 
With chosen structure the first measured lines are adjusted as
$$2(1,1-0,3)=\color{blue}{4.276}\color{black}{[4.438]}$$
$$2(1,1-0,3)+(1,1-1,2)=\color{blue}{7.616\;}\color{black}{[7.654]}$$
$$(1,1-2,0)+(1,0-2,0)=\color{blue}{9.612\;}\color{black}{[9.641]}$$
$$(1,1-2,0)+(0,1-1,0)=
\color{blue}{10.23}\color{black}{[10.3]}$$
Selected structure of 12C brings suitable  fit of Hoyle resonance but   
unusual structure 12C nucleus. In arxiv.org the lot of articles 
about Hoyle state can be found, for example see \cite{Hoyle} 
and references therein. But the author 
cannot says to itself 
that structure of 12C is known clearly and unambiguously.  
    $$\color{red}{13C}$$
Feasible configuration for ground state is 
$$2(0,1)+(0,2)+4(1,0)+6(0,3)=$$
\[\color{blue}{97.806}\color{black}{[97.108]}\]
while for lines it is 
$$5(0,3-1,2)=
\color{blue}{3.045}\color{black}{[3.089]}$$
$$6(0,3-1,2)=
\color{blue}{3.654}\color{black}{[3.684]}$$
$$(0,3-2,1)+(1,1-2,1)=
\color{blue}{3.864}\color{black}{[3.855]}$$
$$(0,3-2,0)+2(1,0-2,0)=
\color{blue}{6.875}\color{black}{[6.864]}$$
This structure of 13C covers all observable on the today lines but even for first line are different variants. 
    $$\color{red}{14C}$$
For C-isotopes theoretical energy levels of the nucleons differ in small while
their spectral data are not coincide. First measured line in 14C is above 6MeV
that exceed essentially first lines of all nuclides in this as well as others chains.
If 6MeV line indeed is the lowest then feasible is cluster structure of 14C, 
we take the one as
$$2(0,1)+4(1,0)+4(1,1)+4(0,3)=$$
\[\color{blue}{=106.34}\color{black}{[105.248]}\]
The lower lines are adjusted as
$$3(1,1-0,2)+(1,1-0,3)=
\color{blue}{6.073}\color{black}{[6.093]}$$
$$(0,1-2,0)=
\color{blue}{6.565}\color{black}{[6.589]}$$
$$2(0,1-0,2)=
\color{blue}{6.754}\color{black}{[6.728]}$$
$$2(0,1-1,0)=
\color{blue}{6.881}\color{black}{[6.902]}$$
$$(1,0-2,0)+(1,0-2,1)=
\color{blue}{7.016}\color{black}{[7.012]}$$
$$(1,1-3,0)+(1,1-3,1)=
\color{blue}{7.320}\color{black}{[7.341]}$$
Remark that (0,3) state, which is as spectator, itself can generate not 
worse fit of lower lines. 

For comparison see \cite{baba} and references therein.   
    $$\color{red}{15C}$$
There (0,2) and (1,0) are almost degenerated, for above mention reasons 
the (1,0) state is filled up and suitable 
choice of ground state is
$$6(0,2)+4(1,0)+4(1,1)+(1,2)=$$
\[\color{blue}{=108.876}\color{black}{[106.502]}\]
The fit of lower lines is following
$$(1,2-2,1)=
\color{blue}{0.776}\color{black}{[0.7401]}$$
$$(0,2-1,2))=
\color{blue}{3.122}\color{black}{[3.103]}$$
$$(1,2-2,2)+(0,2-0,3)=
\color{blue}{4.247}\color{black}{[4.220]}$$
$$(1,2-2,1)+(0,2-2,1)=
\color{blue}{4.674}\color{black}{[4.657]}$$
$$(1,2-2,2)+(0,2-1,2)=
\color{blue}{4.84}\color{black}{[4.781]}$$

    $$\color{red}{16C}$$
For 16C following ground state is selected as 
$$2(1,0)+6(1,0)+3(1,1)+5(0,3)=$$
\[\color{blue}{=112.792}\color{black}{[110.753]}\]
         Adjusting schema for lower
lines is
$$2(1,0-0,2)+(0,3-0,4)=
\color{blue}{1.752}\color{black}{[1.766]}$$
$$2(0,3-1,2)+(0,3-1,3)=
\color{blue}{3.055}\color{black}{[3.027]}$$
$$3(0,3-2,1)=
\color{blue}{3.942}\color{black}{[3.986]}$$
$$(1,1-1,2)+(1,1-2,1)=
\color{blue}{4.079}\color{black}{[4.088]}$$
$$2(1,0-0,2)+(1,1-1,2)+(1,1-2,1)=
\color{blue}{4.146}\color{black}{[4.142]}$$
The result is acceptable.

For next isotopes the database is insufficient for modeling ground states configuration. In this model the 24C is last nucleus.  

\subsection{Z=7 nuclei}        
$$\color{red}{14N}$$
From experience with previous nuclei it is assumed that two lower states 
are empty, selected  configuration of ground state is
$$(0,2)+2(1,0)+2(1,1)+(1,2)=\color{blue}{=107.048}\color{black}{[104.658]}$$
Note that beginning from 10B even some lower states are reversed and main 
quantum number became usefulness.
Lower acceptable transitions are:
$$(1,2-2,0)+(1,2-2,1)=\color{blue}{2.323}\color{black}{[2.312]}$$
$$2(1,2-2,1)=\color{blue}{3.821}\color{black}{[3.948]}$$
$$(1,2-2,1)+(1,2-1,3)=\color{blue}{4.842}\color{black}{[4.915]}$$
$$(1,2-0,4)+(1,2-1,3)=\color{blue}{=5.159}\color{black}{[5.105]}$$
The comparison with previous fits show that as ground state as set of 
transitions will be somewhat rearranged to reach more natural order of lines 
but we hold this structure.

The some of possible and not observable transitions were skipped but it is 
acceptable result.
    $$\color{red}{15N}$$
For 15N following ground state is selected
$$3(0,2)+2(1,0)+4(1,1)+6(0,3)=$$
\[\color{blue}{=116.339}\color{black}{[115.492]}\]
Adjusting schema for lower lines is following 
$$2(0,2-0,3)=\color{blue}{5.251}\color{black}{[5.270]}$$
$$(1,1-2,1)+2(1,0-1,1)=\color{blue}{5.308}\color{black}{[5.298]}$$
$$(1,1-1,3)+(1,0-2,0)=6.356\color{blue}{}\color{black}{[6.313]}$$
But with switched off electrostatic interaction this fit is unreliable 
because the many variants are possible even for first line choice. 
    $$\color{red}{16N}$$
There lower transitions are small that can be explained by presence of the 
nucleons on an upper state. For this reason chosen configuration of ground 
state is
$$2(0,1)+8(0,2)+6(0,5)=$$
\[\color{blue}{=120.276}\color{black}{[117.989]}\]
Correspondingly, the all lower lines are transtions from (0,5) state, namely  
$$2(0,5-3,1))=\color{blue}{0.135}\color{black}{[0.120]}$$
$$4(0,5-3,1))=\color{blue}{0.270}\color{black}{[0.298]}$$
$$6(0,5-3,1))=\color{blue}{0.406}\color{black}{[0.397]}$$
$$2(0,5-3,1))+(0,1-0,2)=\color{blue}{3.368}\color{black}{[3.353]}$$
There is other possibility with compact ground state 
but in this case the transitions are exotic. 
    $$\color{red}{17N}$$
Acceptable configuration for ground state is 
$$2(0,2)+9(1,1)+3(0,3)+3(1,2)=$$
\[\color{blue}{=124.189}\color{black}{[123.965]}\]
while transitions can be following 
$$(1,2-1,3)=\color{blue}{1.459}\color{black}{[1.373]}$$
$$2(1,2-2,0))+(1,2-1,3)=\color{blue}{1.826}\color{black}{[1.849]}$$
$$(1,2-2,2)=\color{blue}{1.903}\color{black}{[1.906]}$$
$$(1,1-2,1)=\color{blue}{2.604}\color{black}{[2.526]}$$
$$(1,1-1,3)=\color{blue}{3.128}\color{black}{[3.128]}$$
For comparison another variant is printed
$$2(1,0)+(1,1)+6(0,3)+4(1,2)=$$
\[\color{blue}{=126.641}\color{black}{[123.965]}\]
$$(1,0-1,1)=\color{blue}{1.354}\color{black}{[1.373]}$$
$$(1,1-2,0)=\color{blue}{1.851}\color{black}{[1.849]}$$
Next three lines are alike.
    $$\color{red}{18N}$$
Chosen ground state is 
$$4(1,0)+8(1,1)+(0,3)+4(1,2)+(3,0)=$$
\[\color{blue}{=127.953}\color{black}{[126.695]}\]
while the transitions can be following 
$$(1,2-2,0)=\color{blue}{0.143}\color{black}{[0.114]}$$ 
$$4(1,2-2,0)=\color{blue}{0.573}\color{black}{[0.587]}$$
$$(3,0-3,1)+2(1,2-2,0)=\color{blue}{0.745}\color{black}{[0.747]}$$
$$(3,0-4,1)+2(1,2-2,0))=\color{blue}{1.703}\color{black}{[1.734]}$$
For small first line best fit can be found if the nucleons are on the higher shells but this looks ambiguously.
    $$\color{red}{19N}$$
Selected ground state is 
$$4(0,2)+8(1,1)+3(1,2)+4(2,0)=$$
\[\color{blue}{=132.653}\color{black}{[132.023]}\]
Suitable fit of lines can be following 
$$(0,2-1,0)+(1,1-0,3)=\color{blue}{1.187}\color{black}{[1.11]}$$
$$(1,1-1,2)=\color{blue}{1.672}\color{black}{[1.65]}$$
$$(0,2-0,3)=\color{blue}{2.581}\color{black}{[2.54]}$$
$$(1,1-0,3)+(1,1-2,1)=\color{blue}{3.528}\color{black}{[3.47]}$$
$$4(1,1-0,3)+(0,2-1,0)=\color{blue}{4.189}\color{black}{[4.18]}$$
On the today these are all measured lines. 
     $$\color{red}{20N}$$
For 16N, 18N and 22N the first observed lines are small, near ~0.1MeV; the 
same can be assumed for 20N. In this nucleus theoretical values for 
transition (0,2-1,0)=0.05; in fact these states are degenerated; but small excitations can exist at nucleon transitions between 
upper states. 
In this case the absence of measured lines can be explained 
as: (1,0) state is filed up, (0,2) state is almost 
filled, the rest of nucleons are placed on upper 
states with such small energies that not tuned perturbation 
destroy 20N isotope.   
     $$\color{red}{21N}$$ 
Selected ground state is compact, namely 
$$5(0,2)+12(1,1)+4(2,0)=$$
\[\color{blue}{=139.14}\color{black}{[138.768]}\]
Suitable fit of lines can be following 
$$(1,1-0,3)=\color{blue}{1.127}\color{black}{[1.160]}$$
$$(1,1-2,1)=\color{blue}{2.364}\color{black}{[2.300]}$$
$$2(1,1-1,2)=\color{blue}{3.311}\color{black}{[3.30]}$$
$$(1,1-0,3)+(0,2-0,3)=\color{blue}{3.602}\color{black}{[3.600]}$$
$$(0,2-0,4)=\color{blue}{4.115}\color{black}{[4.170]}$$
that embraced today measured lines 
     $$\color{red}{22N}$$
 Supposed structure of ground state is following 
 $$12(0,2)+8(1,1)+(0,3)+(0,4)=$$
\[\color{blue}{=143.79}\color{black}{[140.05]}\]
Suitable fit of lines is following 
$$(0,4-1,3)=\color{blue}{0.172}\color{black}{[0.183]}$$
$$(0,4-1,4)=\color{blue}{1.050}\color{black}{[1.014]}$$
$$(0,3-2,2)=\color{blue}{1.921}\color{black}{[1.93]}$$
Those are all today measured lines. 

There are best fit as ground state as lines, chosen 
structures correspond one neutron halo picture for 22N. 

For next - 23N, 24N, 25N - isotopes any line is not 
presented in the database and binding energies 
are uncertain. In this model 27N is last nuclide. 

\subsection{Z=8 nuclei}        
$$\color{red}{16O}$$
Previously selected ground state for 16O is taken 
$$2(1,0)+4(1,1)+2(2,0)=$$
\[\color{blue}{=130.271\;}\color{black}{[127.619]}\]
Here are other configurations with best fit of binding energy but the  
selecting one cause the natural order of lower excitations as well as 
alpha-clustering in the ground state.

The spectrum of 16O is unusual for big value of first 
line - more of 6MeV; previously chosen scheme for adjusting transitions to observable lines is applying, namely 
$$ 2(1,1-0,3)+(1,1-1,2)=\color{blue}{6.0}\color{black}{[6.049]}$$ 
$$ 2(2,0-2,1)+(1,1-1,2)=\color{blue}{6.182}\color{black}{[6.129]}$$  
$$2(2,0-2,1)+(1,1-2,0)=\color{blue}{6.967}\color{black}{[6.917]}$$
$$2(2,0-1,3)+(1,0-1,1)=\color{blue}{7.033}\color{black}{[7.116]}$$ 
It is easily to find the fit of next lines, however, even 
for printed ones few variants exist. Another common approaches  
as well as references can be found in \cite{Itagaki}
     $$\color{red}{17O}$$
There assumed ground state is following
$$6(0,2)+5(0,3)+6(2,1)\color{blue}{=133.48\;}\color{black}{[131.76]}$$
Lower lines are fitted as
$$(0,3-1,2)=\color{blue}{0.945}\color{black}{[0.870]}$$
$$(0,3-1,2)+(0,3-2,1)=\color{blue}{3.004}\color{black}{[3.055]}$$
$$2(0,3-1,2)+(0,3-0,4)=\color{blue}{3.899}\color{black}{[3.842]}$$
$$(0,3-1,2)+(0,2-1,2)=\color{blue}{4.536}\color{black}{[4.553]}$$
Another variants are possible.
     $$\color{red}{18O}$$ 
Acceptable scheme for ground state and lower lines is following
$$4(0,2)+2(1,0)+4(0,3)+4(1,2)+4(2,1)\color{blue}{=140.328\;}\color{black}{[139.88]}$$
and
$$(0,3-0,4)=\color{blue}{2.005}\color{black}{[1.982]}$$
$$(1,2-2,1)+(0,3-1,3)=\color{blue}{3.562}\color{black}{[3.554]}$$
$$(1,2-2,1)+2(1,0-1,1)=\color{blue}{3.637}\color{black}{[3.633]}$$
$$(0,3-1,2)+(0,3-2,2)=\color{blue}{3.957}\color{black}{[3.920]}$$
But for this nucleus many acceptable structures can be found.
     $$\color{red}{19O}$$ 
Because of small observable line the structure of this nucleus is difficult to handle. Two transitions can be 
suitable for fitting first line, namely
$$(0,6-1,5)=\color{blue}{0.114}\color{black}{[0.096]}$$
$$(4,2-5,0)=\color{blue}{0.095}\color{black}{[0.096]}$$
In first case it acceptable for proton transitions but 
if in (0,6) state is single nucleon this transition 
disagree with observed parity. In second case it is acceptable for neutron transition but the change of 
orbital momentum is enough for suppression this line. 
Second variant is chosen so 19O is one-neutron halo 
nucleus. 

Assumed ground state is 
\[3(0,2)+2(1,0)+2(1,1)+5(0,3)+2(1,2)+2(2,1)+(4,2)\]$$\color{blue}{=144.422\;}\color{black}{[143.763]}$$
while lower transitions are
$$(4,2-5,0)=\color{blue}{0.095}\color{black}{[0.096]}$$
$$2(1,1-0,3)=\color{blue}{1.472}\color{black}{[1.471]}$$ 
$$2(2,1-3,0)=\color{blue}{2.374}\color{black}{[2.371]}$$
Although the number of possible variants growth with nucleon 
number  in the nuclei we continue the consideration of nuclei 
structures up to Z=11 chain. 
     $$\color{red}{20O}$$ 
For this nucleus chosen configuration is 
\[4(0,2)+8(0,3)+8(1,2)\]$$\color{blue}{=151.851\;}\color{black}{[151.371]}$$
that is extra compact because only few lowest lines 
are fitting.

Acceptable fit of lower lines is 
$$2(0,3-1,2)=\color{blue}{1.752}\color{black}{[1.673]}$$
$$2(1,2-1,3)+(0,2-1,0)=\color{blue}{3.580}\color{black}{[3.570]}$$
$$2(1,2-2,2)=\color{blue}{4.055}\color{black}{[4.072]}$$

     $$\color{red}{21O}$$ 
Acceptable fit of ground state, again because of first excitation choice, is following 
\[(0,2)+2(1,0)+6(1,1)+8(0,3)+3(1,2)+(2,2)\]$$\color{blue}{=156.931\;}\color{black}{[155.177]}$$
That looks as one-halo nucleus.

The fit of lower lines is
$$(2,2-3,2)=\color{blue}{1.232}\color{black}{[1.220]}$$
$$2(0,3-1,2)+(1,0-1,1)=\color{blue}{2.143}\color{black}{[2.133]}$$
$$(1,2-0,4)+(0,3-0,4)=\color{blue}{3.031}\color{black}{[3.026]}$$
$$2(1,2-2,1)+(1,2-0,4)=\color{blue}{3.106}\color{black}{[3.073]}$$
In the case if assumed first excitation is
$$(1,0-1,1)=\color{blue}{1.320}\color{black}{[1.220]}$$
the halo will be absent. 
     $$\color{red}{22O}$$ 
Compact ground state is selected, namely
\[2(1,0)+8(1,1)+12(0,3)\]$$\color{blue}{=162.830\;}\color{black}{[162.028]}$$
Today measured lines are adjusted as
$$2(0,3-1,2)+(0,3-2,1)=\color{blue}{3.234}\color{black}{[3.199]}$$
$$2(0,3-0,4)+(0,3-1,2)=\color{blue}{4.527}\color{black}{[4.582]}$$
$$(0,3-1,3)+(0,3-2,2)=\color{blue}{4.906}\color{black}{[4.909]}$$
$$(0,3-1,3)+(0,3-2,3)=\color{blue}{5.807}\color{black}{[5.800]}$$
$$2(0,3-1,2)+(0,3-0,4)+(1,1-1,3)=\color{blue}{6.543}\color{black}{[6.509]}$$
$$2(1,1-2,1)+(0,3-2,1)=\color{blue}{6.938}\color{black}{[6.936]}$$
For next nuclei in this chain the lines are absent, 
binding energy was measured up to 27O; in this model 
30O is last nuclide.

 \subsection{Z=9 nuclei}           
$$\color{blue}{18F}$$
The configuration of this nucleus is taking as
$$(1,1)+3(0,3)+3(1,2)+2(0,4)=$$
\[\color{blue}{=138.578};\color{black}{[137.369]}\]
Lower lines can be fitted as 
\[(1,1-0,3)=\color{blue}{0.971}\color{black}{[0.937]}\]
\[2(0,4-2,1)=\color{blue}{1.067}\color{black}{[1.041]}\]
\[(1,2-2,0)=\color{blue}{1.089}\color{black}{[1.080]}\]
\[(0,4-1,3)=\color{blue}{1.118}\color{black}{[1.121]}\]
Another structure of ground state as well as transitions assignment 
can be best choice; nevertheless this is acceptable result. It seems, 
the compactness of lower measured 
lines produce the hitch for usual models of this nucleus. In the 
Proc. Roy. Soc. (London) A (1955) the works of J. P. Elliott and 
B.H. Flowers concerning to spectrum of 18F are first successful 
result.

The compactness of 18F levels is suitable for checking any  
theoretical model.
     $$\color{red}{19F}$$ 
This nucleus is similar to 19O but chosen fit of 
lower transitions differ from 19O set, namely 
\[(1,6-2,5)=\color{blue}{0.111}\color{black}{[0.109]}\]
\[(1,6-3,4)=\color{blue}{0.250}\color{black}{[0.197]}\]
\[2(1,3-2,2)=\color{blue}{1.346}\color{black}{[1.3456]}\]
\[(1,6-3,4)+(1,2-2,1)=\color{blue}{1.494}\color{black}{[1.4587]}\]
\[(1,1-1,2)=\color{blue}{1.59}\color{black}{[1.554]}\]
For second line best fit is
$$(1,5-2,4)=\color{blue}{0.1963}\color{black}{[0.1971]}$$
but in this case the contradiction with custom assignment for parity exist 
if in (1,5) state is single nucleon and (2,4) state is empty.
Ground state configuration can be following 
\[4(1,0)+2(1,1)+2(0,3)+8(1,2)+2(1,3)+(1,6)\]$$\color{blue}{=149.358\;}\color{black}{[147.801]}$$

$$\color{blue}{20F}$$
There selected ground state is 
\[3(0,2)+9(0,3)+(1,2)+(2,0)+(0,4)+(2,1)+(1,3)+3(3,0)\]$$\color{blue}{=155.714\;}\color{black}{[154.403]}$$
while lower transitions are
$$(1,3-2,2)=\color{blue}{0.658}\color{black}{[0.656]}$$
$$(0,4-1,3)+(2,1-1,3)=\color{blue}{0.853}\color{black}{[0.8227]}$$
$$(1,2-0,4)=\color{blue}{0.995}\color{black}{[0.983]}$$
$$(0,3-1,2)=\color{blue}{1.074}\color{black}{[1.0568]}$$

$$\color{blue}{21F}$$
Acceptable fit of ground state and lower lines is following
\[2(0,2)+2(1,0)+8(0,3)+4(2,0)+2(2,1)+(1,3)+(3,0)+(0,5)\]$$\color{blue}{=163.618\;}\color{black}{[162.504]}$$
$$(0,5-1,4)=\color{blue}{0.272}\color{black}{[0.2799]}$$
$$2(2,1-1,3)+(3,0-3,1)=\color{blue}{1.112}\color{black}{[1.101]}$$
$$(3,0-3,1)+(1,3-1,4)=\color{blue}{1.730}\color{black}{[1.7304]}$$
$$(2,0-2,1)+(2,0-1,3)=\color{blue}{1.752}\color{black}{[1.7548]}$$
As for previous isotopes diluting structure of ground state is taken because 
the spectral data are abundant. 
$$\color{blue}{22F}$$
There acceptable configuration of ground state is 
\[2(0,2)+4(1,0)+5(0,3)+6(1,2)+(1,3)+(3,0)+3(1,4)\]$$\color{blue}{=168.207\;}\color{black}{[167.7346]}$$
Available measured lines can be fitted as
$$(1,4-3,1)=\color{blue}{0.089}\color{black}{[0.071]}$$
$$(1,4-2,3)=\color{blue}{0.319}\color{black}{[0.310]}$$
$$(1,4-3,2)=\color{blue}{0.720}\color{black}{[0.709]}$$
$$2(1,4-3,2)=\color{blue}{1.440}\color{black}{[1.420]}$$
$$(1,3-2,2)+(0,3,1,2)=\color{blue}{1.623}\color{black}{[1.627]}$$
$$(0,3-1,2)+2(1,4,2,3)=\color{blue}{1.648}\color{black}{[1.632]}$$
$$(0,3-0,4)=\color{blue}{2.039}\color{black}{[2.006]}$$
$$(1,3-2,2)+2(0,3-1,2)=\color{blue}{2.633}\color{black}{[2.571]}$$
$$(1,4-3,1)+(1,4-3,2)+(0,2-2,1)=\color{blue}{2.848}\color{black}{[2.830]}$$
$$(1,4-2,3)+2(1,2-1,3)=\color{blue}{3.376}\color{black}{[3.376]}$$
$$(0,2-1,2)=\color{blue}{3.645}\color{black}{[3.581]}$$
These are all measured lines.

Other choice for first line is possible, namely
$$(0,7-1,6)=\color{blue}{0.077}\color{black}{[0.071]}$$
but with switched off electrostatic interaction the selection is 
unreliable as for first line as for ground state. 
     $$\color{blue}{23F}$$
Lower lines can be fitted as 
$$2(1,3-2,2)+(1,3-1,4)=\color{blue}{2.376}\color{black}{[2.383]}$$
$$2(1,3-2,3)=\color{blue}{3.080}\color{black}{[2.920]}$$
$$(1,3-2,3)+(1,3-3,2)=\color{blue}{3.459}\color{black}{[3.385]}$$
$$2(1,3-3,2)=\color{blue}{3.838}\color{black}{[3.833]}$$
$$(1,3-2,3)+(1,3-2,4)=\color{blue}{3.896}\color{black}{[3.887]}$$
$$2(1,3-1,4)+(1,3-2,3)=\color{blue}{4.020}\color{black}{[3.986]}$$
About ground state structure we can only said that in (1,3) state are 
at least three  nucleons. 
     $$\color{blue}{24F}$$
Ground state configuration can be following 
\[(0,2)+(1,0)+6(1,1)+10(0,3)+5(2,1)+(1,3))\]$$\color{blue}{=179.916\;}\color{black}{[179.112]}$$
while two available measured lines can be fitted as 
$$(1,3-2,2)=\color{blue}{0.534}\color{black}{[0.521]}$$
$$2(0,2-1,1)=\color{blue}{1.896}\color{black}{[1.831]}$$
It is testing job.
     $$\color{blue}{25F}$$
Ground state can be following 
\[(1,0)+6(1,1)+12(0,3)+6(1,2)\]$$\color{blue}{=183.869\;}\color{black}{[183.875]}$$
Four measured lines can be fitted as 
$$2(1,1-1,2)=\color{blue}{3.318}\color{black}{[3.300]}$$
$$(0,3-1,2)+(0,3-0,4)=\color{blue}{3.762}\color{black}{[3.700]}$$
$$2(0,3-1,2)+(0,3-2,2)=\color{blue}{4.459}\color{black}{[4.430]}$$
$$(1,1-2,1)+(0,3-2,2)=\color{blue}{5.470}\color{black}{[5.450]}$$
     $$\color{blue}{26F}$$
As example, single measured line can be as 
 $$2(0,4-1,3)=\color{blue}{0.656}\color{black}{[0.665]}$$
but best fit can be found.
     $$\color{blue}{27F}$$
Two measured lines can be fitted as 
$$(2,0-2,1)=\color{blue}{0.764}\color{black}{[0.777]}$$
$$2(0,3-1,2)=\color{blue}{1.254}\color{black}{[1.281]}$$
Third possible line can be 
$$4(0,3-1,2)=\color{blue}{2.507}\color{black}{[?]}$$
$$2(0,3-0,4)=\color{blue}{2.506}\color{black}{[?]}$$
The two last line were printed because once the value 2.50 MeV was 
found in the literature but it is absent in the database.  

In this model the last nucleus in Z=9 chain is 32F.
\subsection{Z=10 nuclei}        
$$\color{blue}{20Ne}$$
Previously selected scheme for as ground state as excitations is 
conserved, namely
$$4(0,3)+2(1,2)+2(2,0)+2(0,4)=$$
\[\color{blue}{=165.334}\color{black}{[160.645]}\]
So for alpha-clustering in this nucleus the spherical symmetry is enough.

Lower excitations can be fitted as
\[(2,0-1,3)=\color{blue}{1.698};\color{black}{[1.634]}\]
$$(0,4-1,3)+(0,4-2,2)=$$
\[=\color{blue}{4.227}\color{black}{[4.247]}\]
$$(2,0-1,3)+(2,0-2,2)=$$
\[=\color{blue}{4.960}\color{black}{[4.967]}\]
$$2(0,4-2,1)+(2,0-0,4)+(2,0-2,2)=$$
\[=\color{blue}{5.598}\color{black}[5.621]\]
There some of possible and not observable excitations pass over.

Because only lower part of the spectrum was adjusted the other, best,  
configurations as ground state as transitions are possible. 
    $$\color{blue}{21Ne}$$
Chosen ground state and lower transitions 
are following
$$8(0,3)+4(1,2)+6(0,4)+3(2,1)=$$
\[\color{blue}{=169.406}\color{black}{[167.4059]}\]
\[2(2,1-1,3)=\color{blue}{0.373}\color{black}{[0.350]}\]
\[(0,4-2,1)+2(0,4-1,3)=\color{blue}{1.774}\color{black}{[1.746]}\]
\[(1,2-2,1)+(0,4-2,2)=\color{blue}{2.778}\color{black}{[2.788]}\]
\[2(2,1-2,2)+(1,2-0,4)=\color{blue}{2.799}\color{black}{[2.794]}\]
\[3(2,1-2,2)=\color{blue}{2.867}\color{black}{[2.866]}\]
This is acceptable results as well as another variants can be found.
       $$\color{red}{22Ne}$$
There are closely spaced lower lines, suitable fit of the ones 
can be following 
\[2(0,4-1,3)=\color{blue}{1.285}\color{black}{[1.274]}\]
\[(0,4-1,4)+(0,4-2,2)=\color{blue}{3.380}\color{black}{[3.357]}\]
\[(0,3-1,2)+2(0,4-0,5)=\color{blue}{4.482}\color{black}{[4.452]}\]
\[(0,2-2,1)=\color{blue}{5.142}\color{black}{[5.146]}\]
\[2(1,0-1,2)=\color{blue}{5.330}\color{black}{[5.331]}\]
\[(0,3-0,4)+2(0,4-0,5)=\color{blue}{5.382}\color{black}{[5.363]}\]
Correspondingly, suitable configuration for ground state can be 
$$(0,2)+2(1,0)+2(0,3)+6(1,2)+4(2,0)+7(0,4)=$$
\[\color{blue}{=178.89}\color{black}{[177.770]}\]
The fit of almost degenerated lines is challenge for all 
existing models of nuclei.  
       $$\color{red}{23Ne}$$
The fit of lower lines can be following 
\[(0,4-2,1)+(0,4-1,3)=\color{blue}{1.010}\color{black}{[1.016]}\]
\[(0,4-2,1)+(1,2-2,1)=\color{blue}{1.708}\color{black}{[1.701]}\]
\[3(0,4-1,3)=\color{blue}{1.855}\color{black}{[1.822]}\]
\[(0,4-2,1)+(0,4-1,3)+(0,4-2,2)=\color{blue}{2.364}\color{black}{[2.315]}\]
\[(0,4-1,4)+(0,4-1,3)=\color{blue}{2.561}\color{black}{[2.516]}\]
\[2(0,4-0,5)=\color{blue}{3.226}\color{black}{[3.220]}\]
\[3(0,4-1,3)+(0,4-0,5)=\color{blue}{3.468}\color{black}{[3.431]}\]
In fact, the transitions from (0,4) state comprised lower lines; 
this left uncertain ground state configuration.      
       $$\color{red}{24Ne}$$
Correspondence between lower lines and transitions can be following  
\[2(2,2-2,3)=\color{blue}{1.979}\color{black}{[1.981]}\]
\[2(0,4-1,4)=\color{blue}{3.820}\color{black}{[3.868]}\]
\[4(2,2-2,3)=\color{blue}{3.958}\color{black}{[3.972]}\]
\[(0,3-0,4)+(0,3-1,3)=\color{blue}{4.796}\color{black}{[4.766]}\]
\[2(0,3-2,1)=\color{blue}{4.882}\color{black}{[4.880]}\]
Assumed configuration of ground state is 
$$4(1,0)+2(0,3)+7(1,2)+3(0,4)+4(2,2)=$$
\[\color{blue}{=193.211}\color{black}{[191.840]}\]
       $$\color{red}{25Ne}$$
There are six measured lines which acceptable fit is 
\[(1,3-2,3)=\color{blue}{1.655}\color{black}{[1.702]}\]
\[(1,3-3,2)=\color{blue}{2.099}\color{black}{[2.090]}\]
\[2(1,3-2,3)=\color{blue}{3.312}\color{black}{[3.316]}\]
\[(2,0-2,1)+(0,2-0,3)=\color{blue}{3.323}\color{black}{[3.324]}\]
\[2(2,1-2,2)+(2,1-2,3)=\color{blue}{3.890}\color{black}{[3.891]}\]
\[(2,0-2,1)+2(1,3-2,3)=\color{blue}{4.014}\color{black}{[4.070]}\]
There are two additional lines of unknown origin, those also can 
be covered naturally as 
\[(0,3-0,4)+(0,3-1,3)=\color{blue}{4.718}\color{black}{[4.070]}\]
\[3(0,3-0,4)=\color{blue}{6.240}\color{black}{[6.280]}\]
However, corresponding restrictions on 25Ne structure are not 
enough for exact knowledge of the one.  
       $$\color{red}{26Ne}$$
For available five definite and one indefinite lines the correspondence to 
transitions can be followed  
\[(1,2-2,0)+(1,2-1,3)=\color{blue}{2.022}\color{black}{[2.018]}\]
\[(1,2-2,0)+2(1,2-1,3)=\color{blue}{3.559}\color{black}{[3.517]}\]
\[(1,2-1,3)+(1,2-2,2)=\color{blue}{3.710}\color{black}{[3.690]}\]
\[(1,2-2,0)+(1,2-2,1)+(1,2-2,2)=\color{blue}{3.858}\color{black}{[3.815]}\]
\[(1,2-2,0)+(1,2-2,1)+(1,2-1,3)+(1,2-2,2)=\color{blue}{5.394}\color{black}{[5.360]}\]
\[4(1,2-2,2))=\color{blue}{8.694}\color{black}{[\simeq 9]}\]
As for 23Ne the transitions from single state comprised available lines. 

Chosen ground state 
$$10(0,3)+16(1,2)=\color{blue}{202.857}\color{black}{[201.52]}$$
is extremely compact that can be caused by small number of measured lines. 
       $$\color{red}{27Ne}$$
For this nucleus two lines were measured. Acceptable fit of the ones can be  
\[2(2,3-3,2))=\color{blue}{0.759}\color{black}{[0.765]}\]
\[(0,3-1,2))=\color{blue}{0.965}\color{black}{[0.885]}\]
       $$\color{red}{28Ne}$$
For this nucleus the measured lines can be adjusted as 
\[(2,1-0,4))=\color{blue}{0.032}\color{black}{[0.0+X]}\]
\[4(2,1-0,4)=\color{blue}{0.1265}\color{black}{[1.127+X]}\]
\[(0,4-1,3)+(0,3-1,2)=\color{blue}{1.3}\color{black}{[1.304]}\]
\[3(0,4-1,3)+2(0,3-1,2)=\color{blue}{3.016}\color{black}{[3.010]}\]
\[2(0,3-0,4))=\color{blue}{3.942}\color{black}{[3.904]}\]
As for 23Ne and 26Ne the transitions from two closely spaced states 
covered measured lines.

Ground state can be following
$$4(0,2)+2(1,0)+9(0,3)+4(1,2)+2(2,0)+3(0,4)+4(2,1)=$$
\[\color{blue}{=208.542}\color{black}{[206.892]}\]
       $$\color{red}{29Ne}$$
Three measured lines can be adjusted as 
\[(1,4-2,3)=\color{blue}{0.222}\color{black}{[0.232]}\]
\[(0,5-2,3)+(1,4-2,3)=\color{blue}{0.614}\color{black}{[0.622]}\]
\[(0,5-0,6)=\color{blue}{0.930}\color{black}{[0.631]}\]
Because the states with near average binding energy per nucleon 
are (1,1) and (0,3) states not compact structure of 29Ne is 
expected.  
       $$\color{red}{30Ne}$$
Two measured lines can be adjusted as 
\[(1,2-2,1)=\color{blue}{0.842}\color{black}{[0.792]}\]
\[(1,2-2,1)+(1,2-1,3)=\color{blue}{2.258}\color{black}{[2.235]}\]
If these lines are the lowest the 30 Ne structure will be compact. 

In this model 35Ne is last isotope. 

Interesting topic is there. In the subsection "The quanta of electric charge" 
it was shown that in nonlinear Coulomb field exist lot of 
elementary electric charges. Taking the module of electron charge, 
"e", for unity the next elementary electric charge is "9e" - this 
was not observed in the area of elementary particles. But Z=9 
nuclei can carry an tracks about "9e" charge.  

For this reason the consideration of light nuclei structure was 
extended up to Z=11 without counting electrostatic interaction 
inside of nuclei. We are thinking that for Z=10 isotopes, 
excluding 20Ne, the excitations of protons can be similar to 
excitations of upper electron in the atoms - it means 
caused by transitions from single state. Chosen fit for 23, 26 and 
28 Ne isotopes show somewhat similar. However, similar ladders 
of transitions occurred below of Z=10.

\subsection{The notions}
There are some question. 

Spin-orbital or orbital interaction shift nucleon levels  
and how both mechanisms can be distinguished theoretically as 
well as in the experiment.  

The description of electrostatic interaction inside of nuclei is 
the task. 

In this model, as well as in the custom ones, the origin of partly 
filled and empty nucleon states is unclear; these features are 
dictated by experiment.   

The exactness of experimental works about nuclei structure exceed 
the one in theoretical works. 

In whole, nuclei theories still are at primary stage.  
\subsection{Remark on the nuclear fusion}
Nuclear fusion problem cannot be rounded. Of the corner  
stone of the nuclear synthesis theory the  attractive 
short-ranged nature of nuclear forces is the main. 
Accordingly, on the big distances the Coulomb force 
exceeds the nuclear ones and so prevents melding of 
the nuclei if the medium temperature is not too big. 

However, nonlinear pionic field in the nuclei, which was found in this article and which is grounded firmly, is of attractive long-ranged type. Correspondingly, free external proton with zero energy will slip on the slope of pionic well but for $Z>1$ the automatic capture of the proton cannot happen because on the slope of pionic well there is an obstacle, which is shifted Coulomb barrier. The bar or reject the proton or, because the superposition of both potentials produce additional small potential well, the photon will be emitted and stopped half-captured proton waits for absorption by nucleus. The probability of the fusion depends on the position and the strength of the barrier. Shifted Coulomb barrier does not appear for theoretical reason. At consideration of light nuclei structure it was quite noticeable that electrostatic interaction into nuclei is suppressed that was treated as strong polarizability of pionic field. For this reason the electrostatic potential can have a singularity in some point $R=R_c$, in this case the one of simplest expressions for Coulomb potential is
\[\color{blue}{\varphi_c(R)=c_1\frac{\alpha(Z-1)}{R-R_c}+\frac{c_2}{(R-R_c)^2}}\]
where $c_1\ll 1$ if $R\ll R_c$, on big distances the usual situation must be restored so $c_1=1$ if $R\gg R_c$. 

For Z=1 nuclei reasonable thinking is that  external electric charge polarize the pionic field of the nucleus and so create shifted electrostatic barrier. 

\noindent Also other situations can be realized. For example, in previous subsection it was argued for presence of one knot in the one particle (proton)  wave function of the triton, $\color{blue}{F\sim(R-R_c)^1}$; the appearance of this knot is  disconnected with electrostatic interaction. However, any zero of wave function can be the suppressor of some singularity in the potential energy. So it is possible that the knot of triton wave function has form $\color{blue}{F\sim(R-R_c)^D}$ where constant $D>1;$ this occur when in the point $R=R_c$ is electrostatic barrier. 
Such situation prevents the consideration of proton wave function square as density of electric charge. 

\noindent As alternative, the electrostatic potential can be taken averaged such as 
\[\color{blue}{\varphi_c(R)\sim1/{|R-\overline{R}|}}\]
or
\[\color{blue}{\varphi_c(R)\sim\big<{{\Psi}\big|\frac{1}{|R-r|}\big|\Psi}\big>_r}\]
etc. The singularity of electrostatic potential is avoidable easily.  For presence of shifted Coulomb bar in Z=1 nuclei many arguments can come out.

Note that in this model the short-ranged pionic force can be found too. Indeed, general solution for potential of free w-field, $\color{blue}{\Delta u(R)\vec e_R=0}$, is
\[\color{blue}{u(R)=c_1/R^2+c_2R}\]
and so pionic field tension is
\[\color{blue}{\varphi'\sim\frac{1}{R^2}exp(-a/R-R^2/b^2)}\]
Within this section it was taken $\color{blue}{c_2=0}$ for conviction that on the big distances the analyticity has to be and so the expanding in $\color{blue}{1/R}$ series can be done for any physical quantity. You can like short-ranged forces but for the ones the Coulomb barrier is shifted also that make the custom situation for building of fusion reactor worse.

For physicists working in the area of nuclear synthesis these circumstances require the attention.

\section{Gluonic field} 

\subsection{Free gluonic field} 

By physical meaning the gluonic 
field is primary field that binds the 
quarks in a hadron. Similarly to 
electromagnetic field this is four 
vector field with gradient symmetry 
and with four potential 
$\color{blue}{G(x)=G_0 \gamma_0 +G_n \gamma_n}$ 
which is restricted by Lorentz gauge condition. 

Physical differences with 
electromagnetic field are following. 
Electromagnetic field may exist in 
three forms: as charged electric field, 
as not charged magnetic field and as transverse 
waves while gluonic field always exist as 
charged field. On infinity the 
electrostatic field 
disappears, on small distances 
both the linear and nonlinear Coulomb 
potentials have singularity. Gluonic 
forces, as it is thinking today commonly, 
do not vanish on infinity. Gluonic
potential has no singularities in 
center of field,  moreover, here  
the forces vanish that is known as 
asymptotic freedom of strong 
interactions. As for any physical 
quantity these
properties of gluonic field and the
field existence itself are grounded 
on data and their theoretical interpretations.
  
We regard the gluonic field as 
classical object in static spherical 
symmetrical state. The potential of the 
one is $\color{blue}{G \gamma_0=g(R)}$ 

Simplest state of any field is the 
free field.  In classical physics, 
the lagrangian of free field always 
is square form of field tensions. 
If we take such lagrangian 
\[ \color{blue}{L_0 \sim (\overrightarrow\nabla g)^2 }\] 
\noindent then  we get the Coulomb-like potential 
\[\color{blue}{g(R)=g_1 +\frac{g_2 }{ R}}\] 
which has infinite self-energy on 
small distances. 
Due to asymptotic freedom of strong 
interaction just the free gluonic 
field act in area of small distances 
and so the Coulomb-like potential 
is not being the  
potential of free gluonic field. We 
must accept that lagrangian of free 
gluonic field has more complicated 
form. If the one depends  upon 
field tension only, then another, not 
Coulomb-like, solution of variation 
task exists. It is 
\[\color{blue}{(\overrightarrow\nabla g(R))^2=constant}\] 
and it has no matter how complicated 
is the lagrangian of free field. 
In this case the potential of free 
gluonic field is following 
\[\color{blue}{g(R)=g_1 +g_2 R}\] 
where the constant $\color{blue}{g_2}$ 
determinate the scale of strong forces. 
It is fundamental quantity similar 
to electric charge in electromagnetic 
interaction. Remark, this is not 
faultless because it is general 
result and the unknown fields may exist. 

As application example, regard the 
bound states of a particle in this 
field. Gluonic field tie up only the
quarks but for simplicity regard a 
scalar particle in this field. For 
particle with mass $\color{blue}{m}$, 
energy $\color{blue}{E}$  and orbital 
moment $\color{blue}{l}$ 
the Klein-Gordon equation for radial part of wave function is following 
\[\color{blue}{F''+\frac{2}{ R}F'=\left [\frac{l(l+1)} { R^2}+m^2 -(E-g_0 -bR)^2 \right ]F    }\] 
Effective potential energy goes to 
minus infinity on big distances so 
this field is the unrestricted source of 
kinetic energy. If we do not believe 
in existence of the one then 
virtual inertial forces need take into account. 

Yet, we regard free gluonic field. 
Radial part of particle wave function 
takes as follow 

\[\color{blue}{F(R) \sim R^l \prod_0^N (R-R_n) exp(AR +\frac{1}{ 2} BR^2)}\] 
Here restriction $\color{blue}{Im E<0}$ is needing because the full wave function contain multiplier 
$\color{blue}{exp(-iEt)}$. 

The solutions exist if the condition 
\[\color{blue}{m^2 =-ib(2N+2l+3)}\] 
is valid. So in this field only 
resonances are being and the square 
of their masses have linear 
dependence upon own spin. 

From experiment such connection 
between the mass and spin of 
resonances is known few ten years. Firstly the 
Regge-pole then string models of 
strong interaction were established taking 
this connection as base \cite{o}. However, 
the simplest tools of semi-classical 
physics enable to sight on 
high-energy physics phenomena. 
 
\subsection{On a quark confinement}

Gluonic interactions have unusual 
property known as confinement - the 
quark as well as a gluon, the 
quantum of gluonic field, - are not 
observable in free states. 
For gluon this is natural 
because the free gluonic field has 
infinite self-energy, this is similar 
to situation with scalar photons 
which are not being seeing because the 
free electrostatic field does not exist. 
Because the mass
of a particle is finite 
the confinement of quarks
is striking phenomenon. In quantum field theory 
special mechanisms evoked for 
confinement of the quarks, \cite{f} and \cite{g}, they
are obscure still. 
At that time, it is hard
to have the doubts that the some dynamics 
for confinement is not presented
in quantum mechanics. Here the example
of confinement in quantum mechanics 
is given.

Let us admit that gluonic field in not 
free state is usual physical field, then 
on infinity the potential of the one 
can be written as
\[\color{blue}{G(R\to\infty)=G_0+\frac{G_1}{ R}+ ...}\]
For simplicity put $\color{blue}{G_0=0
}$ and the potential energy,  
$\color{blue}{V(R)}$, of a quark in gluonic
field cut off as

\[\color{blue}{V(R)=\frac{g_1}{ R}+\frac{g_2}{ R^2}
+\frac{g_3}{ R^3}+\frac{g_4}{ R^4}};\]
Using the Schroedinger equation, 
that is reasonably for heavy quark,
find the states of the 
'quark+gluonic field' system. For easy 
viewing the dependence of the system
spectrum at all,
not one by one, quantum numbers the 
'exact' solutions are searching. So the
potential $\color{blue}{V(R)}$ is
considered as being true on whole R-axis.

It is conveniently, replacing $\color{blue}{F'=fF}$,
convert the Schroedinger equation to
Riccati equation and get 
\[\color{blue}{f'+f^2+\frac{2}{ R}f=
\frac{l(l+1)}{ R^2}+2m\varepsilon +
2m\Big[\frac{g_1}{R}+\frac{g_2}{ R^2}+
\frac{g_3}{R^3}+\frac{g_4}{ R^4}\Big ]}\]
where $\color{blue}{c=1,\;\hbar=1}$ and
$\color{blue}{m,\;\varepsilon,\;l}$ are
the mass, binding energy and orbital
moment of the quark. 
In this representation 
the form of wave function is
well noticeable. It is
\[\color{blue}{f=D+\frac{B}{ R}+\frac{A}{ R^2}+
\sum_{n=0}^{n=N}\frac{1}{{R-R_n}}}\]
or
\[\color{blue}{F=CR^B\prod_{n}
{(R-R_n)}exp\big(DR-\frac{A}{ R}\big)}\]
where $\color{blue}{C,B,R_n,D,A}$
are constants.

The case $\color{blue}{D<0}$ reproduce 
the usual situation, we put 
$\color{blue}{D=0}$. This restriction 
at once carries the condition
$\color{blue}{\varepsilon=0}$, and so the 
motion of a quark in gluonic field is
free anywhere.

Nevertheless, the quark is binding. Indeed,
in case $\color{blue}{A>0}$ with condition $\color{blue}{(B+N+3/2)<0}$
the wave function is square-integrable 
and free motion of the quark take 
place 
in some middle area, on infinity as 
well as in center of the field the 
quark is unobservable.
Uncertain physical meaning this 
situation has. May be the system   is inaccessible for external 
strong interaction 
and so it is 
out of the hadron family.  However, 
the same is not unlikely for any field 
and such wild situations are possible
because for small perturbations the 
invisibility is typical property of 
any quantum system. 
Another interpretation is that the 
quark is a spectator.
External perturbations do not change the
quark energy, they modify the field energy.
Here is the similarity to excitation of an 
electron on internal, not filled, shells
in the atom if the energy of external 
electron is the constant. Of course, 
these do not mean the confinement. 
For realization of the one 
the spectrum of the system, it coincide with
the gluonic field spectrum, must grow
without limit when the numbers $\color{blue}{N,\;l}$ increase.
This is possible.

For calculation of two unknown constants
$\color{blue}{A,\;B}$ we have four 
restrictions from the central 
$\color{blue}{R^{-1},\;R^{-2},\;R^{-3},\;R^{-4}}$ 
sin\-gu\-la\-ri\-ties. 
The va\-lues of $\color{blue}{}R_n$ con\-stants are fixed by 
$\color{blue}{(R-R_n)^{-1}}$ 
sin\-gu\-la\-ri\-ties.
Consequently, the parameters of gluonic potential
are not all free, the two restrictions 
are on the ones.

All restrictions are:

\[\color{blue}{A^2=2mg_4;\;\;\;AB=mg_3;}\]
\[\color{blue}{2A\sum_n \frac{1}{R_n}=B(B+1)-l(l+1)-2mg_2};\]
\[\color{blue}{(B+1)\sum_n \frac{1}{ R_n}
+A\sum_n\frac{1}{ R_n^2}=-mg_1}\]
\[\color{blue}{\sum_{k,\;k\neq n} \frac{1}{{R_n-R_k}}+\frac{A}{{R_n^2}}+\frac{B+1}{ R_n}=0}\]

Summing the last equations we get
\[\color{blue}{A\sum_n \frac{1}{ R_n^2}+(B+1)\sum_n \frac{1}{ R_n}=0}\]
Hence $\color{blue}{g_1=0}$, that is 
impossible for Coulomb field, but for 
gluonic field it is. Like situation 
is for intermolecular interactions 
because the potentials are similar. 

Multiplied the last equations by 
$\color{blue}{R_n}$  
and summing we get
\[\color{blue}{\frac{N(N-1)}{ 2}+A\sum\frac{1}{ R_n}+
(B+1)N=0}\] 

The parameters $\color{blue}{A>0,\;g_2>0,\;g_4>0}$
are regarding as unknown constants 
and so $\color{blue}{B=B(N,l),\;g_3=g_3(N,L);}$

After these simplifications the B-coefficient is
\[\color{blue}{B=-N-\frac{1}{2}-
\sqrt{(l+\frac{1}{ 2})^2+mg_2}}\]
For fixed $\color{blue}{N}$ and big
orbital moment the B-coefficient is
\[\color{blue}{B(l\to \infty)=-l},\]
this solution is unphysical 
for usual boundary conditions.

At first approach the excitations of 
gluonic field can be calculated as the
average value of the field 
self-energies.  
Because of asymptotic freedom the main 
order of contributions to self-energies is
\[\color{blue}{\int_R^{\infty}|g'(R)|^2R^2dR\sim R^{-3}}\]  
The $\color{blue}{N=0}$ states are
simplest for integration of R-degrees 
\[\color{blue}{\overline{R^{-1}}=
\frac{2\|B\|-3}{ 2A}\to\frac{l}{ A}};\]
\[\color{blue}{\overline{R^{-2}}\to
\displaystyle{\frac{l^2}{ A^2}}};\] 
etc. The 
$\color{blue}{g_3\overline{R^{-3}}}$ 
and $\color{blue}{g_4\overline{R^{-4}}}$ 
terms have equal, $\color{blue}{\sim l^4}$,
order and the suitable choice of 
$\color{blue}{g_4}$ constant  makes
the spectrum of field growing when orbital
moment of quark increase. 
Similar situation holds for the
$\color{blue}{N\neq 0}$ states.

Hereby it is shown that in qu\-ntum 
me\-cha\-nics the con\-fi\-ne\-ment phe\-no\-me\-non
is possible, the boundary conditions are 
leading cause of its appearance, the 
growing of a field 
potential on infinity is unnecessary.
Instead the some parameters of the field
 are not constants, a quantization of 
a field arise. The quark is spectator 
which does not take part in 
interaction. For electromagnetic and 
pionic fields the situation is 
opposite - the fields are 
the spectators with constant energy. 
These are because the free fields have
different properties and so different 
boundary conditions exist. It will have 
interest if in some system the roles 
depend upon value of excitation energy.

The picture is much simpler of 
existing in quantum field theory that
does not mean that it is more far from 
reality.

\subsection{Some fourvector fields.}
For effortlessly we took the couple of an abstract fourvector 
physical field with potential $\color{blue}{G(x)}$ and w-field 
which is generated by fourvelocity $\color{blue}{U(x)}$ of that  
physical field, just $\color{blue}{U(x)}$ is potential of w-field. 
The details about w-field can be found in the upper parts of this 
article.

Simplest models for coupling of these two fields are outlined. 
For shortness, below we referring to abstract fourvector physical 
field as gluonic field, of course the one is uncolored. 

The interaction lagrangian for coupled w- and gluonic 
fields can be written as  
\[\color{blue}{L_{int}\sim G\cdot J}\]
where jet fourvector $\color{blue}{J(x)}$ is linear function of 
$\color{blue}{U(x)}$ - it is basic restriction on the current. 
Simplest expressions for gluonic field current  are 
following: 

Cornel-type, $\color{blue}{J=U}$

Pion-like, $\color{blue}{J=[(\nabla \cdot U)G]}$

EM-like, $\color{blue}{J=F_g(x)U(x)F_g(x),\;\;F_g=\nabla \wedge G}$

Only stationary, spherical symmetric solutions of field equations 
are searching.
\subsubsection{Cornel-type gluonic field.}
I seems more simple interaction lagrangian is impossible to find. 
For w-field being in free state the fields equations and their  
solutions are following
\[\color{blue}{\Delta u_0=0,\;\;u_0(R)=b+a/R}\]
\[\color{blue}{\Delta G_0=ku_0,\;\;G_0(R)=c_1+c_2/R+c_3R+c_4R^2}\]
In the case $\color{blue}{c_1=0,\;c_4=0}$ it is Cornel potential 
which is used in hep-theory. However, it is unphysical potential 
because the self-energy of this field is infinite.  
\subsubsection{Pion-like gluonic field}
Repeating the calculations which were done at searching the 
potential of nonlinear pionic field we get 
 \[\color{blue}{\vec u=\frac{a}{R^2}\vec e_R;\;G_0=g_1+g_0exp(-\frac{a}{R})};\]
where $\color{blue}{a,\;g_1,\;g_0}$ are integration constant. 
This field has finite self-energy and so is physical field. 
However, the  self-energy can be as positive as negative number 
dependently on $\color{blue}{g_1/g_0}$ constant value. 
In the first case stable fermion can be formed by field, in second 
case this is impossible. In addition, in first case, if at least 
one integration constant depends on an real numbers, the lot of 
stable fermions exist, at least on the paper. 

The existence of this field can be testing in  nuclear 
physics while possible particles can be identified in hep-physics. 

\subsubsection{Coulomb-like gluonic field}
Similarly, repeating the calculations which were done at searching 
the potential of nonlinear Coulomb field we get nontrivial solution 
of field equations as
\[\color{blue}{u_0=-s+2\sum_{n}\frac{1}{s-s_n}}\]
\[\color{blue}{\dot s=c\prod_{n}(s-s_n)^2 exp(-\frac{s^2}{2});} 
\] 
\[\color{blue}{s_n=2\sum_{i\neq n}\frac{1}{s_n-s_i}}\] 
here all quantities are dimensionless; $\color{blue}{c}$ is 
integration constant; upper dot means differentiation in the 
variable $\color{blue}{a/R}$ while $\color{blue}{s_n}$ are Hermit 
numbers which index number runs $\color{blue}{[-N,-N+1,...,N-1,N]}$ 
set. More details about this solution can be found in the section 
"Nonlinear electromagnetic field". 

At least two stable branches of the field are here. 

The one is $\color{blue}{s<-|s_N|}$. The states of this 
branch are physical if it is possible define positive constant  
$\color{blue}{\delta_{\infty}}$ which restrict the potential in 
 the area of big distances. In this case on the small distances 
the potential growth up to point were  
$\color{blue}{s=-|s_N|-\delta_{\infty}}$, after this point the 
potential is continued as constant. This field can form the 
set of stable fermions. It is analogue of leptons formed by 
nonlinear Coulomb field.  

The second branch is $\color{blue}{s>|s_N|}$. The states of 
this branch are physical if it is possible define positive 
constant  $\color{blue}{\delta_{0}}$ which restrict the potential 
in the area of small distances. In this case, on small distances 
the potential is continual function equal to $\color{blue}{s=|s_N|+\delta_{0}}$ 
 up to an point were it is sewing with Gaussian 
potential which slow growth up to infinity and so ensure 
confinement the particles moving in this potential. This 
gluonic field forms stable fermions which cannot be found in 
free state.  

Nonlinear field quality is the distribution in the space as own 
mass as own electric charge, the last can be produced by nonlinear 
Coulomb field as well as gluonic field  itself - why not? 
In this case the correlation between elementary electric charges 
is the question. 
Maybe both fields are different faces of one field?

For charged field the constants 
$\color{blue}{\delta_{0},\;\delta_{\infty}}$ can be defined  
via electric charge of particle formed by field. For not charged 
particles such possibility is not visible now. 

\subsubsection{Short outlook} 
Either pion-like or coulomb-like gluonic fields can be the main  
force acting in the nuclei. This must be checked. In particular, 
the fermion in fourvector field have two not degenerated ground 
states, the task for investigation. 

Below the quotation from recent work \cite{47} is printed:

"As remarked by Weisskopf [7], several early nuclear models 
suffered from deviating more strongly from data when what appeared 
to be additional correctional elements of physics were added. This 
could be interpreted as being due to some of that physics already
being included in the initial model."  

This statement is true for today, the nuclear theory still is 
on elementary stage. 

Both nonlinear Coulomb field and Coulomb-like gluonic field 
contain the branches with confined potential and so hidden 
leptons, the light twin of the quarks, can be found. 

Where the bosons are? In nonlinear EM field the bosons  
exist, they are photon itself and nonlinear electromagnetic waves. 
Those states are space-like, time depending, without spherical 
symmetry objects. Other nonlinear fields also can have such 
qualities and so can create the bosons.    

This sketch is for young theorists and not young 
experimenters working in corresponding areas of physics.

\section { Clifford algebra} 
This is addition for reader who is not familiar with this algebra. 

Any algebra is richer variety compare with vector space. In algebra the sum and  multiplication of the elements with different algebraic structure are defined. 

Is it possible the extension of  
vector space variety to algebra? 
W. K. Clifford finds the answer in 
1876 year. For this doing it is 
enough regard 
the coordinate vectors as matrices. 

In physics, the space algebra L3 and the 
space-time algebra L4 are the essentials.  
Let us regard their properties briefly. 
Remark, from relativity principle it 
has no matter which coordinate system 
is using. But it 
became as standard to divide  a 
vector on components. Which troubles 
this dividing create easy is seeing 
on example switching interaction 
of the electron with external magnetic field in quantum mechanics.  We avoid such way. Then the 
flat coordinate system is using in 
general case (of course, the existence of the gravitation 
which deformed the space is ignored). Only when numerical 
calculations are doing the suitable coordinates are taking. 

In space algebra the coordinate vectors (orts, basic vectors) $\color{blue}{\vec e_n}$, are equal to two dimension Pauli matrices with following properties 
\[\color{blue}{\vec e_n=\sigma_n}\] 
\[\color{blue}{\vec e_i \vec e_k+\vec e_k\vec e_i=0;\;\;i\neq k}\]
\[\color{blue}{\vec{e_n}^2 =1; \; { n=1, 2, 3 }}\] 
\[\color{blue}{\vec e_1\vec e_2 \vec e_3=i 1}\] 
The last matrix change sign at parity transformations, the ones 
are $\color{blue}{\vec e_n \to -e_n}$,  so the imaginary unite 
of the complex numbers algebra at that time is the pseudoscalar of space algebra. 
Then general element in the space 
algebra is the sum of scalar, pseudoscalar, vector, and pseudovector. 

The gradient operator in L3 algebra is following 

\[\color{blue}{\overrightarrow\nabla =\vec e_n \partial_n}\]

Few examples of calculations in space algebra. 

$\color{blue}{\vec a\vec b=a_n b_k\vec e_n\vec e_v=\vec a\cdot\vec b +i\vec a 
\times \vec b}$\par 
$\color{blue}{\overrightarrow\nabla(\vec a\vec b)=(\overrightarrow\nabla\vec a)\vec b-\vec a(\overrightarrow\nabla\vec b)+ 
2(\vec a\cdot\overrightarrow\nabla)\vec b}$\par 
$\color{blue}{\overrightarrow\nabla R^N\vec e_z=NR^{(N-1)}\vec e_R\vec e_z=NR^{(N-1)} 
(cos\theta -i sin\theta\vec e_{\varphi}) }$\par 

In algebra of space-time the coordinate vectors,  $\color{blue}{u_\mu}$, are equal to four dimension Dirac matrices, $\color{blue}{u_\mu=\gamma_{\mu}}$, with following properties 
\[\color{blue}{u_{\mu} u_\nu+u_\nu u_\mu=0;\;\;\mu\neq\nu;\;\;\mu=(0,1,2,3)}\]
\[\color{blue}{u_0^2=1; \; u_s^2=-1; \;  {s=1, 2, 3\;  }}\] 
\[\color{blue}{u_0 u_1 u_2 u_3=i_c}\] 
The last matrix change sign at 
inverse of space as well as time directions, standard denomination of 
this matrix is $\color{blue}{i \gamma_5}$. We use almost the denomination of G. Casanova because here are two pseudoscalars  which 
coincide at passing to space algebra. 
Remark, the existence of two 
pseudoscalars $\color{blue}{i, i_c}$ 
in the space-time algebra commonly is 
missing as implicit standard. In  
fact the space and the space-time 
algebras are the complex varieties.  
General element in the space-time 
algebra is the sum of scalars, pseudoscalars, four vectors, pseudo-four-vectors and bevectors.  

If $\color{blue}{A,B}$ are two four 
vectors then bevector $\color{blue}{F}$ 
is external multiplication of the ones  

\[\color{blue}{F=A\wedge B=(AB-BA)/2}\] 
The matrices  
$\color{blue}{e_n=u_n u_0}$  are the four di\-men\-sion an\-ti-dia\-go\-nal re\-pre\-sen\-tation of Pauli matrices. So any bevector has other form 

\[\color{blue}{F=\vec V+i_c\vec H}\] 
where $\color{blue}{\vec V, \; {  } \vec H}$ are the  space vectors in four 
dimensional representation. This property makes easy the crossing between the space and the space-time algebras. 

The gradient operator in L4 algebra is following 
\[\color{blue}{\nabla=u_0\partial_0 -u_k \partial_k}\] 
\[\color{blue}{\displaystyle{\partial_0=\frac{1 }{ c}\partial_t}}\] 

With common convention about the phases of 
physical quantities the operator of four 
impulse is  
\[\color{blue}{\hat p =i \hbar \nabla }\] 
Note that the definition of this 
operator with opposite sign is using widely.

Few examples of calculations in this algebra 

\[\color{blue}{\nabla A=\nabla u_0 u_0 A=(\partial_0 -\overrightarrow\nabla)(A_0-\vec A)=\nabla \cdot A+\nabla \wedge A}\] 
\[\color{blue}{\nabla\cdot A=\partial_0 A_0 +\overrightarrow\nabla\cdot\vec A}\] 
\[\color{blue}{\nabla\wedge A=-\partial_0\vec A -\overrightarrow\nabla A_0 +i_c\overrightarrow\nabla \times\vec A}\]

For more details see any textbook on 
Clifford algebra, for example \cite{e} and \cite{s}, the links are in \cite{a}.

\section{Coherence condition} 

This is some extension of standard variation formalism.

Formally, this is trivial thing that is more simply understood in examples. 

Regard the scalar field with potential 
$\color{blue}{s(x)}$ in one dimension 
space $\color{blue}{x>0}$. The 
lagrangian of the field take as following
\[\color{blue}{ 
L=\frac{s'^2}{2}+s s'^2}\] 
The variation of the lagrangian is
\[\color{blue}{\delta S=s'\delta s'+s'^2 \delta s+2ss'\delta s'}\] 
It is the sum of few terms and so more 
than one solution of variation task exist. 
Regard some of these solutions. 

In case $\color{blue}{s''=0}$ 
the coherence condition is following
\[\color{blue}{\int(s'^2\delta s+2ss'\delta s') dx=0}\]
where $\color{blue}{s=a+bx}.$ 
The variations in class of linear functions are 
\[\color{blue}{\delta s=\delta a +x\delta b}\]
where  $\color{blue}{\delta a,{ }\delta b}$ are free numbers. Then 
\[\color{blue}{\int\left[b^2(\delta a+x\delta b)+2b(a+bx)\delta b\right ] dx=0}\] 
Hence $\color{blue}{b=0}$ and by  
physical meaning this is the vacuum
 state of the field. 

Another solution is 

$\color{blue}{s''=s'^2}$ 

$\color{blue}{s=a-ln(b+x)}$ 

Coherence condition for this state
\[\color{blue}{\int \left[a-ln(b+x)\right ]\frac{dx}{(b+x)^3}=0}\] 
connect between themselves the integration constants. 

Next state is 

$\color{blue}{s''(1+2s)+2s'^2=0}$ 

$\color{blue}{s+s^2=a+bx}$ 

with coherence condition
\[\color{blue}{b^2\int (\delta a+x\delta b)\frac{dx}{(1+2s)^3}=0}\] 
In this state $\color{blue}{b=0}$ and 
it is another vacuum state because 
the potential constants of states 
differ. 

Generally used equation is 

\[\color{blue}{s''+2(ss')'=s'^2}\]
without any restrictions for integration constants. 

Another example is the lagrangian of scalar static field 
in three dimensions, namely
\[\color{blue}{L=(\vec\nabla \varphi)^2+m^2\varphi^2+2\lambda (\varphi-\varphi_0)^2}\]
The Yukawa solution, which emerge after variation of two first terms, is
\[\color{blue}{\varphi_{y}=\frac{c}{R}exp(-mR)}\]
In this case, holding parameters $\color{blue}{m,\;\lambda,\;\varphi_0}$
fixed the coherence condition is
\[\color{blue}{\delta \int(\varphi-\varphi_0)exp(-mR)RdR=0}\]
so $\color{blue}{c=2\varphi_0/m}$. It is as if additional boundary condition was imposed what determinate the integration constant. 

After these examples the shape up of variation procedure become clear. 
We build some lagrangian. Then, taking any part of the one and following usual variational procedure, we get some system of field equations and solve the ones. Searching after we make the variation of missing part of the lagrangian on the set of solutions of field equations. There are many possibilities, such in last example we can regard constant $\color{blue}{\lambda}$ as variable quantity 
then solutions exist at some connection between parameters. 

Any problem with searching the restrictions imposed on integration 
constant can be solved simply - it is enough regard those constant 
as charges, similarly as it is for electric charge. In this case 
the variation of the charge is equal to zero that ensure solution 
of variation task. In fact this variant is using along the article. 
 
It is unclear for me, because of too simple procedure, is this creation new or old. In any case the section 'Nonlinear electromagnetic field' of this article 
showed that coherence condition is strong tool because it opens the road for bootstrapping. Unfortunately this tool is unknown for most physicists. 

\section{Wave function and boundary conditions} 

For bound states of a particle in 
spherical symmetrical field the 
radial part of wave function take as following 

\[\color{blue}{ 
F=C R^B \prod_{0}^{N} (R-R_i)exp(DR)} 
\] 
where $\color{blue}{C, B, R_i, D}$ 
are the constants. In some cases the some 
of wave function zeros can be not single.

This form of wave function is settled 
on Sturm-Liouville oscillation 
theorem and can be considering 
as a generalization, or as a
simplification, or as an extension of 
the one. 
Remark, in general case the coefficient C is finite, without 
zeros anywhere, analytical function on 
real R-axis which runs to constant on 
infinity. For physicist the infinity means 
the big, compare with some scale, distance. 
If the Taylor series of a function has 
arbitrary radius of convergence on real 
R-axis and on infinity the function is 
equal to constant then this function is 
constant anywhere. 

The analyticity principle forced such 
type of wave function for Coulomb-like 
potentials; although, for example, in case of 
oscillatory potential 
the wave function on infinity has other 
behavior, 
$\color{blue}{\sim exp(DR^2)}$. 
So the main source of information 
about 
a system is contained in polynomial part of 
wave function. For some types of  
differential equations such solutions were known before the quantum 
mechanics appeared. 

At first look this is trivial thing but 
the direct solution of differential 
equation is fruitful because the working
with algebraic equations is more easy. 

Let us determinate the additional boundary 
conditions for wave function in case when 
the potential of a field has a singularity 
in area of small distances. The exact 
mathematical solution of any equation 
is not the exact physical solution of 
the one. Have we deal with linear or 
nonlinear Coulomb field, or with any 
other field, always the area of small 
distances is unknown land. 
In this area all known 
and unknown forces are working. For example, 
the usual Coulomb potential has in center the
singularity, $\color{blue}{e/R}$, which 
is non-physical because 
the one is not existing in nature, so it 
is true to regard the Coulomb potential
on infinity only but not as the one having the 
singularity in center. 
For microscopic system 
the quantities that are visible 
on infinity have physical sense. 
In other words, for such 
systems the boundary conditions 
are invoked on infinity. 

For this doing the Schroedinger equation 
divide on wave function, this means that 
the logarithmic derivative of wave function 
is using, and multiply the expression 
on $\color{blue}{R^{X}}$, where $\color{blue}{X}$ 
is natural number. Then run the distance to 
infinity. Because on infinity the 
C-parameter is constant we 
get $\color{blue}{X+1}$ algebraic 
equations, the ones fixed the 
unidentified parameters of wave function.  

Which must be the number of these restrictions? 
It is must be equal to number of 
unknown parameters. For example, count 
the ones for upper wave function. 
Here are the N unidentified wave 
functions zeros, constants D, B 
and energy E. For this case the 
degree of the multiplier is 
$\color{blue}{X=N+2}$. 

However, we may include in this list 
the parameters of the potential. If 
differential equation determinate all 
unknown constants it is well. If for 
some parameters the solutions did not 
exist then corresponding constants 
are regarding as external quantities. For almost all 
of this paper, the parameters of the 
potential are considering as the externals. 

Taking those boundary conditions 
we avoid the uncertainties connected 
with knowledge of physical quantities 
in not physical area of small distances. 
And this looks, because of simplicity, 
as trivial but only looks. Too widely 
the contradicting correlations between 
the parameters of the potential and 
wave function used to use. From these 
boundary conditions follow that a 
states with more wave function zeros 
contain more information about 
internal structure of the object. 
In other words, the precise 
measurements of energy levels can be 
an replacement to high energy scattering experiments. 

Of course, the things did not have to be so simple. For theoretical 
sciences it is typical to be in Godel area.

\section{As summary} 

This article is grounded on the "bootstrap" idea and 
contains few news. The 
coherence condition and the direct 
solution of Dirac equation are 
technical tools. The w-field 
conception is physical assumption 
and it is working. Remark that any 
field has this w-field as shadow. 
We may regard this model as a 
description of virtual states in 
classical physics, especially if the 
local four impulse take as potential 
of w-field. However, in quantum field 
theory the local  four impulse is the
variable of integration but not the 
potential of a field. 

The methods for elimination of the 
divergences of classical theory may 
be different \cite{p}.
For electromagnetic field the 
continual 
extension of classical field theory 
is almost trivial. For nucleus 
forces in simplest case the situation 
is even simpler than for 
electromagnetic field. However, for 
mechanical medium the model 
is not extension, it is another way and 
by this or other manner this needs 
doing because the mechanical interaction travel 
with finite velocity. 

Also in general case the density of 
energy in any physical field is not 
zero so one more shadow, however 
scalar, field may exist. Therefore, 
no one feedback may be in any 
physical field, just the free 
gluonic field example uses this 
circumstance. J guess these will have 
interest for physicist and will be 
useful.

\section{Acknowledgment} 

The author would like to thank Dr. 
Andre A. Gsponer for given 
possibility upload this article in 
arXiv.org.

\end{document}